\documentclass[structabstract]{aa}
\usepackage{graphicx}
\usepackage{txfonts}
\def\gtrsim{\mathrel{\hbox{\rlap{\hbox{\lower4pt\hbox{$\sim$}}}\hbox{$>$}}}}
\def\lesssim{\mathrel{\hbox{\rlap{\hbox{\lower4pt\hbox{$\sim$}}}\hbox{$<$}}}}
\newcommand{\hi}{H{\sc i} }
\newcommand{\kms}{km s$^{-1}$}
\newcommand{\msun}{M$_{\odot}$$\,$}
\newcommand{\cmsq}{cm$^{-2}$$\,$}
\newcommand{\expmin}{$^{\prime}$ }
\newcommand{\nhi}{N$_{HI}$ }

\newcommand{\al}{AA}

\title{\hi  clouds in the proximity of M33}
  \author{M. Grossi,
          \inst{1}
          \and
          C. Giovanardi\inst{1} \and E. Corbelli\inst{1} \and R. Giovanelli\inst{2} \and M.P. Haynes\inst{2} A.M. Martin\inst{2} \and A. Saintonge\inst{3} \and J.D. Dowell\inst{4}      }

\institute{INAF-Osservatorio Astrofisico di Arcetri, L.go E. Fermi 5, 50125, Florence\\
              \email{grossi@arcetri.astro.it} \and
              Center for Radiophysics and Space Research, Cornell University, Ithaca, NY 14853
              \and  Institute for Theoretical Physics, University of Zurich, CH-8057 Zurich, Switzerland
              \and Department of Astronomy, Indiana University, Bloomington, IN 47405}

\begin{document}

\date{}

%\pagerange{\pageref{firstpage}--\pageref{lastpage}} \pubyear{2008}

%\maketitle
{}

\abstract{}{Neutral hydrogen clouds are found in  the Milky Way and Andromeda halo both as large complexes and smaller isolated clouds.  Here we present a  search for \hi clouds in the halo of M33, the third spiral galaxy of the Local Group.}{We have used two complementary data sets:  a 3$^{\circ} \times 3^{\circ}$ map of the area provided by the Arecibo Legacy Fast ALFA (ALFALFA) survey and deeper pointed observations carried out with the Arecibo telescope in two fields  that permit sampling of the north eastern and south-western edges of the \hi disc. } {The total amount of \hi around M33 detected by our survey is $\sim 10^7$ M$_{\odot}$. At least 50\% of this  mass is made of  \hi clouds that are related both in space and velocity to the galaxy. We discuss several scenarios for the origin of these clouds focusing on the two most interesting ones: $(a)$ dark-matter dominated gaseous satellites, $(b)$ debris from filaments flowing into M33 from the intergalactic medium or generated by a previous interaction with M31. Both scenarios seem to fit with the observed cloud properties. Some structures are found at anomalous velocities, particularly an extended \hi complex previously detected by Thilker et al. (2002). Even though the ALFALFA observations seem to indicate that this cloud is possibly  connected to M33 by a faint gas bridge, we cannot firmly establish its extragalactic nature or its relation to M33.} {Taking into account that the clouds associated with M33 are likely to be highly ionised by the extragalactic UV radiation, we predict that the total gas mass associated with them is $\ge 5\times 10^7$~M$_\odot$. If the gas is steadily falling towards the M33 disc it can provide the fuel needed to sustain a current star formation rate of  0.5 \msun yr$^{-1}$. }


\maketitle

\section{Introduction}

The  origin of \hi High Velocity Clouds (HVCs) is a long-standing problem since their discovery around the Milky Way about 40 years ago (Muller et al. 1963, Oort 1966). Different scenarios have been addressed to explain the nature of these clouds, and here we briefly summarise the ones that are of
particular interest to the interpretation of the data presented herein.

\begin{itemize}

\item The "galactic fountain" model predicts that the clouds have a local origin and they form in the halo from hot gas ejected by supernova explosions in the disc. As the gas flows into the halo it radiatively cools down  and then  falls back onto the disc appearing as high-velocity moving  gas (Shapiro \& Field 1976, de Avillez 2000).

\item Other interpretations suggest that at least a fraction of the observed HVCs have an extragalactic origin. One possibility is that such clouds are the remnant of gas stripped by previous interactions with smaller neighbors. The most direct evidence of such a process is given by the Magellanic Stream between the Milky Way and the Magellanic clouds (Mathewson et al. 1974) which contains about 2 $\times 10^8$ \msun of \hi (at a distance of 55 kpc) (Putman et al. 2003).

\item Another hypothesis for the origin of the HVCs assumes that \hi clouds may constitute the gaseous counterparts of the "missing" dark satellites predicted by galaxy formation theories (Blitz et al. 1999, Braun \& Burton 2000). The structure formation scenario in the Lambda Cold Dark Matter ($\Lambda$CDM) paradigm predicts a larger number of dark matter dominated satellites around massive galaxies, such as  the Milky Way or Andromeda, compared to what is currently observed (Klypin et al. 1999). Although  new very faint Local Group (LG) dwarfs have been discovered in the last few years with the Sloan Digital Sky Survey (SDSS) and other wide field optical surveys (see Simon \& Geha 2007 and references therein), a discrepancy between theory and observations still exists, unless one assumes that such dark halos have not been able to form stars.  They should then contain mainly ionised and neutral hydrogen bound in their dark matter potential wells. Therefore HVCs (particularly the compact HVC class) have been considered in the last few years as possible tracers of the missing dark matter halos.

\item Hot intergalactic medium condensing into galaxies is another possibility to accrete gas and fuel star formation. At low redshift, low mass  galaxies can get part of their gas through a cold accretion mode (Binney 1977, Katz \& White 1993, Fardal et al. 2001, Murali et al. 2002, Keres et al. 2005). In this case the gas is not heated to the virial temperature of the halo but accretes at $T\sim 10^4$ K and establishes radiative equilibrium with extragalactic ionising radiation. In Cold Dark Matter (CDM) models this gas resides in filamentary structures and its accretion rate depends on the environment, being higher for isolated galaxies in low density regions. The gas might enter the virial radius with high speed, of order 100-300~km~s$^{-1}$ and it might  be shocked close to the galaxy disc where it radiates most of its excess energy. Alternatively the gas might convert part of its infall velocity into rotational velocity (see Keres et al. 2005 for a more detailed discussion). In the case of M33 it is unclear whether any of the gas is accreted in this way,  being a galaxy of intermediate mass, and the detection of any warm neutral gas in the halo will help quantify this possibility. This scenario is not very different from the original Oort suggestion (Oort 1970), that HVCs around galaxies are related to residual gas left over from their formation and  gradually accreted by the host galaxy.

\item Finally, in the case of galaxies orbiting a more massive one, such as M33 which is a satellite of M31, gas may be removed from the disc when the galaxy is close to  the pericenter, and then, as the distance increases, the gas might fall back onto the disc. The orientation of the M33 gas warp in the direction of M31 and the shift of the center of the orbits in its outermost parts (Corbelli \& Schneider 1997) might indicate that the M31 disturbances are not negligible on the gaseous disc of its smaller companion, even though the smooth appearance of the M33 stellar distribution constrains  the history of their interaction (see Loeb et al. 2005). Numerical simulations (Jiang \& Binney 1999) show that the warp itself might be generated by a slow gas-accretion process since gas infall reorients the outer parts of the halo.

\end{itemize}

Searching for gas in the halo of %HVCs and \hi complexes in
nearby galaxies can help to better understand the properties of HVCs  and to discriminate between these different scenarios. A population of \hi clouds possibly located in the vicinity of M31  out to a projected distance of 50 kpc
has been recently discovered (Thilker et al. 2004, Westmeier et al. 2005). Some of the \hi features  appear to have a tidal origin, being located  in proximity of the giant stellar stream of M31 and of its satellite NGC 205, while others are spatially isolated with radial velocities that differ from any known dwarf companion.

\begin{figure*}
\begin{center}
\includegraphics[width=10.cm]{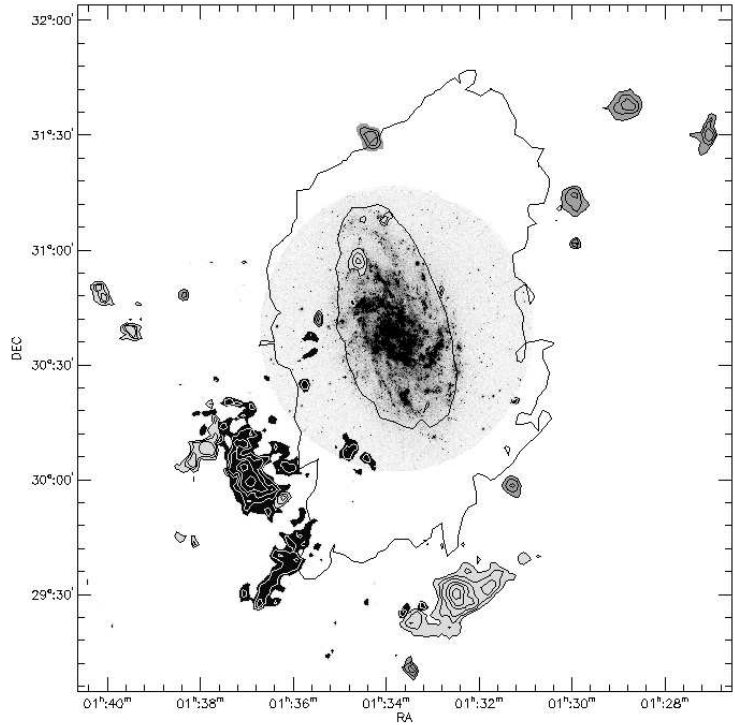} %\newline
\end{center}
\label{map} \caption{The distribution of all the \hi clouds detected around the disc of M33 within the ALFALFA cube. The density contours range from N$_{HI} = 2 \times 10^{18}$ \cmsq, to $3 \times 10^{19}$ \cmsq. {\em Type 1} clouds with $V > -180$ \kms are shown in light grey, {\em Type 1} clouds with $V < -180$ \kms are in dark grey, while {\em Type 2} clouds are plotted in black. The extent of the M33 \hi disc is shown with two contours at a column density of 5 $\times 10^{19}$ \cmsq and $1 \times 10^{21}$ \cmsq.}
\end{figure*}

\begin{figure*}
%\begin{center}
\includegraphics[width=8.cm]{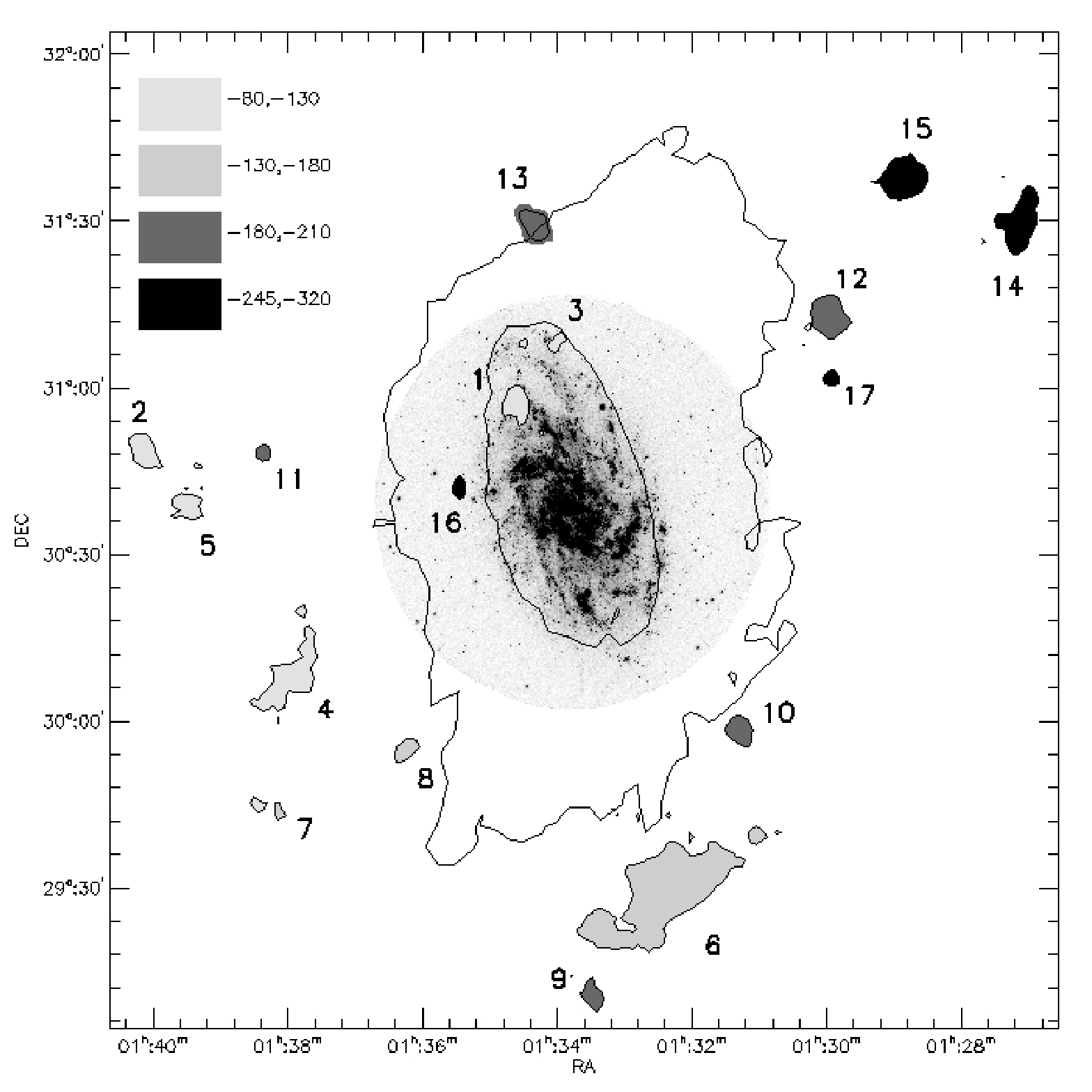}
\includegraphics[width=8.cm]{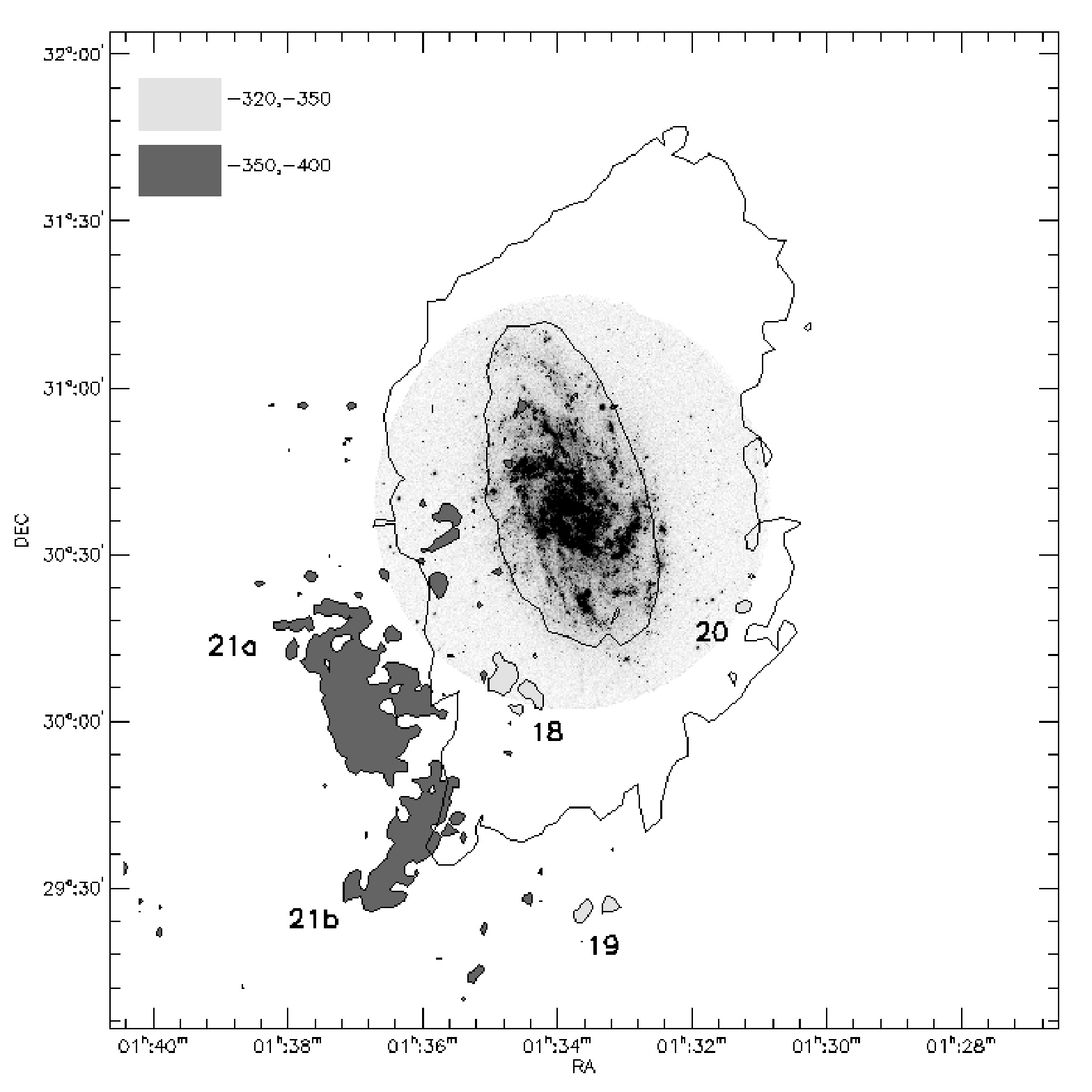}
\caption{{\em Left}: The distribution of {\em Type 1} clouds which have radial velocities compatible with the rotation of the M33 disc  ($-80$ \kms $<V< -320$ \kms). {\em Right}: The distribution of {\em Type 2} clouds whose radial velocities are below -320 \kms .
 The extent of the M33 \hi disc is shown with two contours at a column density of 5 $\times 10^{19}$ \cmsq
 and $1 \times 10^{21}$ \cmsq.}
%\end{center}
\label{map_vel}
\end{figure*}

Contrary to the Milky Way and Andromeda, M33 is fairly isolated, and its recent evolution  seems to have been rather smooth since there is no evidence for  recent mergers; the disc does not appear to be disturbed, there is no prominent bulge, and there are no traces of substructure in the stellar population of its outer regions (Ferguson et al. 2006). At the same time, M33 is a blue galaxy with a much higher star formation rate per unit surface compared to the more massive companion M31.  Finding evidence of extraplanar gas, or gas in the very outer disc, can help constrain the evolutionary scenarios for this galaxy.
Recent results of a chemical evolution model of M33 point out that, to sustain its star forming activity and to explain its shallow metallicity gradient, M33 needs to accrete gas at a rate of about 1 \msun yr$^{-1}$ (Magrini et al. 2007). The 21-cm survey around M31 by Westmeier et al. (2005) was also extended to M33, and it led to the discovery of only one HVC. Moreover, the presence of an arch of carbon stars at the edge of the optical disc  (Block et al. 2004) supports an inside-out formation scenario and suggests that M33 is still in the process of accreting matter which sporadically enhances star formation.

Therefore to search for additional HVCs and to inspect the \hi content of the halo of M33   we have used two complementary sets of data: 1) a $3^{\circ} \times 3^{\circ}$ 21-cm map of the M33 region (centred on the optical position of the galaxy) extracted by the Arecibo Legacy Fast ALFA survey  (ALFALFA; Giovanelli et al. 2005) carried out with the Arecibo telescope; 2) a  map of two $3^{\circ}.5 \times 1^{\circ}.5$ regions in the south-east and north-west part of the galaxy enabling a study of the edge of the \hi disc. The latter data have also been taken with the Arecibo telescope but with an integration time of 6$^{m}$ per pointing ($\sim$10 times longer than ALFALFA) and  higher spectral resolution (5 times ALFALFA).

The main thrust of this paper will be that of testing hypotheses that place the newly discovered clouds at the distance of M33.
We describe their main properties and we discuss the possible origins of these complexes  and the implications  for the evolution of the gaseous disc.

The paper is organised as follows: in Section 2 we describe the data sets we used for this study, in Section 3 we give a brief overview of the results, in Sections 4, 5 and  6 we describe  the clouds detected in both data sets, in Section 7 we present the global properties of the clouds, in Section 8 we discuss some possible formation scenarios and we estimate the ionisation fractions and total gas masses of these clouds, and in Section 9 we present our final conclusions.

\section{Observations}

\subsection{The ALFALFA survey}

Part of  our data set is provided by  the ALFALFA survey, an ongoing blind 21-cm survey  with the Arecibo telescope, to cover $\sim$ 7000 deg$^2$ of the high galactic latitude sky. We will only briefly describe the ALFALFA observations and data reduction procedure, since a more detailed description can be found elsewhere (see Giovanelli et al. 2005, 2007).

The ALFALFA observing strategy consist in a two-pass drift scan technique. The seven beam ALFA receiver is positioned along the local meridian and it records fourteen spectra (two polarizations per beam) every 1 s as the sky drifts over the dish. Each region of the sky is observed at two separate epochs to facilitate the separation of the signal from radio-frequency interference and spurious detections. All the drift scans over a certain region of the sky are then gridded to produce  three dimensional cubes with grid points separated by 1$^{\prime}$ both in RA and DEC. The 100 MHz  bandwidth is divided in 4096 channels providing a velocity coverage between -2000  and 18000 \kms. A data cube consists of 4 grids each stretching over 1024 channels. The M33 cube has been chosen to be $3^{\circ} \times 3^{\circ}$ in size to ensure the coverage of the entire gaseous disc, and given that the systemic velocity of the galaxy is -180 \kms we have restricted our analysis to the first grid ($-2000 <$ V$_{hel} <  3300$ \kms). The corresponding velocity resolution is $\sim$ 5 \kms (increasing to 10 \kms after Hanning smoothing). The average rms in the cube is around 2.5 mJy per channel. This implies a 1$\sigma$ \hi column density sensitivity of $5 \times 10^{17}$ \cmsq over the velocity resolution of 10 \kms.

\subsection{Higher sensitivity ALFA observations}

\begin{table*}
\caption{Catalog of \hi detections in the ALFALFA cube. The columns denote: the cloud identifying number in the catalog, the J2000 right ascension and declination, the heliocentric radial velocity, the velocity FWHM of the 21-cm line, the integrated flux, the \hi mass, the cloud radius in kpc, the projected distance from the centre of M33 in kpc, the virial mass and the Type of the cloud (1 or 2) according to the relative radial velocity with respect to the systemic of M33 (see Section 3).}
\begin{center}
\begin{tabular}{lcccccccccc}
\hline \hline
ID & RA & DEC & V$_{hel}$ & $\Delta$V & S$_{HI}$  & M$_{HI}$ & R & d & M$_{vir}$  & {\em Type} \\
   & J2000 & J2000 & \kms & \kms & Jy \kms & $10^5$M$_{\odot}$ & (kpc)  & (kpc) &  $10^7$M$_{\odot}$ & \\
\hline \hline
%ID &    RA     &    DEC   &   v  & Dv & fl   &  M    &  R    & d  & MHI3s & R3s & Mvir   \\
\al1  & 01:34:36.9 & 30:59:35 &  -83 & 22 & 0.63 &  0.95  & 0.81 & 5.3 & 5.6 & 1\\
\al2  & 01:40:13.2 & 30:50:57 & -107 & 31 & 0.67  & 1.07  & 0.76  & 20.3 & 11.8 & 1\\
\al3  & 01:34:02.9 & 31:12:12 & -111 & 23 & 0.21  & 0.34 & 0.54  & 8.0 & 4.3 & 1\\
\al4  & 01:37:56.4 & 30:10:15 & -122 & 29 & 3.46  & 5.27 & 1.43   & 14.9  & 18.7 & 1\\
\al5  & 01:39:31.4 & 30:41:30 & -128 & 25 & 0.57  & 1.24 & 0.91  & 18.0 & 8.2 & 1 \\
\al6  & 01:32:30.6 & 29:26:56 & -148 & 26 & 12.2  & 24.1  & 1.60   & 18.6 & 16.0 & 1\\
\al7  & 01:38:11.8 & 29:47:11 & -155 & 19 &  0.28 & 0.35 & 0.63   & 19.0  & 2.9 & 1\\
\al8 & 01:36:15.0 & 29:58:10 & -158 & 15 & 0.35  & 0.55& 0.82  &12.7  & 1.8 & 1\\
\al9 & 01:33:27.8 & 29:14:49 & -185 & 22 &  0.38 & 0.64 &  0.58 & 21.0 & 4.2 & 1\\
\al10  &01:31:15.2  &30:01:14  &-188 & 13 & 0.56  & 0.66 & 0.7  & 12.6 & 0.8 & 1   \\
\al11 & 01:38:26.0 &30:50:37  &  -189 & 13 & 0.25 &  0.40 &0.45 & 14.7 & 0.6 & 1\\
\al12  & 01:29:54.6  & 31:16:36 & -194 &15 & 1.08 &1.87 & 1.24  & 15.3  & 2.8 & 1 \\
\al13  & 01:34:16.7 & 31:31:25 &  -196 &25 & 1.45 & 2.60 &1.32  & 12.7 & 12.1 & 1  \\
\al14  &01:27:51.3 & 31:31:21 & -246 & 24 & 0.92 &1.88&0.90   & 22.5 & 7.7 & 1      \\
\al15  &01:28:46.3  & 31:41:02&-264 & 26& 1.25  & 2.54  & 1.12  & 21.9  & 11.3 & 1\\
\al16  &01:35:28.9  & 30:43:17 &-292 & 15& 0.21 & 0.36   &0.53   &  5.2 & 1.0  & 1\\
\al17  & 01:29:52.0  & 31:04:23 & -298 & 14 &0.24  & 0.38  &0.50  & 13.9 & 0.7 & 1\\
\al18   & 01:34:39.9 & 30:09:20 & -326 & 23& 1.21 & 1.11 &1.31 & 17.8  & 7.7 & 2\\
\al19   & 01:33:26.6 & 29:29:42 & -327 & 24&1.56  & 1.30  &1.62 &  7.9  & 14.4 & 2\\
\al20  & 01:31:12.5  & 30:24:04 & -341 & 25& 0.61  & 0.80  & 1.03  &  9.2 & 9.7 & 2\\
\al21a &   01:36:56.6  & 30:02:43   &  -372 &   27   &     6.91  & 11.35  & 3.01  & 13.5  & 33.6 & 2\\
\al21b &   01:36:07.2  & 29:42:39   &  -383 &   25   &     4.62  & 7.7  &2.8  & 15.9  & 26.7 & 2\\
\hline \hline
\end{tabular}
\end{center}
\end{table*}

We have mapped two fields around M33 with the ALFA array on the Arecibo telescope: one to the north-west (field A) and the other to the south-east (field B) (Figure \ref{carlo_clouds}). The size of each field is approximately 5 square degrees. The bottom left corner of field A is at $\Delta$RA = -20$^{\prime}$ and $\Delta$DEC = 0$^{\prime}$ with respect to the center of the galaxy, while the top right corner of field B is at $\Delta$RA = - 20$^{\prime}$ and $\Delta$DEC = -20$^{\prime}$. The data have been obtained in December 2006. The channel spacing is $\sim$ 1 \kms increasing to 2 \kms after Hanning smoothing. With an integration time of 6$^{m}$ per pointing we reach an average rms noise of 1.5 mJy per channel for the 7 ALFA beams. Our observing strategy consisted  in covering the survey area in Sparse Mode, a sampling scheme which allows to map approximately 1/3 of the total survey area (Cordes et al. 2003).
Three adjacent pointings of ALFA are required in a dense sampling scheme to tile the sky with gain comparable to half the maximum gain (Freire 2003), while in Sparse Mode  only one pointing out of these three is made.
This scheme has the advantage that a larger solid angle is covered per unit time, and it has been adopted for a first analysis of the region.  We have carried out 70 pointings in field A and 52 pointings  in field B. Instead of offsetting the telescope to a blank area of the sky for the OFF observations we have taken the OFF pointings in the same field, 7$^m$ apart in RA. Therefore half of the pointings within each field are related to the ON observations, and half to the OFF observations. Calibrations have been performed using the pipeline available at the Arecibo site. To obtain a uniform map of the area despite  the incomplete sky coverage,  the data  have then been gridded using the IDL procedure GRIDDATA, that interpolates scattered data values to a regular grid using the Kriging interpolation method, with grid points separated by 1\expmin both in RA and DEC. The three dimensional cubes related to each field have then been inspected to search for \hi clouds and they have been compared to the ALFALFA data set.

\begin{table*}
\caption{Catalog of \hi detections for the two fields observed with deeper ALFA pointings. The columns denote: the cloud identifying number in the catalog, the J2000 right ascension and declination, the heliocentric radial velocity, the velocity FWHM of the 21-cm line, the integrated flux, the \hi mass, the cloud radius in parsecs, the projected distance from the centre of M33 in kpc and the Type of the cloud (1 or 2) according to the relative radial velocity with respect to the systemic of M33. Here the radius R corresponds to the size of the Arecibo beam in parsecs since the clouds are not resolved. }
\begin{center}
\begin{tabular}{lccclccccc}
\hline \hline
ID & RA & DEC & V$_{hel}$ & $\Delta$V & S$_{HI}$  & M$_{HI}$ & R & d & {\em Type}\\
   & J2000 & J2000 & \kms & \kms & Jy \kms & $10^4$ M$_{\odot}$ & (pc) & (kpc) &\\
\hline \hline
A22  &   01:35:20.4  & 29:25:58    &     -94  &        24 &   0.42      &  7.0      &   $\leq$ 450 & 18.4 & 1  \\
A23  & 01:29:12.2  & 31:13:41  &-266 & 7 &  0.07 & 1.1 & $\leq$ 450 & 16.9 & 1\\
\hline \hline
\end{tabular}
\end{center}
\end{table*}

\section{Overview of the results}

The ALFALFA data and the higher sensitivity observations taken with the ALFA array at the Arecibo telescope reveal a population of  \hi clouds  and  complexes around M33 (Figure 1, Table 1). The galaxy has a systemic heliocentric velocity of -180 \kms; from the amplitude of the rotation curve (Corbelli 2003), one can infer that the maximum velocity a cloud can have to be gravitationally bound to the galaxy is within $\pm$140~km~s$^{-1}$  of the systemic value. Since the detected objects are  within the range $-80$ \kms $<$ V$_{helio}$ $< -400$ \kms, we can distinguish two types of features: {\em Type 1} clouds   are those at velocities comparable with the rotation of the disc, between -80 \kms and -320 \kms; {\em Type 2}  have radial velocities between -320 \kms and -400 \kms and they might not be gravitationally bound to the galaxy. {\em Type 1} clouds can be either extended and spatially connected to the M33 disc or compact and 'discrete' at least down to the resolution of our data sets.
The spatial distribution of the majority of the clouds appears to follow the orientation of the outer \hi disc of M33 (Figure 1), and their kinematic distribution is correlated with its rotation (Figure 2), since  the most negative velocities occur in the NW region and the less negative ones in the SE area of the outer disc.

The surface density maps and the spectra of all the objects detected are shown in the Appendix. In a few cases the spectra of the clouds cannot be fully separated from the bright disc of M33. Due to the confusion with Galactic emission,  it is impossible to isolate clouds with $V > -80$ \kms .

The main properties of the clouds are displayed in Table 1 and 2 for the two different data sets respectively.
For the ALFALFA data (Table 1), the \hi masses  have been calculated from the  following relation $M_{HI} = m_H d_{M33}^2 \tan^2 \phi N_{HI}^{tot}$, where $m_H$ is H atom mass, $d_{M33}$ is the distance to M33, $\phi$ is the angular size of 1 pixel (1$^{\prime}$), $N_{HI}^{tot}$ is the integrated column density above the lowest significant contour density level, according to the rms in each zeroth-moment map of the clouds (see Table A.1). The \hi column densities were derived from the zeroth-moment of the spectra under the assumption that the optical depth of the gas is negligible.
Assuming a distance to M33 of 840 kpc (Freedman et al. 1991) the \hi masses range from $\sim 10^4$ \msun to a few times $10^6$ \msun, with spectral line widths between 15 and 30 \kms .

The sizes of the clouds have been determined by fitting an ellipse to the lowest significant contour density level (see Table A.1), and then taking the geometric mean of the semiaxes. The degree of isolation with respect to M33 was determined on the basis of the lowest value of N$_{HI}$.
Table 1 displays also the projected distance of the clouds from the centre of M33, $d$, the viral masses, and whether they are classified as {\em Type 1} or {\em Type 2}.

In Table 2 we summarise the observational parameters of the additional detections in the higher sensitivity data set.
In this case the clouds are not resolved, therefore we have calculated the neutral hydrogen masses from the relation M$_{HI} = 2.35 \times 10^5 d_{M33}^2$ S$_{HI}$, and the upper limit on the radius R is given by the size of the Arecibo beam in parsecs.

Even though the ALFA beam  has significant sidelobes ($\sim 9^{\prime} \times 13^{\prime}$ at the 3\% level) the ALFALFA data do not show  any clear correlation between the \hi clouds and the brightness of the nearby gaseous disc of M33. However as a further check on the reality of the detections the data was deconvolved using a CLEAN-like (H$\mathrm{\ddot{o}}$gbom 1974) algorithm.  In this procedure the effective ALFALFA beam is modeled on a source-by-source basis so as to account for its position-dependent nature.  The modeling consists of combining maps of the seven ALFA beams that were obtained as part of the A1963 project (Hoffman et al. 2007) with simulations of observations and gridding of a point source located at the same location as the source of interest.  Once the effective beam pattern is known, the data in correspondence of each cloud  are cleaned on a channel-by-channel basis until either a flux density or iteration limit is reached (see Dowell et al. 2008 for a more complete description of the methods used). The cleaning procedure confirmed the reality of all the detections listed in Table 1. However in the remainder of this paper we  use and show the raw data (without applying the deconvolution algorithm) because the procedure produces an increase in the noise level.

\section{Type 1 clouds}

{\em Type 1} clouds have velocities closer to the systemic velocity of M33, therefore they are more likely to be associated with the galaxy and possibly gravitationally bound to it. These clouds appear to be either isolated or spatially connected to the disc, however in the latter case, in order to be included in the catalog, their 21-cm emission must be spatially separated from that of the disc by a minimum spectral range of  10 \kms.

\begin{figure}[h]
\begin{center}
\includegraphics[width=6.5cm]{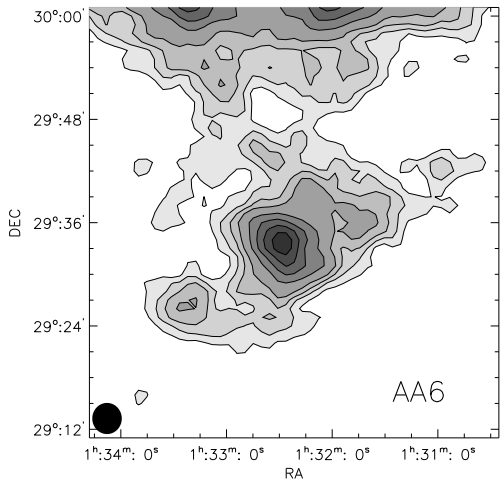}
\caption{Column density map of AA6 (M33HVC1  in Westmeier et al. 2005). Contours are drawn from 3.5, 5.5, 7.5, 10, 15, 20, 25, 30 $\times 10^{18}$ \cmsq. The map includes emission within the velocity range -143 to -169 \kms.} \label{cloud_braun}
\end{center}
\end{figure}

\begin{figure}[h]
\begin{center}
\includegraphics[width=6.5cm]{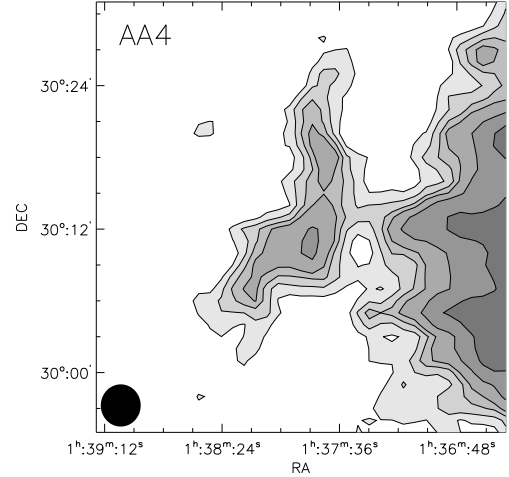}
\end{center}
\caption{Column density map of AA4 integrated over the velocity range -102 to -133 \kms. The cloud contours are drawn at 3, 4.5, 6, 7.5, 10 $\times 10^{18}$ \cmsq, while the border of the M33 disc continues at  15 $\times 10^{18}$ \cmsq.} \label{eastern_filament}
\end{figure}

\subsection{Clouds spatially connected to the disc}

The largest structure is M33HVC1 found by Westmeier et al. (2005) in the southern edge of the disc (AA6  in Table1). The cloud is at about 1$^{\circ}$ south of the centre of the galaxy and it appears to be connected to M33 by a tidal bridge. It has a central core with a peak column density of 3 $\times 10^{19}$ \cmsq and it extends to the south east where there is another substructure at a lower column density ($\sim 1 \times 10^{19}$ \cmsq) (Figure \ref{cloud_braun}). This in Westmeier et al. (2005) appeared as a separate gas clump. The   \hi mass above a column density of 3.5 $\times 10^{18}$ \cmsq is 2.4 $\times 10^6$ \msun, the highest of the whole sample of clouds.

No clear optical counterpart related to the \hi structure has  been identified by McConnachie et al. (2004)  using the Isaac Newton Telescope (INT) survey fields of the area. However, recently an extended globular-like cluster ($\sim 20$ pc) has been discovered with the Subaru telescope in the southern edge of the disc of M33 at a distance of $\sim$ 25 arcmin from the peak of the 21-cm emission  (Stonkute et al. 2008). The stripping of a dwarf satellite  is one of  the possible formation scenarios of extended clusters as it has been suggested for similar objects found in M31 (Huxor et al. 2005). Therefore the possibility that the extended cloud is related to a tidal interaction with a dwarf-like satellite remains open.

\begin{figure}[h]
\begin{center}
\includegraphics[width=8cm]{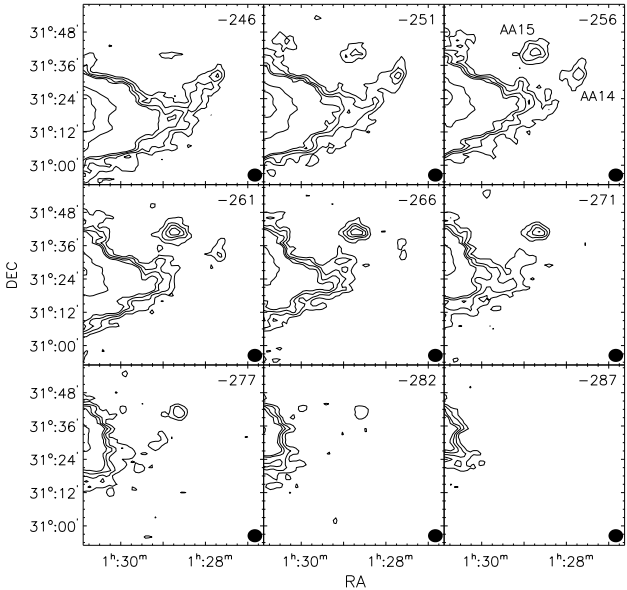}
\caption{Channel maps of the area around AA14 and AA15. Velocity ranges from -246 \kms to -287 \kms in steps of 5 \kms as indicated in the top-right corner of each frame. Contours are at, 7,13,19,25,100,200 mJy. The figure shows the extent of the \hi plume which connects  cloud AA14 to the gaseous disc of M33. The figure shows also the location of AA15 which appears to be the separated from the \hi plume down to the sensitivity level of the ALFALFA data set.} \label{cloud7_chans}
\end{center}
\end{figure}

AA4 is another extended structure with a large \hi mass  found in the eastern edge of the disc at a velocity range between -130 and -100 \kms (see Figure \ref{eastern_filament}). The cloud has a filamentary shape and it extends along the south-north direction from ($\alpha, \delta$) = (01:38:00, 30:00) for approximately 30 arcmin in declination. A smaller  cloud presumably associated with this  feature is found at ($\alpha, \delta$) = (01:38:35, 29:48) (AA7). The cloud shows a gradient in radial velocity along the south-north direction. At $V < -130$ \kms the cloud merges with the emission from the disc. The  \hi mass at half peak brightness amounts to $5.3 \times 10^5$ \msun. Figure \ref{eastern_filament} shows the column density map of AA4  over the velocity range where the emission from the cloud can be separated from that of the M33 disc (from -102 to -133 \kms). Other clouds which connect to the disc are found (AA12, AA13, AA14). They have masses below $10^6$ \msun, and average column densities below $10^{19}$ \cmsq. Their properties are listed in Table 1, and their surface density maps and spectra are shown in Figures A.3 and A.4.
Among these clouds, AA14 is particularly interesting, because it constitutes the final condensation of an extended \hi  plume connected to the disc of M33. It lies at the edge of the extended \hi disc (Figure \ref{cloud7_chans}) where Corbelli, Schneider \& Salpeter (1989) found a disturbed velocity field and an irregular  \hi distribution. The cloud is at about 15 arcminutes from AA15, also shown in Figure \ref{cloud7_chans}, which will be discussed in the next subsection.

\begin{figure}[h]
\begin{center}
\includegraphics[width=6.5cm]{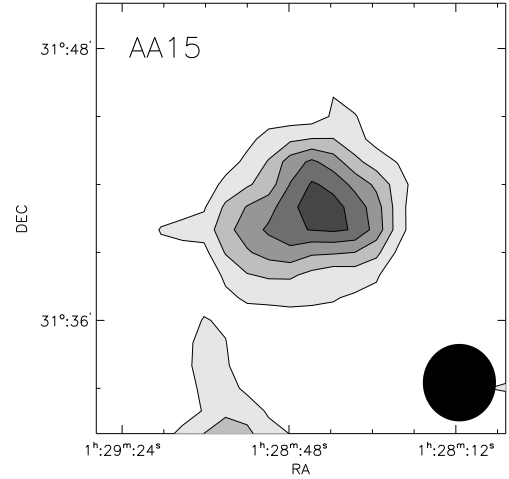}
\end{center}
\caption{Column density map of AA15. Contours are at 3.5, 6, 8.5, 11, 13.5 $\times 10^{18}$ \cmsq. The velocity ranges from -246 \kms to -292 \kms.} \label{cloud7}
\end{figure}

\subsection{'Discrete' clouds}

\begin{figure}[h]
\begin{center}
\includegraphics[width=6.5cm]{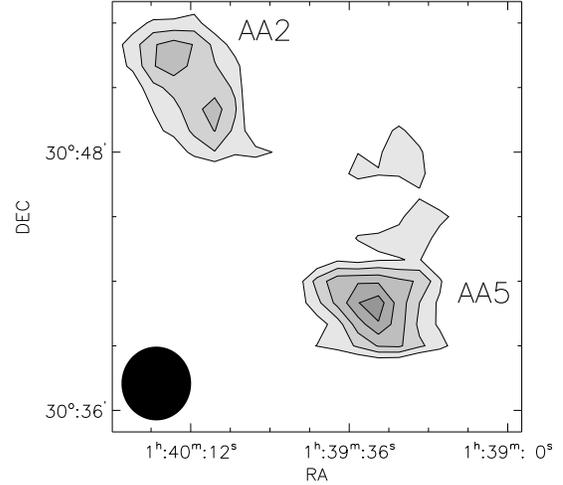}
\end{center}
\caption{Column density map of AA2 (top-left) and AA5 (bottom-right). Contours are drawn from
 3.5 to 8.5  $\times 10^{18}$ \cmsq in steps of 1 $\times 10^{18}$ \cmsq. The map includes emission within
 the velocity range -91 to -153 \kms.} \label{cloud9-10}
\end{figure}

\begin{figure}[h]
\begin{center}
\includegraphics[width=8cm]{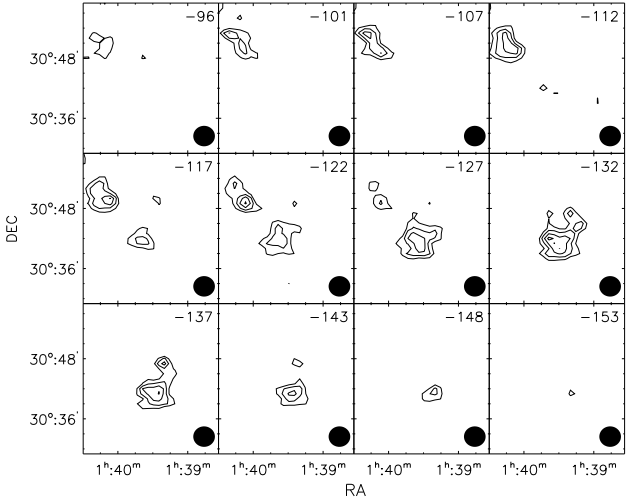}
\caption{Channel maps of AA2 and AA5. Contours are at 7, 10, 13, 16 mJy and the velocity ranges from -96 \kms to -153 \kms in steps of 5 \kms as indicated in the top right corner of each frame.} \label{cloud9-10_chans}
\end{center}
\end{figure}

\begin{figure}[h]
\begin{center}
\includegraphics[width=8cm]{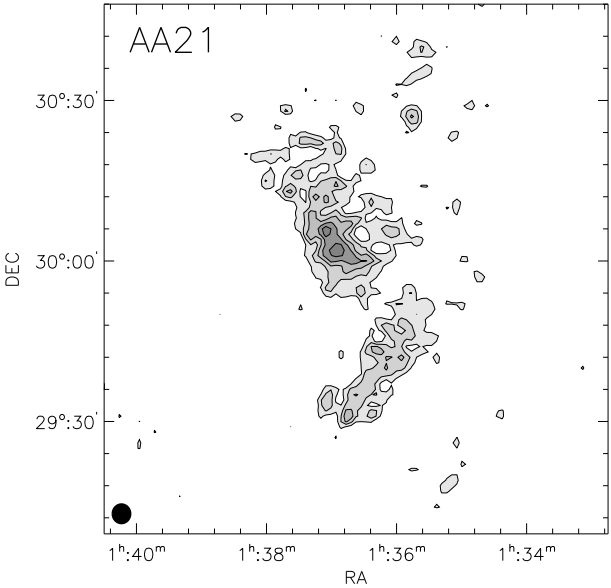}
\caption{Column density map of AA21, the \hi complex with anomalous velocity to the south
 east of M33. Contours are drawn at 3 to 10.5 $\times 10^{18}$ \cmsq in steps of 1.5 $\times 10^{18}$ \cmsq. The \hi emission was integrated over the velocity range -354 to -395 \kms.}
\label{HI_complex}
\end{center}
\end{figure}

We have identified a population of discrete clouds whose \hi emission appears to be spatially well separated  from that of the disc of M33. The majority of them are small barely resolved systems, with M$_{HI} \lesssim 10^5$ \msun. Notable
among these are AA2, AA5 and AA15.

AA15 is located at the edge of the north-western side of the disc ($-245 $ \kms $< V < -285$ \kms). The cloud has been detected in both our data sets and it shows a spherically-symmetric appearance (Figure \ref{cloud7}) with possible evidence for a radial velocity gradient in the east-west direction (see Figure \ref{cloud7_chans}). The cloud has a radius of $\sim 1$ kpc and a mass of 2.5 $\times 10^5$ \msun.
A diffuse and low column density \hi filament has been detected to the north west of M33 towards M31. According to Braun \& Thilker (2004), this diffuse structure appears to be fueling denser gaseous streams and filaments in the outskirts of both galaxies. AA15, together with AA14  lie along the same direction of this bridge and they can possibly represent the higher density condensations of the gas filaments connecting to M31. It is also possible that the both structures may be related to a previous interaction with M31, a possibility that we will discuss in section 8.

Two other fairly isolated clouds are AA2 and AA5 (Figure \ref{cloud9-10}). At a projected distance of about one degree from the centre of M33, they lie in the eastern side of the disc and they are separated by less than 10 arcmin. There is a velocity gradient between the two clouds, as one can see from Figure \ref{cloud9-10_chans} with AA5 being closer to the systemic velocity of M33 with respect to the other cloud. Faint 21-cm emission can also be seen between the two structures and it is likely that they are the highest column density condensation of the same \hi structure. The total \hi mass associated with the two clouds is $2.3 \times 10^5$ \msun.

\section{Type 2 clouds}

The most striking feature appearing from the ALFALFA data cube is the \hi complex that had been serendipitously discovered by Thilker et al. (2002) with Westerbork Synthesis Radio Telescope (WSRT) observations of M33.
The Arecibo telescope beam is 10 times smaller than that of the WSRT (used for that survey  in auto-correlation mode),
thus we can better resolve the structure of this complex.

\begin{figure}[h]
\includegraphics[width=8cm]{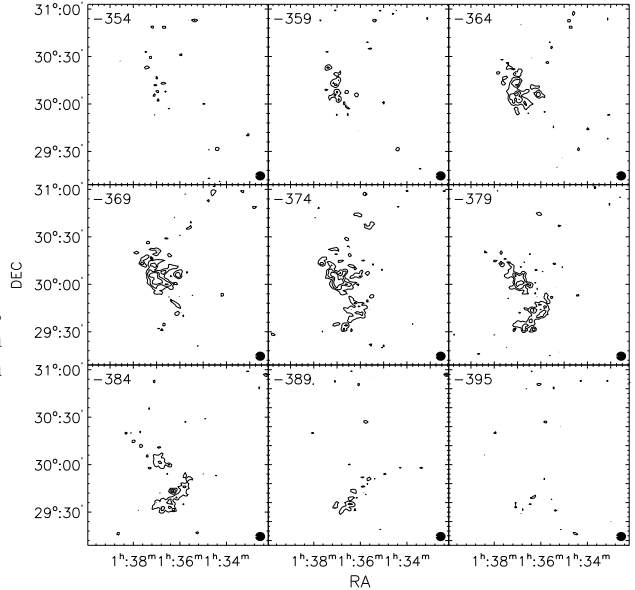}
\caption{Channel maps of complexes AA21a and AA21b. Contours are at 8, 12, 16 mJy, and velocity ranges from -354 \kms to -395 \kms in steps of 5 \kms as indicated in the top-left corner of each frame.}
\label{complex_chans}
\end{figure}

\begin{figure}[h]
\includegraphics[width=4cm]{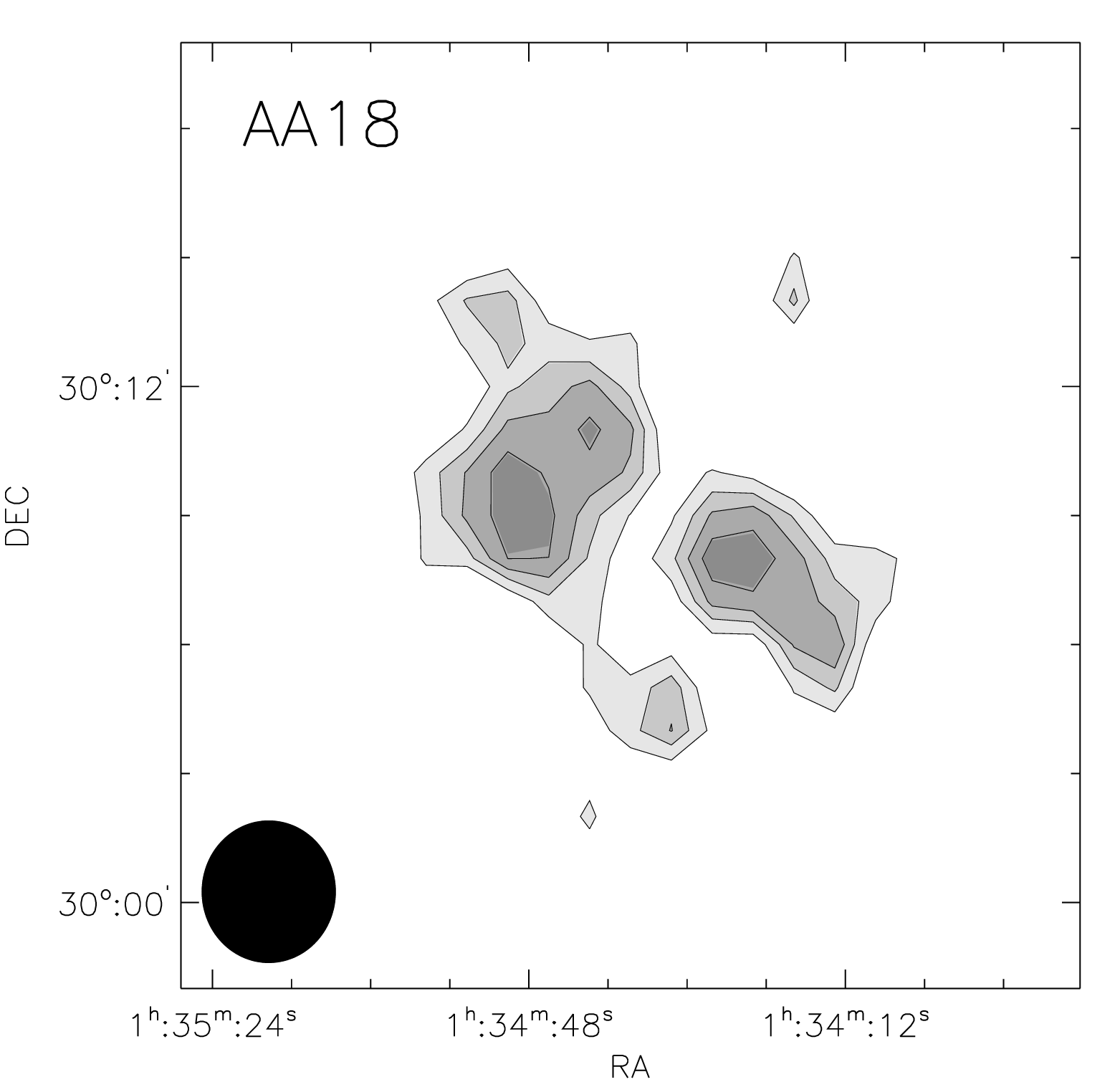}
\includegraphics[width=4cm]{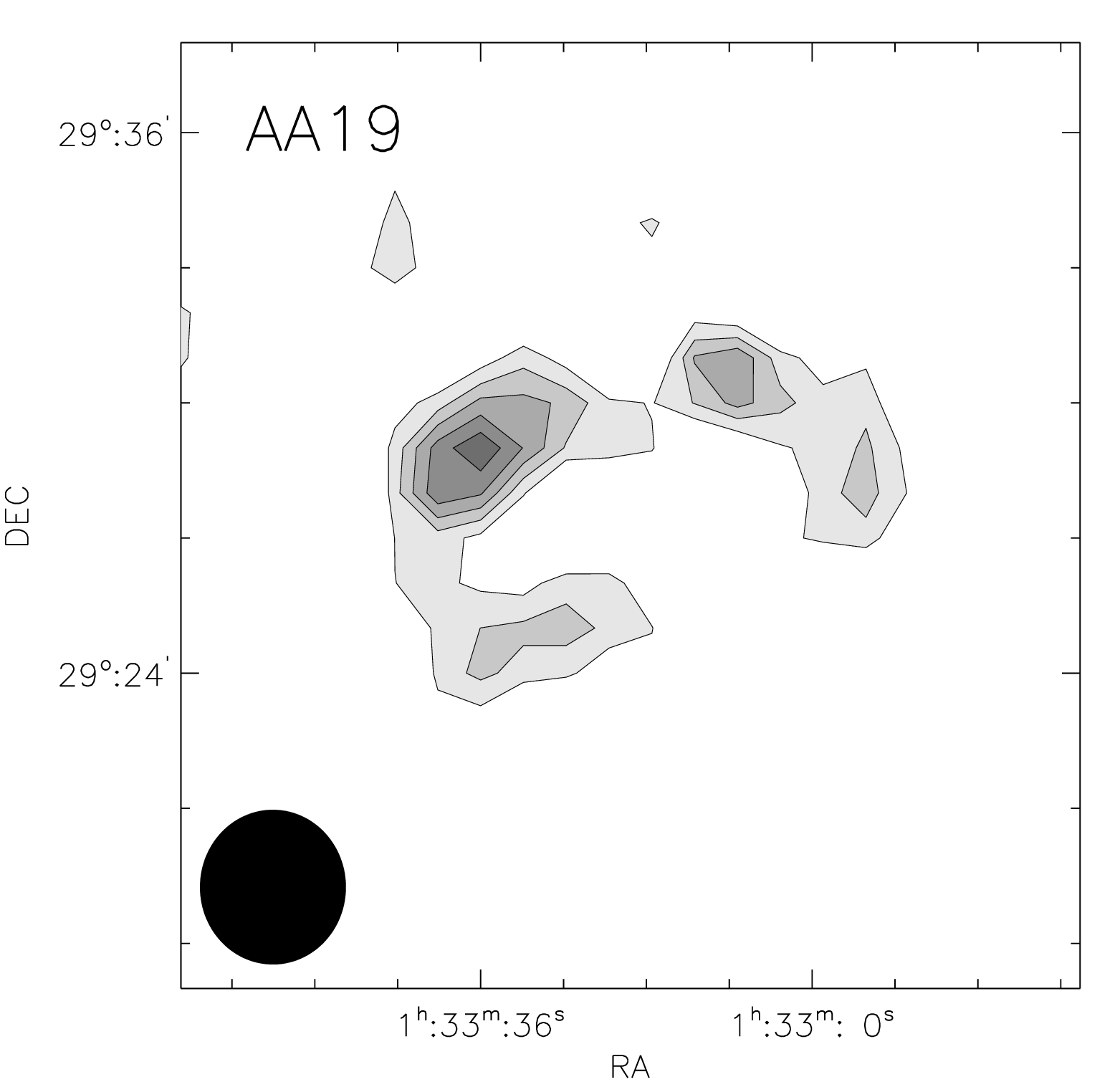}
\caption{Column density maps of the other \hi clouds with anomalous velocity: AA18 (left) and AA19 (right). Contours range from 3 to 5 $\times 10^{18}$ \cmsq in steps of 0.5 $\times 10^{18}$ \cmsq.}
\label{01-02}
\end{figure}

\begin{figure*}
\centering
\includegraphics[width=12cm]{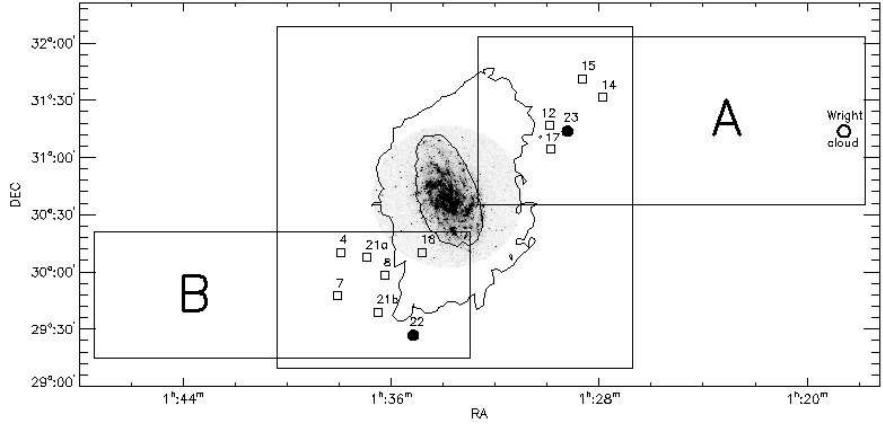}
\caption{A view of the fields observed with higher sensitivity. The $3^{\circ} \times 3^{\circ}$ square is the size of the ALFALFA cube, while the two rectangles indicate the fields where we have carried out deeper pointed observations with the ALFA array. Filled circles correspond to the additional \hi clouds detected with the higher sensitivity data labeled with the ID given in Table 2. The squares indicate the objects detected in both data sets and they are labeled with the ID given in Table 1. The circle at the edge of the northern field gives the position of the peak emission of the tip of the Wright cloud. We also display the Galex image of M33 (near UV filter) and the two contours show the \hi disc at a column density of 5 $\times 10^{19}$ \cmsq and $10^{21}$ \cmsq. }
\label{carlo_clouds}
\end{figure*}

The complex  extends for  one degree in declination with radial velocities within the range --350 \kms $< V_{hel} <$ --400 \kms. Two main substructures (AA21a and AA21b in Table 1)  can be distinguished from the ALFALFA data (Figure \ref{HI_complex}) showing a clear velocity gradient with the northern clump being closer in velocity to the disc ($-350$ \kms $< V_{hel} < -390$ km s$^{-1}$) (Figure \ref{complex_chans}). At the same velocity range we find low column density clumps of gas to the north west of the complex  (\nhi $\lesssim 5 \times 10^{18}$ \cmsq) pointing towards the center of the disc, which may give indication for a possible connection to the disc of M33 (see  Figure \ref{HI_complex}). A faint tail extends  to the east  for about one degree in RA, but this structure is more clearly detected with the deeper ALFA observations (see Figure \ref{HI_complex_mom0_fieldB}).

The mass of the two structures, if placed at 840 kpc, is $1.1\times 10^6$ \msun  and $8\times 10^5$ \msun  respectively, and the column density does not exceed $10^{19}$ \cmsq at the Arecibo beam resolution. The complex is located at $\sim$ one degree from the centre of M33, corresponding to a projected distance of about 15 kpc.

At higher velocities (-310 \kms $<$ V$_{hel} <$ --350 \kms) three previously undetected clouds are found (AA18, AA19, AA20). They are located in the southern area of the  disc, all have similar radial velocity  but they are spatially separated from one another. They are far less extended than the previous \hi complex (with radii below 10\expmin)
and their line widths are between 20 and 25 \kms. Clouds 18 and 19 (Figure \ref{01-02}) have larger masses of around  $10^5$ \msun and show evidence for substructures while cloud 20 is smaller and more compact with a sightly lower \hi mass ($8 \times 10^4$ \msun).
Given their velocities they are at the border between {\em Type 1} and {\em Type 2} classification.

\section{Additional detections with the higher sensitivity ALFA observations}

Figure \ref{carlo_clouds} shows the three regions around M33 that have been observed at Arecibo. The central square corresponds to the region covered by the ALFALFA survey. The two rectangles labeled as A and B indicate the fields where we performed deeper pointed observations with the ALFA array. The integration time was 10 times longer than ALFALFA. As we mentioned in section 2.2 we reach a rms which is at best around 1.3 mJy (at a spectral resolution of 2 \kms), approximately  2 times better than
the ALFALFA data set. This implies a 3$\sigma$ mass limit of 1.3 $\times 10^4$ ($\Delta V$/20 \kms) \msun.

Given the incomplete sampling of fields A and B, to obtain a uniform 21-cm map of both areas, the fluxes at the different pointings have been  interpolated over a regular grid with a spacing of  1$^{\prime}$ both in RA and DEC (see Section 2.2). We have created a three dimensional data cube for each field to search for additional detections with lower \hi masses with respect to the ALFALFA survey ($\simeq  10^4$ \msun).

In Figure \ref{carlo_clouds} we display the clouds that have been identified in both sets of data (squares), and the filled dots correspond to the two additional candidate clouds, one in the southern field (A22) and the other in the northern one (A23).
A22 is located at the border of the low column density disc. Even though its mass (M$_{HI} = 7 \times 10^4$ \msun) is above the sensitivity threshold of ALFALFA, this cloud has not been identified in the survey data cube.
Because of the gaps in our observations, it is not clear from the higher sensitivity data whether we are detecting the edge of the disc or a  cloud, therefore we consider this detection as dubious.
A23 appears to be related to the \hi plume that we have discussed in Section 4.1 both because of its position and velocity. With a mass of $\sim 10^4$ \msun and a very narrow line width, this cloud could not have been detected in the ALFALFA survey.
Both candidate clouds are not resolved, thus we only show their spectra in Figure A.5 without the corresponding contour maps.

Since we only find one low mass cloud within this data set, one can infer that a fully sampled survey of the area with the same sensitivity would not largely increase the number of \hi clouds with M$_{HI} \simeq 10^4$ \msun. We estimate that for an area of 3$^{\circ} \times 3^{\circ}$ a fully sampled database would add approximately 6 clouds in that range of \hi masses.

\begin{figure}
\begin{center}
\includegraphics[width=9.2cm]{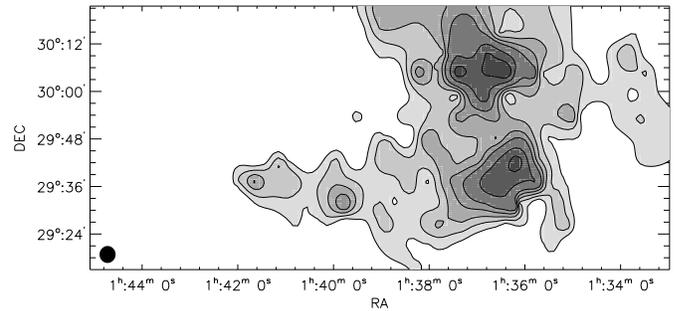}
\end{center}
\caption{\hi column density map of complex AA21 from the higher sensitivity data set. Contours are at 2, 3, 4, 5, 6, 8, 10 $\times 10^{18}$ \cmsq.}
\label{HI_complex_mom0_fieldB}
\end{figure}

In the southern field  we have also identified the extended \hi complex AA21 and we show its contour map in Figure \ref{HI_complex_mom0_fieldB}. As we have already mentioned in  \S 4.1 the two substructures of this complex (AA21a and AA21b) are detected  as one single cloud extending for 8 minutes in right ascension ($\sim$ 2 degrees). Figure \ref{HI_complex_mom0_fieldB} shows the column density map of the complex.
The contours range from 2 to 10 $\times 10^{18}$ \cmsq. Summing the flux of all the pixels within the lowest column density contour we measure a mass of 4 $\times 10^6$ \msun. This value is two times larger than what we have measured
with the ALFALFA survey. This implies that a substantial fraction of the \hi mass resides in low column density
gas. It is also possible that we are overestimating the extension of the cloud since the eastern edge  at RA$>$12$^h$:41$^m$ could be an artifact due to the presence of two strong background radio sources, B2 0138+29B  and
MG3 J014111+2938, with a flux at 1.4 Ghz of 500 mJy and 236 mJy respectively  (Condon et al. 1998).

We do not find any clouds in the outer edges of the fields both in the northern and southern side. The lack of features in the northern field is notable since here is where the Wright cloud is located, whose possible association with  M33 has been considered in the past (Wright 1979). The northern edge of the complex appears at V$_{hel} \sim -380$ \kms, at about 3.5 degrees from the centre of M33. The absence of a clear connection between the disc of the galaxy and the cloud (see Figure \ref{carlo_clouds}) suggests that this is a distinct complex not related to M33.

\section{Cloud virial masses}

As we noted in Section 5 the more massive clouds (M$_{HI} > 2\times 10^5$ \msun) have an elongated or filament-like shape, and in some cases appear to connect to the disc at a certain velocity suggesting  a tidal origin (see AA4, AA6, AA13, AA14). The 'discrete' clouds instead have masses below $10^5$ \msun, radii smaller than 1 kpc and smaller line widths (except for AA2).
If we assume that these clouds are close to the internal dynamical equilibrium and they are self-gravitating, then the total mass is given by M$_{vir} = (5$R$\sigma^2$/$\alpha$G) where $\alpha$ is a parameter which depends on the cloud geometry and the degree of  virialisation (Bertold \& McKee 1992), and $\alpha \sim 1$  if the clouds are spherical and virialised. For M$_{vir} = f M_{HI}$, the velocity dispersion $\sigma$ (i.e. the full width at half maximum divided by $2\sqrt{2\ln2}$ and corrected for the spectral resolution) can be expressed as a function of M$_{HI}$ and the radius of the cloud R

\begin{equation}
\frac{G M_{HI}}{5R} = \frac{1}{f} \frac{\sigma^2}{\alpha}
\end{equation}

\begin{figure}
\includegraphics[width=8cm]{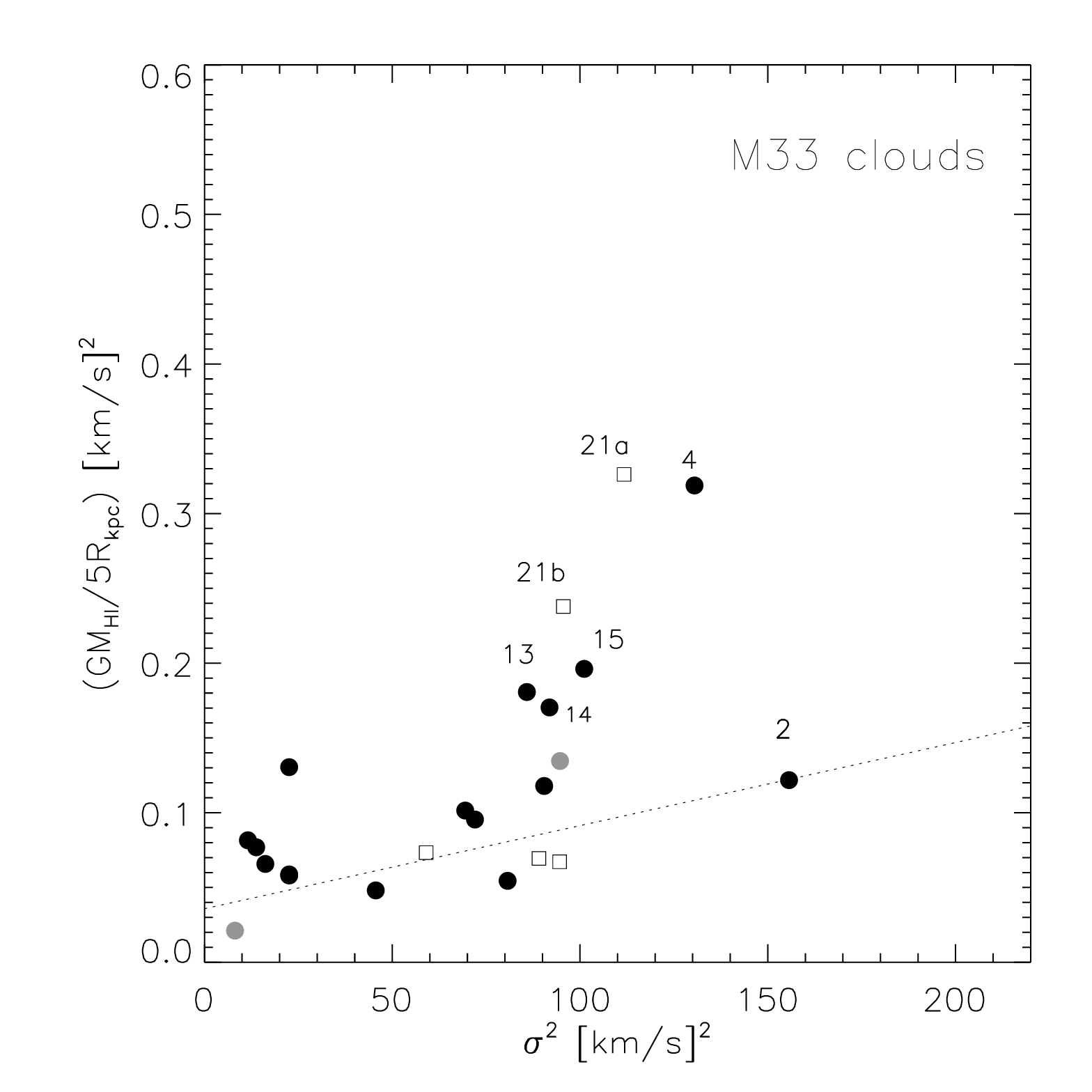}
\includegraphics[width=8cm]{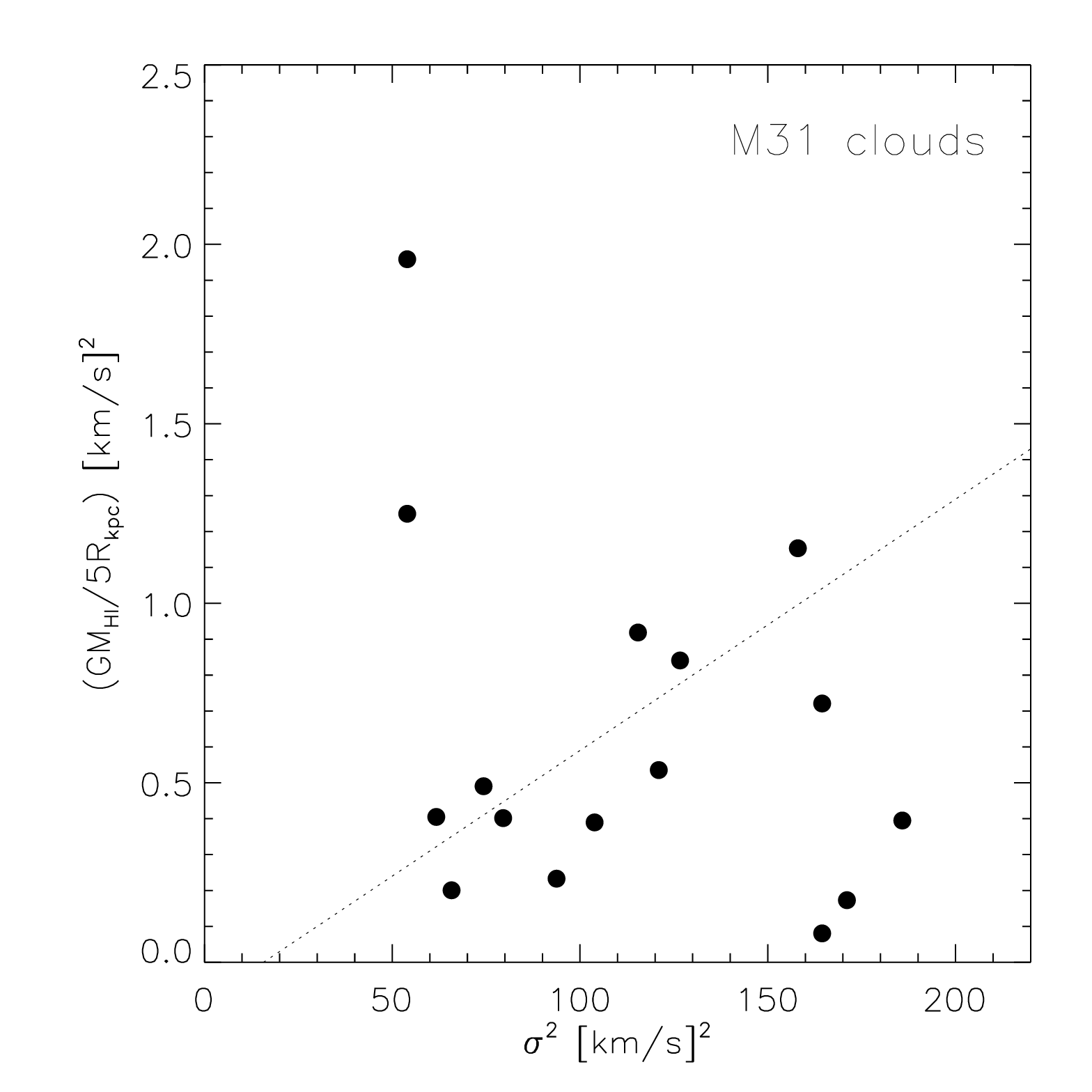}
\caption{{\em Top:} The observed velocity dispersion  versus the ratio of the \hi mass to the radius for  clouds in M33. Filled dots correspond to {\em Type 1} clouds detected in the ALFALFA cube (black) and in the higher sensitivity data set (grey).  Squares indicate the complexes with anomalous velocities. The dotted line corresponds to  a total to \hi mass ratio  $f \sim 2000$. %, while the dotted line indicates $f \sim 400$.
{\em Bottom:} A similar plot for the clouds in M31. The values are taken from Westemeier et al. (2005, Table1). Here the dotted line corresponds to $f \sim 150$} \label{disp_vs_HI}
\end{figure}

Thus,  the relation between the observed values of $\sigma^2$ and the ratio of the \hi mass to the radius ($\frac{G M_{HI}}{5R}$) should be linear if the clouds are self-gravitating and the slope should give the neutral hydrogen mass fraction of the total dynamical mass  ($f^{-1}$). Figure \ref{disp_vs_HI} (upper panel) shows that
the majority of the clouds have ($\frac{G M_{HI}}{5R}$) $\lesssim$ 0.15 (\kms)$^2$ and their location on the plot is fitted by a line corresponding to a total to \hi mass ratio $f \sim 2000$. %, while the others have $f \sim 400$.
The clouds that do not follow this relation are the more extended and massive ones
(AA4, AA13, AA21a, AA21b, AA15 and AA6 which is significantly offset to the top of the figure, beyond the plotted range of the y axis).
The high values of $f$ imply that without an additional mass component, such as dark-matter or ionised gas, the self-gravity of the neutral gas would be inefficient
to keep clouds  bounded.
Moreover the  compact and isolated clouds appear to  be more dark matter dominated
than the more extended ones, as if there were two different population of clouds. However it is also possible that   such a different trend is due the fact that the assumption of virialisation is especially wrong for the more extended clouds if they have a tidal origin.

As a comparison we plot the same quantities  for the Andromeda clouds  (Westmeier et al 2005). A clear trend is not visible  in this case (see the lower panel of Figure \ref{disp_vs_HI}). Only some of the clouds can be fit by a linear relation with $f \sim 150$, while the others show a more scattered  distribution in the $\sigma^2$, $\frac{G M_{HI}}{5R}$ plane compared to the M33 ones.

The  average \hi column densities of the M33 cloud population are below 2 $\times 10^{19}$ \cmsq, which is where sharp \hi edges of spiral discs appear (Corbelli \& Salpeter 1993, Maloney 1993), and they are low enough to expect  that the gas is highly ionised. Thus it is important to evaluate the total gas mass and the ionisation fraction which  depends both on the intensity of the ionising radiation and the gravitational potential of the clouds. The source of ionisation  can be both the local UV extragalactic background radiation and the UV light escaping the star forming disc of M33. In the first case, the theoretical calculation by Haardt \& Madau (1996) is generally adopted, and it is in agreement with the observational limits given by H$\alpha$ surveys of intergalactic \hi clouds (Weiner et al. 2002) and by the \hi truncation of  the extended galactic discs (Corbelli \& Salpeter 1993).

As regards the escaping fraction of the M33 UV flux, Hoopes $\&$ Walterbos (2000) estimate that it is very low. At most 4\% of the stellar ionising photons can escape if one assumes that there are additional sources that can ionise the interstellar gas. Given a SFR of 0.5 \msun yr$^{-1}$ (Magrini et al. 2007), the total number of ionising photons produced by M33 is $\dot{N} = 10^{53}$~photons s$^{-1}$. If the escape fraction is $\epsilon=0.02$,
the ionising flux from M33 equals the extragalactic background on one side of the cloud (F$_{EG}^{UV} = 3.2\times 10^3$~photons~cm$^{-2}$~s$^{-1}$) at a distance  from the  disc $d \sim 70$ kpc

Hence, at 20~kpc the ionising flux escaping the M33 disc can be as much as a factor 12 stronger than the extragalactic radiation field. We shall consider this as an upper limit because the effective escape fraction and the cloud galactocentric distances are unknown.
In the next section we will estimate the ionisation fraction of the clouds assuming  the intensity and energy distribution of the extragalactic ionising background  as given by  Haardt \& Madau (1996), unless stated differently.

\section{Possible cloud formation scenarios}

The origin of the gas clouds detected in the proximity of M33 is uncertain and we have not reached a definitive conclusion yet.
The first issue is to address  whether they are gravitationally bound to M33 or whether they represent a local population of \hi structures in the Milky Way halo.

We have distinguished two type of objects according to their velocity. {\em Type 1} clouds have  -40 \kms $<$ V$_{hel} <$ -320 \kms and we have assumed that they are associated with M33, also because they seem to follow the rotation pattern of the galaxy (i.e. most of the clouds in the northeast part of the galaxy have more negative velocities and those in the southwest have more positive velocities than -180 \kms). However we cannot easily derive  the cloud velocity vectors and their galactocentric radii since they are not a disc population. {\em Type 2} clouds have radial velocities greater than $-40$~km~s$^{-1}$ or less than $-320$~km~s$^{-1}$ and they have been considered unbound to M33. They might just lie in the same sky area  but they might be closer or more distant than M33, or they might be colliding with it at high speed on a hyperbolic orbit.

Among {\em Type 2} objects, the most interesting and puzzling is the \hi complex AA21 showing a difference  of 200 \kms with respect to the systemic velocity of M33. Given its larger extension  compared to the other clouds, and the anomalous velocity   the more straightforward interpretation would be that this is a local \hi complex. If AA21 is in the Milky Way halo, its radial velocity would be about -240 \kms in the Galactic Standard of Rest frame. Four clouds with similar velocities are found in the Leiden/Dwingeloo Survey (LDS) catalog (de Heij et al. 2002) within 10 degrees from M33. Some  clouds with similar high velocities in the area can also be noticed in
the map of Wakker (2004) between the Wright cloud and the Magellanic Stream (see also Wright 1979).
The complex extends over an area which is larger than 1 square degree (see Figure \ref{HI_complex_mom0_fieldB}) and the line width of the spectral profile is around 30 \kms.
The estimated crossing time -- $t_{cr}\propto$ 2R/($\Delta$V) $\simeq$ ~0.6 Myr (d/kpc), where d is the cloud distance  -- gives an estimate of  the timescale for the cloud to double its size.  Either  if we assume it is in the Milky Way halo (between 10 and 50 kpc implying $t_{cr}$ of the order of 6 to 30 Myr), or at M33 distance ($t_{cr} \sim$ 500 Myr) the cloud is not gravitationally stable and it would disperse very quickly, unless it is confined by an external pressure medium (if it is in the MW halo) or by dark matter.

If this is a cloud on a hyperbolic orbit, which is passing near M33,  without being accreted to it, one might find evidence for an interaction between the two systems. As mentioned in section 4.1, the ALFALFA data show  faint clumps of gas to the north-west of  the complex towards the center of M33. The  structure also shows a faint tail on the opposite side which  may suggest a process of tidal stripping, as the  cloud is entering  the gravitational potential of M33. Higher sensitivity observations of  the northern and southern faint structures of this complex are needed to establish whether it is interacting with  M33 or not.

In the following subsections we discuss  five possible scenarios for the origin of {\em Type 1} clouds: $(i)$ gaseous satellites, confined in dark mini-halos around M33, $(ii)$ condensations from hot cosmic filaments, $(iii)$ tidally stripped gas from the M33 disc during a close encounter with M31, $(iv)$ gas stripped from a dwarf galaxy during a closer encounter in the past or a merger, $(v)$ disc gas ejected by supernovae in giant H{\sc ii} regions. {\em Type 1}
clouds can fit in any a scenario from $(i)$ to $(v)$ while {\em Type 2} clouds fit into scenarios $(ii)$, $(iv)$.

To estimate the cloud total gas mass and to better constrain the scenarios on their origin we shall consider two possible configurations. In one case the clouds are confined by their own dark halo, as if they were small satellites of M33 (scenario $(i)$). The other possibility is instead that the  clouds have no associated dark mini-halo but they are inside the gravitational potential of the M33 dark matter halo. This second option will be discussed mainly in the case of  scenario $(ii)$, but it applies also to $(iii), (iv)$, $(v)$.

\subsection{Gaseous counterparts of  dark matter mini-halos }

In the absence of dark matter, the gas would be insufficient to keep the clouds gravitationally bound. As mentioned in the previous section, the crossing time of the gas would be rather short, about $100$ Myr, or half a revolution period, $t_{cr} \propto \sim$ 100~Myr (R/kpc) ($\Delta V/20$ \kms)$^{-1}$, where R is the cloud radius and $\Delta$V the 21-cm line width.

If the clouds are the gaseous counterparts of the missing dark satellites, they will be confined by
their own dark matter halo. Their line width and extent will give an estimate of the virial mass at the  \hi radius (see column 10 in Table 1), but not to the total mass of the satellite, because the 21-cm emission comes from a smaller region of the dark halo where most of the neutral hydrogen is contained.
Therefore the expected halo mass is relatively small, of the order of
few times 10$^8$~M$_\odot$. Numerical simulations of structure formation in the CDM scenario (e.g. Klypin et al. 1999) predict  that the number of dark
minihalos around a  massive galaxy is a strongly decreasing function of their virial mass.

According to Sternberg et al. (2002, hereafter ST02), the number of minihalos within a distance $d$ of the parent galaxy with a mass M$_{vir,p}$,  exceeding a certain rotational scale velocity $\mathrm{v_s}$ which defines the total mass of the minihalo\footnote{For example for $\mathrm{v_s} = 15$ \kms, M$_{vir,h} = 2 \times 10^8$ \msun (see Sternberg et al. 2002).} (M$_{vir,h}$) is

\begin{equation}
N(> \mathrm{v_s,} < d) = 1.06 \times 10^3 \left( \frac{M_{vir,p}}{10^{12} M_{\odot}} \right)
 \left( \frac{d}{1 \mathrm{Mpc}} \right) \left( \frac{\mathrm{v_s}}{10 \,\mathrm{km s^{-1}}} \right)^{-2.75}
\end{equation}

For a maximum distance  of 50~kpc, and a M33  halo mass  M$_{vir,p} = 5\times 10^{11}$~M$_\odot$ (Corbelli 2003), the number of subhalos with mass greater than 10$^{9}$~M$_{\odot}$ is only 4, while it increases to 25 for  M$_{vir,h} > 10^{8}$~M$_{\odot}$, comparable to the number of objects we  detected. Among these, in  Figure \ref{disp_vs_HI} we show that there are 16 clouds with a large ratio of the virial to \hi mass (f $\sim$ 2000), which could more likely represent a population of dark matter dominated satellites.

If we take out the two largest clouds the average  radius is 0.7~kpc and the average \hi mass is 0.9$\times 10^5$~M$_{\odot}$. ST02 have computed models of  clouds confined by a dark matter minihalo for different
cloud parameters and for different values of the pressure of the medium
%\hi radii, \hi and virial masses, assuming either a Burkert (1995) or a Navarro Frank \& White (1996) density profile, as a function of the pressure of  the hot ionised medium
(P$_{HIM}$) which surrounds them (a galactic corona or a hot intergalactic medium). Since M33 does not show  evidence of a corona, we will consider models with a low external pressure $P_{HIM} = 10$~cm$^{-3}$~K. In this condition the minimum halo mass needed to have the gas  bound is $2 \times 10^{8}$~M$_{\odot}$ (see Figure 7 in ST02). If we choose the model closer to our observed values (r$_{HI}=0.6$~kpc, M$_{HI}=2\times10^{5}$~M$_{\odot}$, and a Burkert halo with M$_{vir}=3\times 10^{8}$~M$_{\odot}$),
the cloud total gas mass is 50 times more massive than the neutral component and extends  for more than 3~kpc.

We estimate the
total \hi mass associated with all {\em Type 1} clouds to be
 %except the two most massive ones (AA4 and AA6) is
 $\sim 5\times 10^6$~M$_{\odot}$  which gives a total gas mass of $\sim 10^8$~M$_{\odot}$, most of which is ionised.

Assuming they fall onto the
disc at a speed of 100~km~s$^{-1}$, from an average distance of 20~kpc, the estimated gas accretion rate is about 0.8~M$_{\odot}$~yr$^{-1}$. This is comparable  to the value predicted by the chemical evolution model of Magrini et al. (2007) to sustain a star formation rate of about 0.5~M$_{\odot}$~yr$^{-1}$, as observed in M33.

Note, however, that these estimates only take into account the intergalactic radiation field.
If the escaping fraction of the UV photons from M33 is not negligible, this provides an additional source of ionising radiation which will increase the estimated total gas mass, therefore our estimate is a lower limit to the gas content of the M33 halo.

\subsection{Gaseous filaments connecting to the HI disc}

How galaxies get their gas is still an open question in astrophysics. As the gas enters a galaxy virial radius and cools, it might accrete into the outer disc and propagate through radial infall towards the inner star forming regions, or it might populate halo orbits and spiral in towards the star forming disc. Addressing in detail this issue is beyond the scope of this paper but we will consider here a simplified picture to estimate some basic properties of the clouds in M33 in the case where they do not have their own dark matter halo.

At column densities lower than $5\times 10^{19}$~cm$^{-2}$ the gas neutral fraction may decrease very rapidly for a small variation in the gas column density (Corbelli \& Salpeter 1993, Maloney 1993), we will then assume that the clouds are \hi condensations inside a more extended, ionised  component confined to a plane. This simplified picture may describe either the cosmic filament scenario or clouds which are tidal debris resulting from a close encounter with M31. We also assume that the neutral hydrogen vertical extent is comparable to the observed cloud \hi radius.
If the gas is embedded in a dark matter distribution (the M33 halo), %under reasonable hypothesis,
the vertical acceleration near the plane depends only on the surface density of the mass within the gas layer (Kuijken \& Gilmore 1989; Maloney 1993), which at the radii of interest, is mainly contributed by the dark halo.

Given an incident ionising radiation field  we can then compute the halo dark matter density and the total gas surface density needed to reproduce the observed \hi size and  column density. We use a numerical code which solves for the ionisation, chemical and hydrostatic equilibrium of a plane-parallel slab of gas. The procedure and the numerical code is described in detail by Corbelli, Salpeter $\&$ Bandiera (2001). The density scale height $h$ will depend on the dark-to-gas-mass ratio, $\eta$, according to:

$$h={c_s^2 \over 4G \Sigma_g}\sqrt{1 \over 1+\eta^2}$$

$$\eta={\pi \Sigma_{dm} \over 2 \Sigma_g} $$

where $G$ is the gravitational constant, $c_s$ is the gas sound speed,  $\Sigma_g$ is the gas surface density and $\Sigma_{dm}$  the dark matter surface density. Assuming the Haardt \& Madau (1996) UV background, for a given $\Sigma_g$ and $\eta$,  the numerical code finds the gas neutral fraction as a function of the height above the slab mid-plane, and hence the \hi column density perpendicular to the plane. Therefore we vary $\eta$ and $\Sigma_g$ until we match the observed \hi column density and $h$ equals the observed \hi radius.  From  $\eta$ and $\Sigma_g$ we infer the total gas mass associated with the cloud  and the dark matter density that we can compare to the value required by the rotation curve.

For {\em Type 1} clouds we find that the dark matter densities range from 0.17 to 2.7 $\times 10^{-25}$~gr~cm$^{-3}$ with a mean value of 10$^{-25}$~gr~cm$^{-3}$.
If we assume a typical cloud galactocentric distance of 20 kpc, the halo dark matter density ($\rho_{DM}$) at that radius for both a NFW and a Burkert profile is $\rho_{DM} = 10^{-25}$~gr~cm$^{-3}$ (Corbelli 2003), which is in perfect agreement with the mean value derived for the clouds.
From the ratio of the total gas to \hi surface density we find that the clouds are 90\% ionised and  the total gas mass associated with {\em Type 1} objects is around 5$\times 10^7$~M$_\odot$.  As noted before this is only a lower limit, since this does not include the ionising field contributed by the M33 disc itself.
%the ionisation fraction increases if additional ionising photons escape the star forming disc of M33.

\subsection{Tidal debris from an interaction with M31}

The clouds  appear to be mostly concentrated along a main axis which is oriented towards M31 and this may be an indication  that the clouds  have formed in a previous interaction with Andromeda. The two galaxies are at a projected distances of $\sim$ 200 kpc.  There is evidence for possible past interactions between them: Braun \& Thilker (2004) found a large \hi structure at low column density between the two galaxies, and the warp in the outer \hi disc is skewed towards the direction of M31 (Corbelli \& Schneider 1997). Simulations of the motion of the system M31/M33 (Loeb et al. 2005) on the basis of the measured proper motion of M33 (Brunthaler et al. 2005) predict that there may have been close encounters between the two galaxies in the past which  resulted in the stripping of material from the M33 outer disc, leaving the stellar disc unperturbed. Calculation of the transverse velocity of M31 (van der Marel \& Guhatakurta 2007) would imply an orbit with a semimajor axis a= 127 kpc, a pericenter distance $r_{per}$ = 30 kpc and an orbital period of 2.4 Gyr. With M33 in motion on such an orbit tidal deformations would be expected. Particularly  the clouds to the north eastern side (AA14, AA15, ) are related both in space and velocity to an extended plume of gas  (see Figure \ref{cloud7_chans})   which may give  further support to the tidal origin hypothesis and the whole structure (both the plume and the clouds) may indicate the continuation of  the \hi filament detected by Braun \& Thilker between the two galaxies (2004, see Figure 9 therein).

\subsection{Gas stripped from dwarf galaxies}

Another possibility for the origin of the clouds could be that they are the remnants of material stripped from a dwarf galaxy during a past close encounter.
%Another possibility for the origin of the clouds could be that they are the remnants of material stripped from a dwarf satellites during recent or ongoing mergers.
To date evidences for mergers or accretion are rare for M33 and this galaxy is thought  to have evolved in relative isolation (Ferguson et al. 2006). However this is still a controversial issue since there are claims of the presence of a halo stellar component and of a possible tidal stream in M33 (McConnachie et al. 2006).
The  recent discovery of an extended cluster in the southern side of the disc has been related to the possibility of an accretion event (Stonkute et al. 2008), as it has been suggested for similar objects found in M31 (Huxor et al. 2005). The cluster is at about 25 arcminutes (6 kpc) from the center of cloud AA6, located in projection against the \hi bridge that connects it to the disc of M33. Further investigation of the area is needed to understand whether the origin of this cloud could be related to  a possible merger with a dwarf-like satellite.

\subsection{Galactic fountain}

\begin{figure}[h]
\includegraphics[width=8cm]{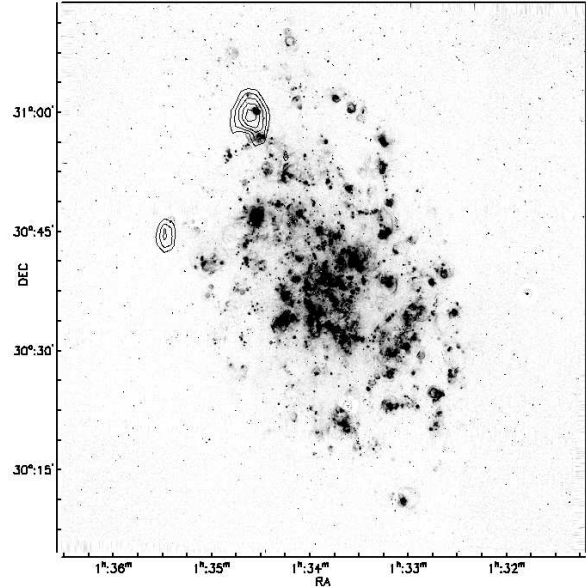}
\label{cloud_SF} \caption{Contour density maps of the two closest \hi clouds to the stellar disc of M33 (AA1, AA16) superimposed on the H$\alpha$ image of the galaxy (Hoopes \& Walterbos 2000). Only AA1 appears (in projection) to be related to a star forming region to the north east of the optical disc. All the other clouds are farther from the star forming disc of the galaxy. }
\end{figure}

A  galactic fountain within M33 seems unlikely to have produced the observed cloud distribution. In this scenario gas
is heated and ionised by supernova explosions, rises above the galactic plane within the halo where it cools down,  condenses and then falls back towards the disc (Shapiro \& Field 1976, Bregman 1980).
%Simulations suggest that the height of the condensation (de Avillez 2000) is at most several kpc above the disc.
Only one cloud (AA1) appears to be correlated (in projection) to  a bright H{\sc ii} region (see Figure \ref{cloud_SF}). The H$\alpha$ flux is 1.6 $\times 10^{-13}$ erg
 s$^{-1}$ cm$^{-2}$  (6000 times higher than the background). From the catalog of  \hi holes in the interstellar medium of M33 (Deul \& den Hartog 1990) we find a correspondence in position between the hole n. 128, the H{\sc ii}
  region and the center of cloud AA1.  The diameter of the hole is $\sim$500 pc and  the estimated swept-up mass is $10^6$ \msun with an age of 30 Myr, values that are compatible with the observed \hi mass and size of AA1.
All  other clouds are beyond the stellar disc, making this scenario appear unlikely to explain the origin of the HVCs in M33. Moreover the galactic fountain hypothesis could not explain large velocities like those of several clouds in this sample

\section{Conclusions}

 \hi clouds seem to be a common feature in the halos of  spiral discs at least in the LG. The Milky Way and Andromeda have been known to contain a population of high velocity clouds (HVCs) and here we show that a similar gaseous component is detected in the environment of M33.

At the distance of 840 kpc, the \hi masses of the clouds  range  between 10$^4$ and few times 10$^6$ solar masses.  They are found within a projected distance of $\sim$ 20 kpc and they appear to be distributed (in projection)  along the major axis of the outer disc  which is warped towards the direction of the Andromeda galaxy. The total \hi mass associated with {\em Type 1} clouds is 5 $\times 10^6$ \msun and it goes up to $7 \times 10^6$ \msun if one includes also {\em Type 2} objects. %, which is  about 10$\%$ of the total \hi mass of M33.

The origin of high velocity clouds is still under debate; in this paper we have discussed some possible scenarios for the formation of HVCs in M33 and how they fit with their observed properties.

We have  explored the possibility that these clouds might be associated with a population of dark-matter dominated satellites. From ST02 approximately 25 dark mini-halos would be expected within 50 kpc from M33 with total masses around $10^8$ \msun. This is comparable to the total number of clouds we have detected (19 excluding those with anomalous velocity). From the comparison  to their mini-halo models we find that this scenario would be  consistent with the observed line widths and radii. Given the overall low \hi column densities (below $10^{19}$ \cmsq)  a large fraction of the gas would be ionised. The ionised component could be up to 50 times larger than the neutral one, with the ionised envelope extending for more than 3 kpc. We derive that the total gas mass associated with all {\em Type 1} clouds would be of the order of $10^8$ \msun.  Assuming that all the gas locked in the halo is falling back towards the disc at an average velocity of 100 \kms from an average distance of 20 kpc, we derive a gas accretion rate of $\lesssim$ 0.8 \msun yr$^{-1}$.

We have also considered other scenarios in which  \hi clouds are  condensations within a hot intergalactic medium (cosmic filaments) or they are either tidal debris of a previous interaction between M33 and M31 or remnants of gas stripped from a dwarf satellite at earlier times.
Quantifying the total gas mass (\hi and H{\sc ii}) within the M33 halo in these cases is more difficult and we can only set a lower limit of $5 \times 10^7$ \msun.

Higher angular resolution observations of the M33 population of \hi clouds will be useful to probe their inner structure, and to search for a higher column density core and internal velocity gradients.

At smaller radial velocities ($V < -320$ \kms) we have resolved a \hi complex that had been serendipitously discovered by Thilker et al. (2002) with Westerbork observations of M33.  The complex is  at about one degree from the centre of the galaxy ($\sim$ 15 kpc), but it is  characterised by a large velocity relative to M33 ($\sim 200$ \kms) which complicates a firm  association to the galaxy. Two main substructures can be distinguished from the ALFALFA data and there appear to be faint clumps of gas pointing towards the center of M33, which may give an indication for a possible interaction with the galaxy (see  Figure 2).  We cannot exclude that this is a cloud not gravitationally bound to M33 and higher sensitivity observations of the region  are needed to test this hypothesis. On the other hand, if this is a local \hi complex, a search for similar clouds in this part of the sky would help to disentangle the origin of this object.

\begin{acknowledgements}
We would like to thank A. Burkert for useful discussions on this manuscript. RG, MPH acknowledge partial support from NSF grants AST–0307661 and AST–0607007, and from the Brinson Foundation. AMM is supported by a National Defense NDSEG Fellowship.
This work is based on observations collected at Arecibo Observatory. The Arecibo Observatory is part of the National Astronomy and Ionosphere Center, which is operated by Cornell University under a cooperative agreement with the National Science Foundation. We wish to thank the Arecibo Observatory staff for the help during the observations and data reduction.
\end{acknowledgements}

\appendix
%\section{}
\section{Column density maps and spectra of the \hi clouds in M33}

 In the Appendix we display the maps of the integrated \hi emission and the corresponding spectra of the  clouds detected in the proximity of M33. The velocity range over which the contour maps have been derived is indicated by the vertical dotted lines in the spectrum plots. The contour levels for each cloud are tabulated in Table A.1. Figures A.1 and A.2 show the maps and spectra of  {\em Type 1} clouds, which have been divided into two subgroups as described in Section 4. Discrete clouds (Section 4.2) are illustrated in Figure A.1, while  the clouds which appear to be spatially connected to the disc of M33 (Section 4.1) are displayed in Figure A.2. Figures A.3 shows {\em Type 2} clouds. Finally Figure A.7 refers to the spectra of the two additional clouds detected with the higher sensitivity data set (Section 6).

\begin{table}[h]
\begin{center}
\caption{Contour levels of the column density maps shown in Figures A.1, A.2 and A.3.}
\begin{tabular}{lc}
\hline \hline
ID & N$_{HI}$ \\
   & 10$^{18}$ \cmsq \\
\hline \hline
AA1 & 3, 4.5, 6, 7.5 \\
AA2 & 3, 4.5, 6, 7.5 \\
AA3 & 2.5, 3, 3.5, 4 \\
AA4 & 3, 4.5, 6, 7.5, 10, 15 \\
AA5 & 3, 4.5, 6, 7.5, 9 \\
AA6 & 3.5, 5.5, 7.5, 10, 15, 20, 25, 30 \\
AA7 & 2, 3, 4 \\
AA8 & 2, 3, 4 \\
AA9 & 3, 4, 5 \\
AA10 & 3, 5, 7 \\
AA11 & 2, 3.5, 5 \\
AA12 & 2, 4, 6, 8 \\
AA13 & 3, 5, 7, 9 \\
AA14 & 3, 4, 5, 6, 7, 8 \\
AA15 & 3.5, 6, 8.5, 11, 13.5 \\
AA16 &  2, 3, 4 \\
AA17 &  3, 4, 5 \\
AA18 &  3, 3.5, 4, 4.5\\
AA19 &  3, 3.5, 4, 4.5, 5\\
AA20 &  2, 3, 4, 5\\
AA21 &  3, 4.5, 6, 7.5, 9, 10.5\\
\hline \hline
\end{tabular}
\end{center}
\end{table}

\begin{figure*}
\includegraphics[width=4cm]{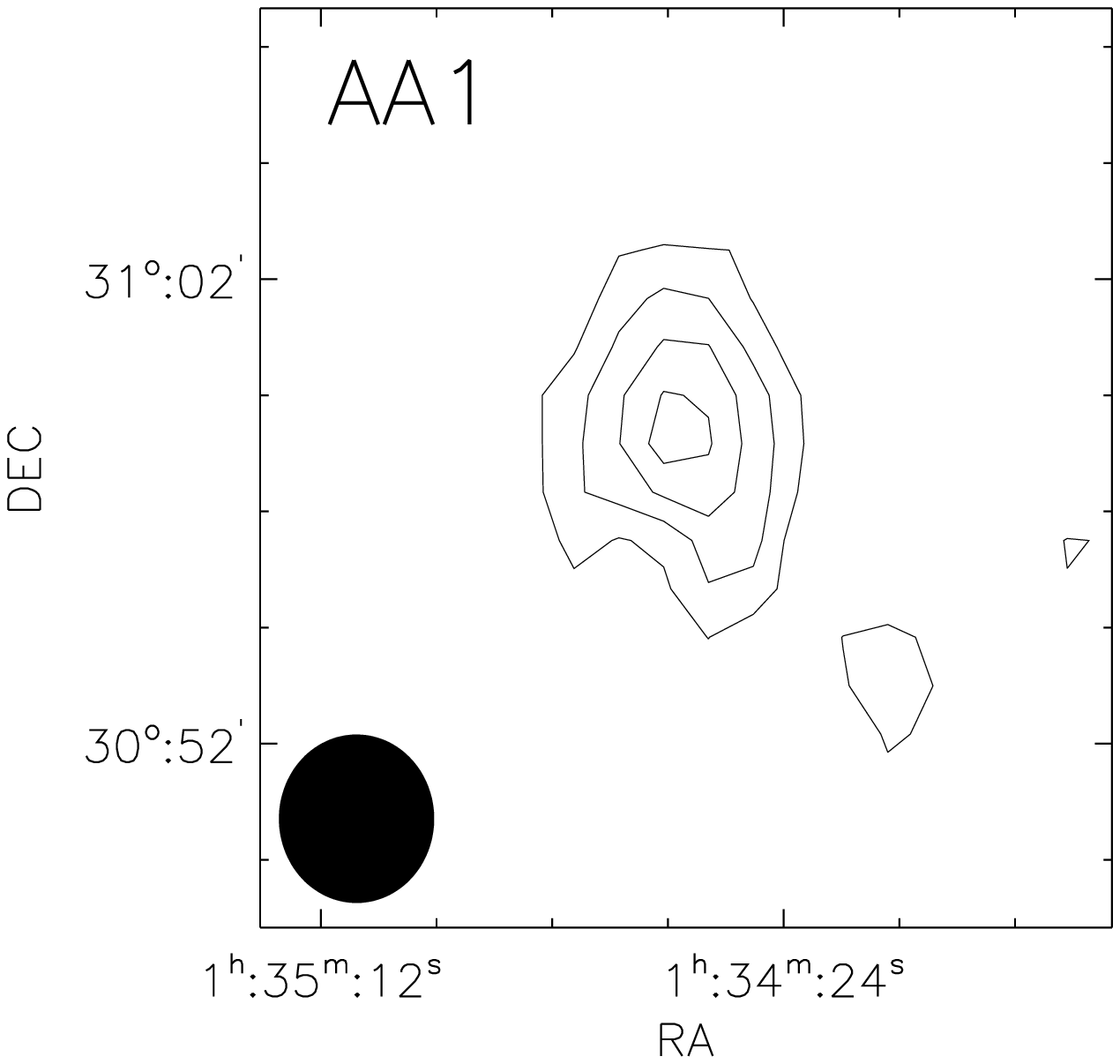}
\includegraphics[width=4cm]{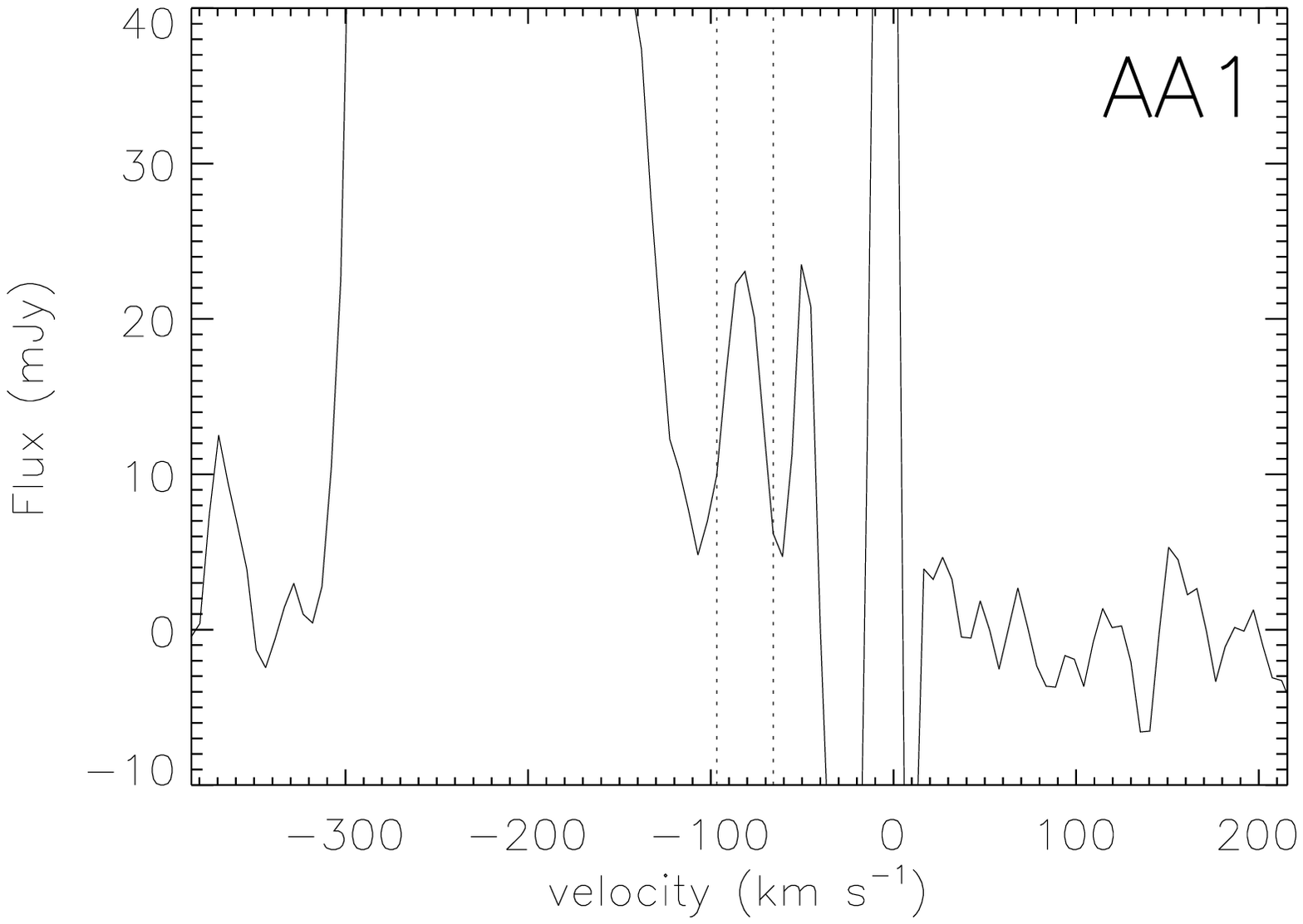}
\includegraphics[width=4cm]{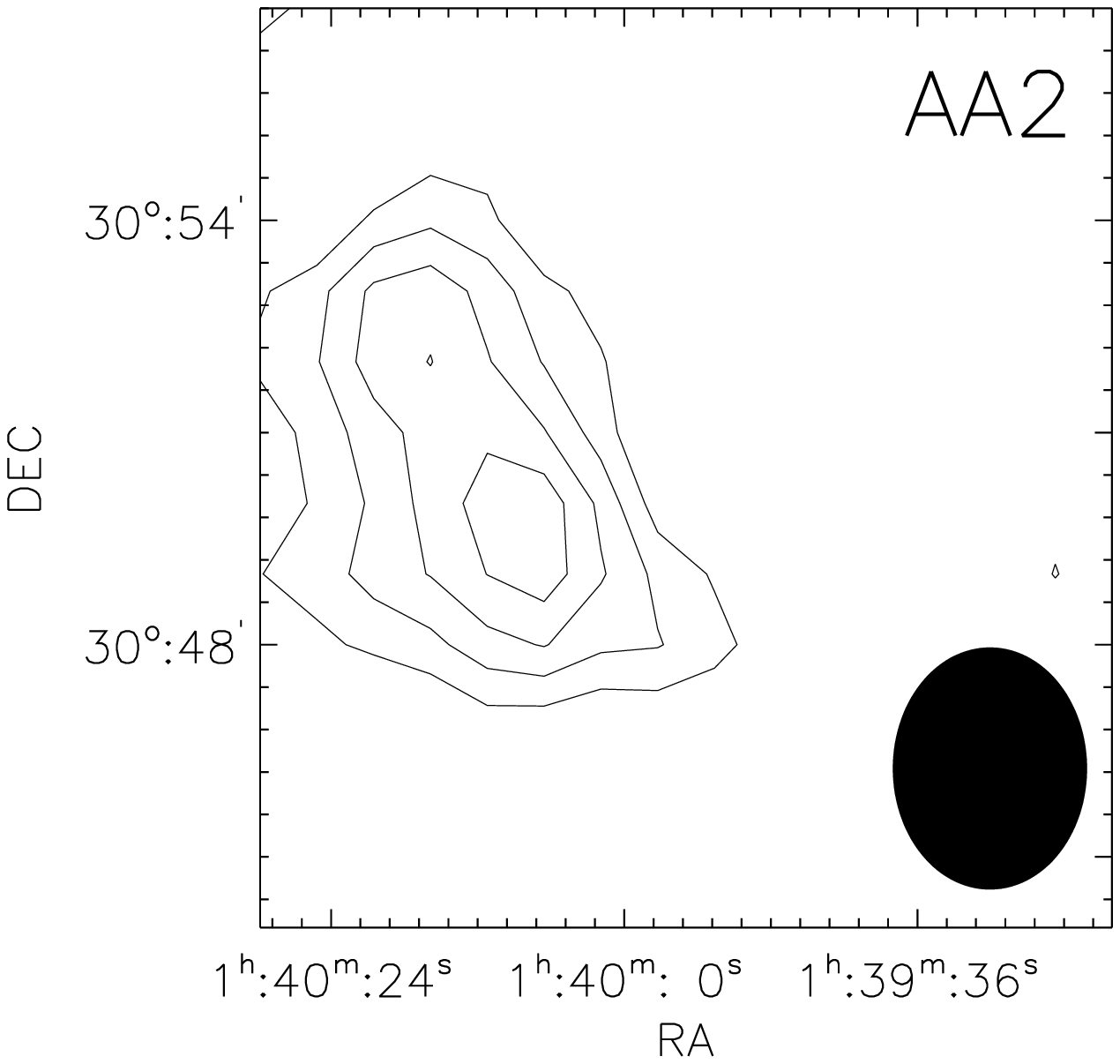}
\includegraphics[width=4cm]{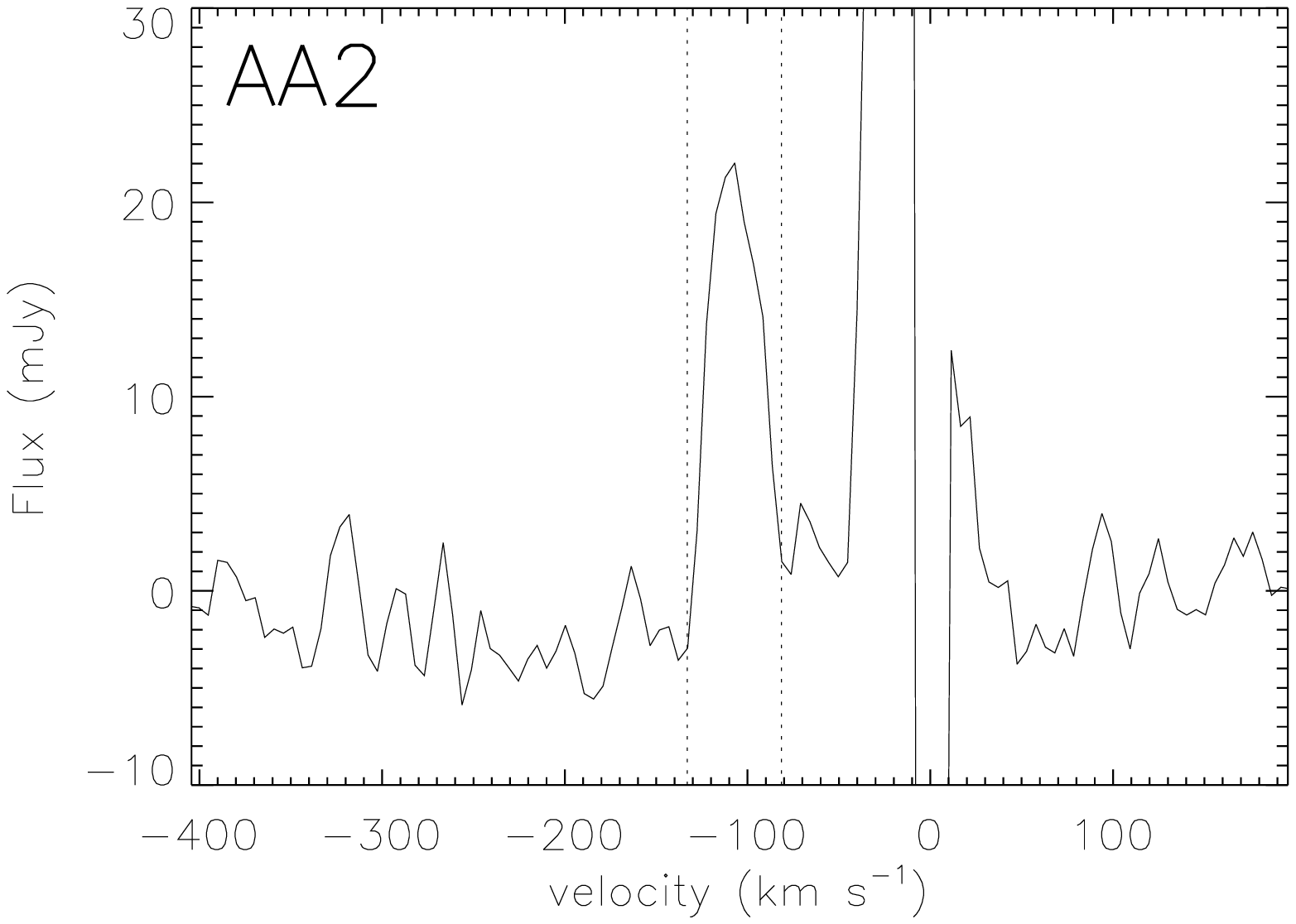}
\includegraphics[width=4cm]{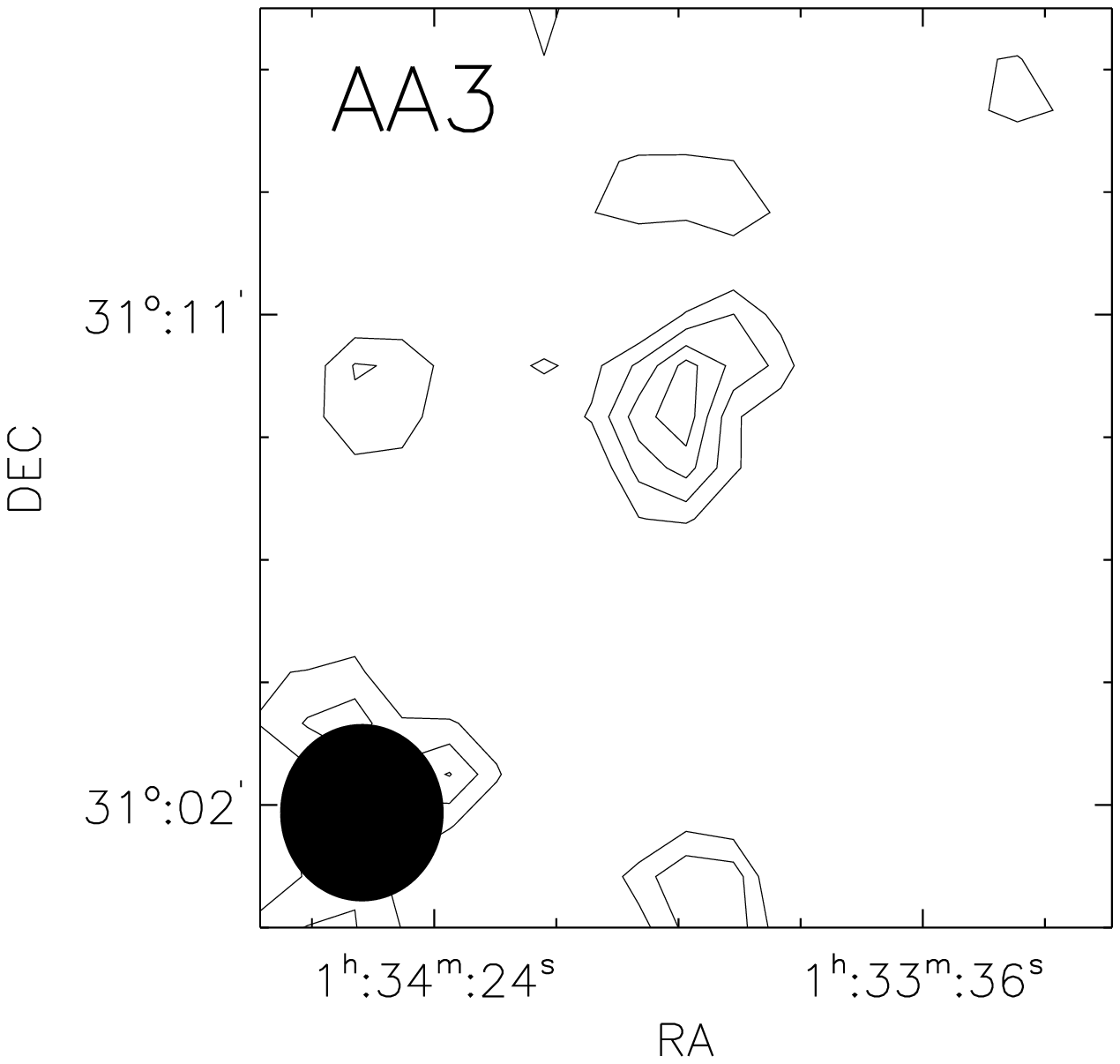}
\includegraphics[width=4cm]{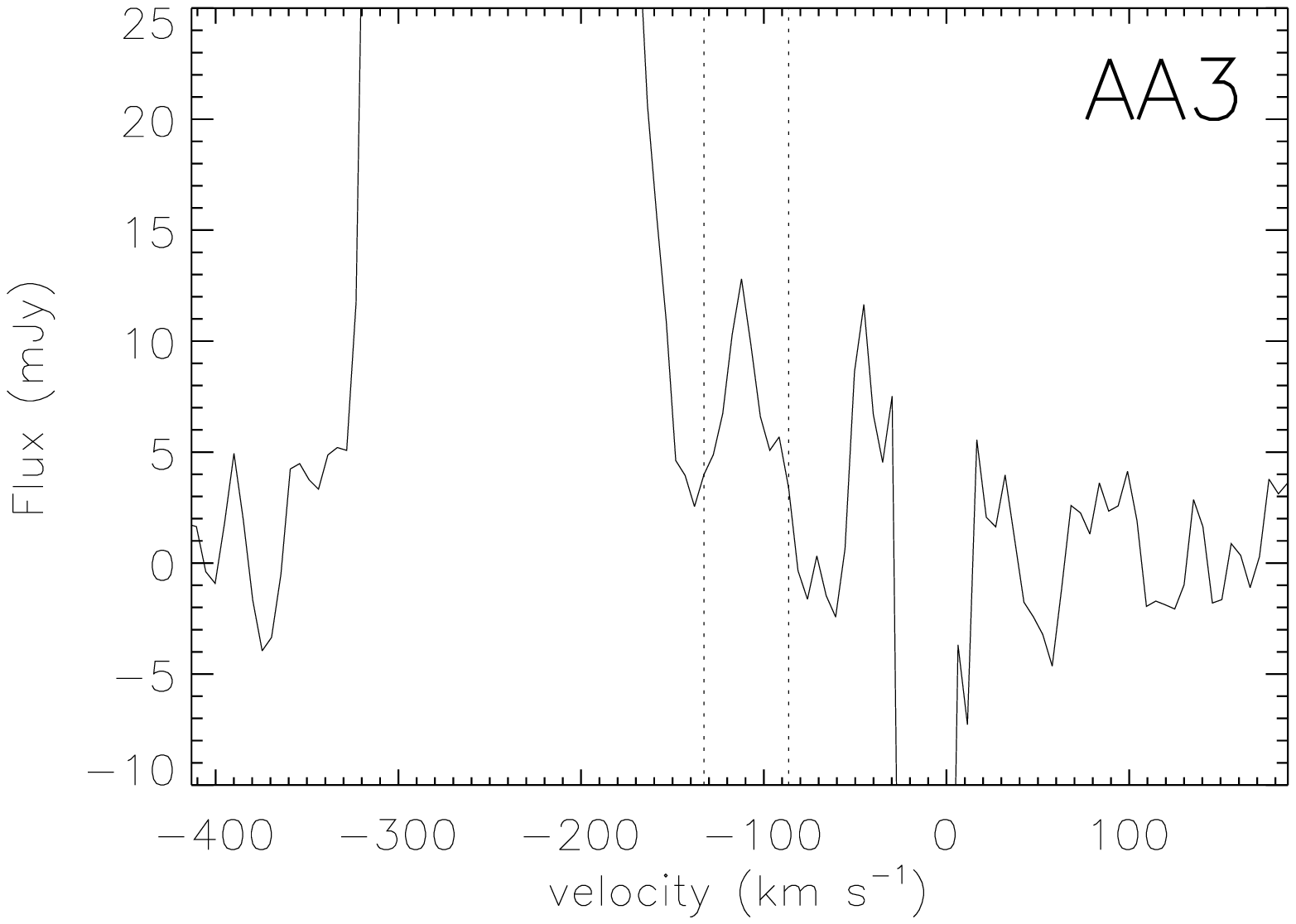}
\includegraphics[width=4cm]{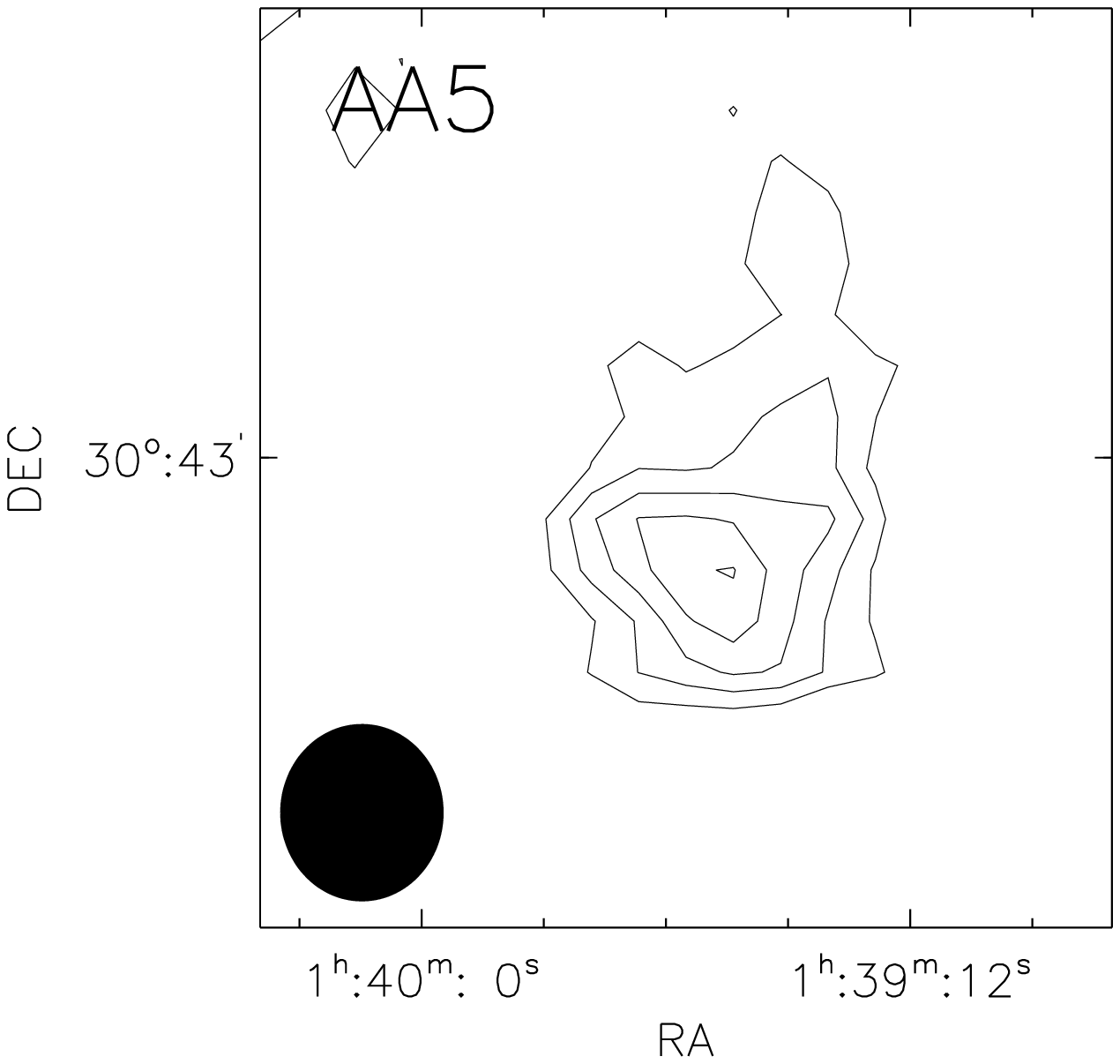}
\includegraphics[width=4cm]{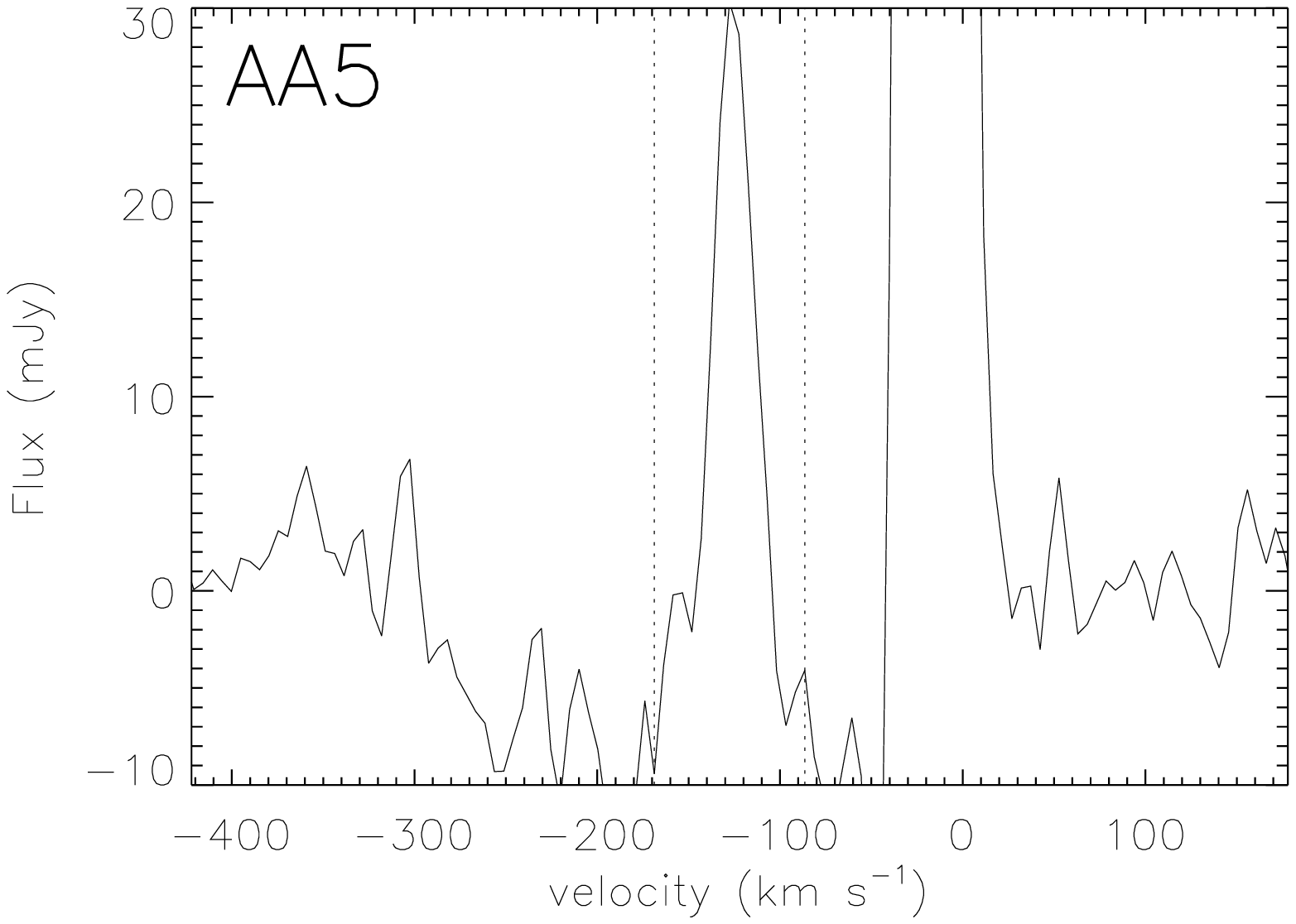}
\includegraphics[width=4cm]{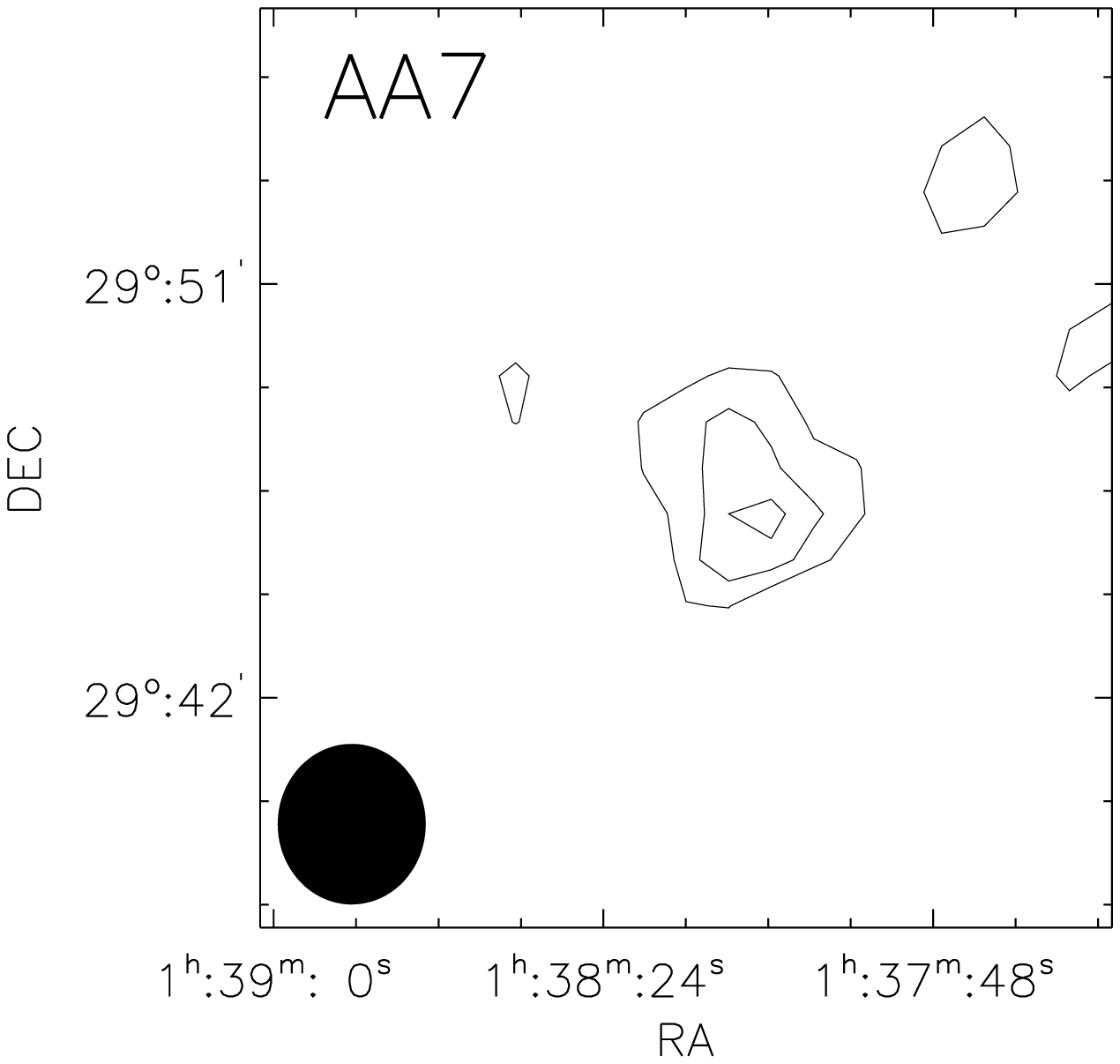}
\includegraphics[width=4cm]{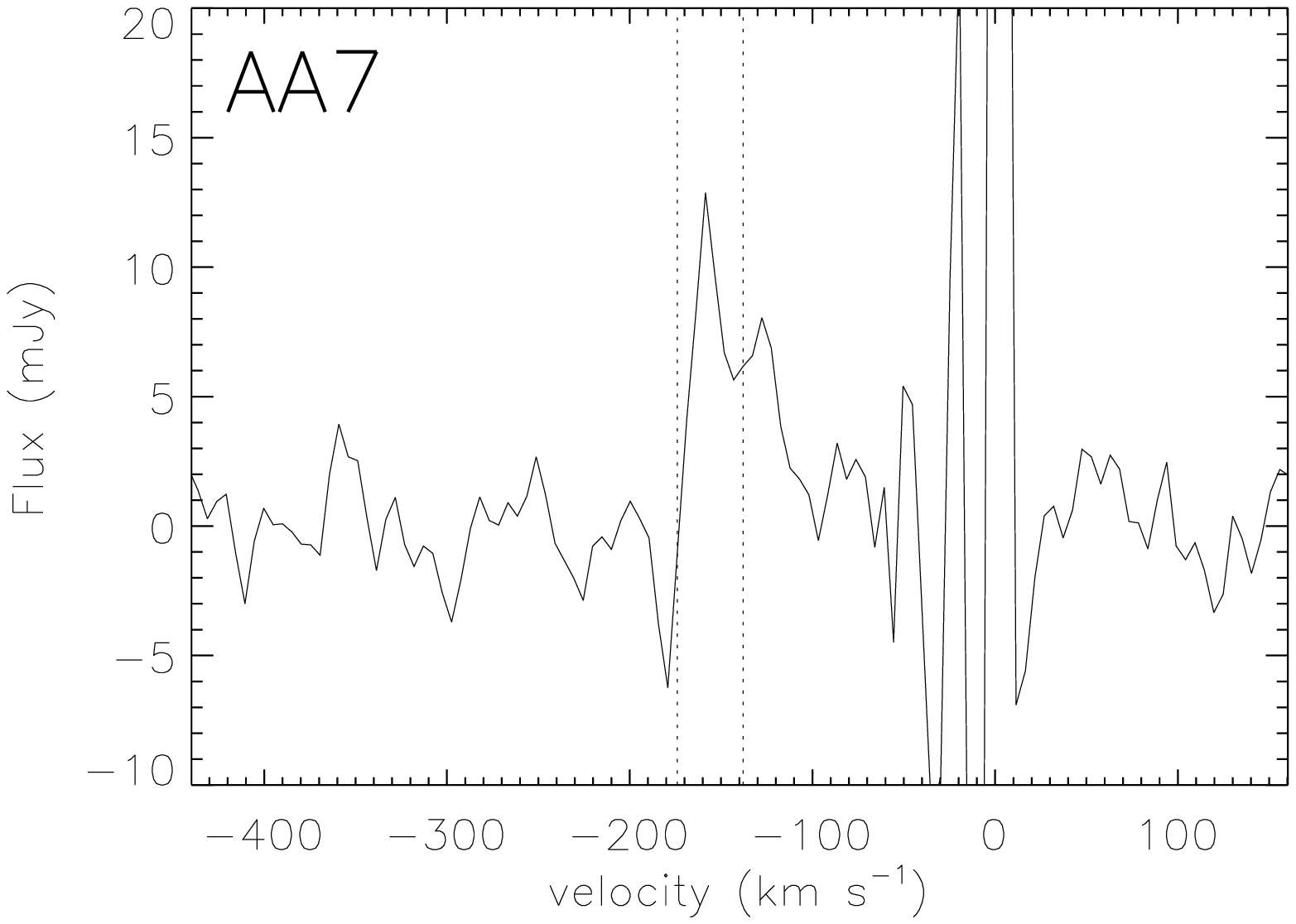}
\includegraphics[width=4cm]{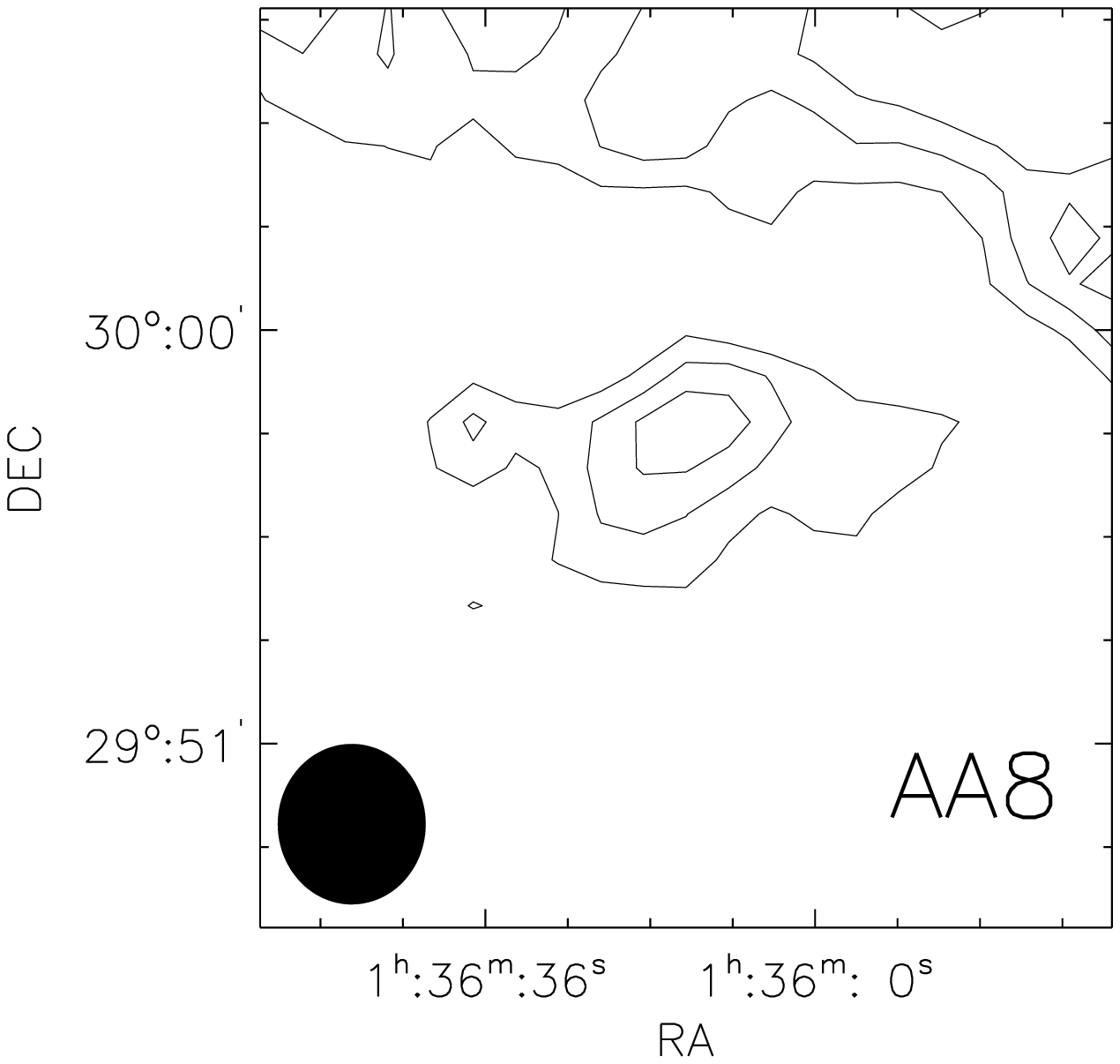}
\includegraphics[width=4cm]{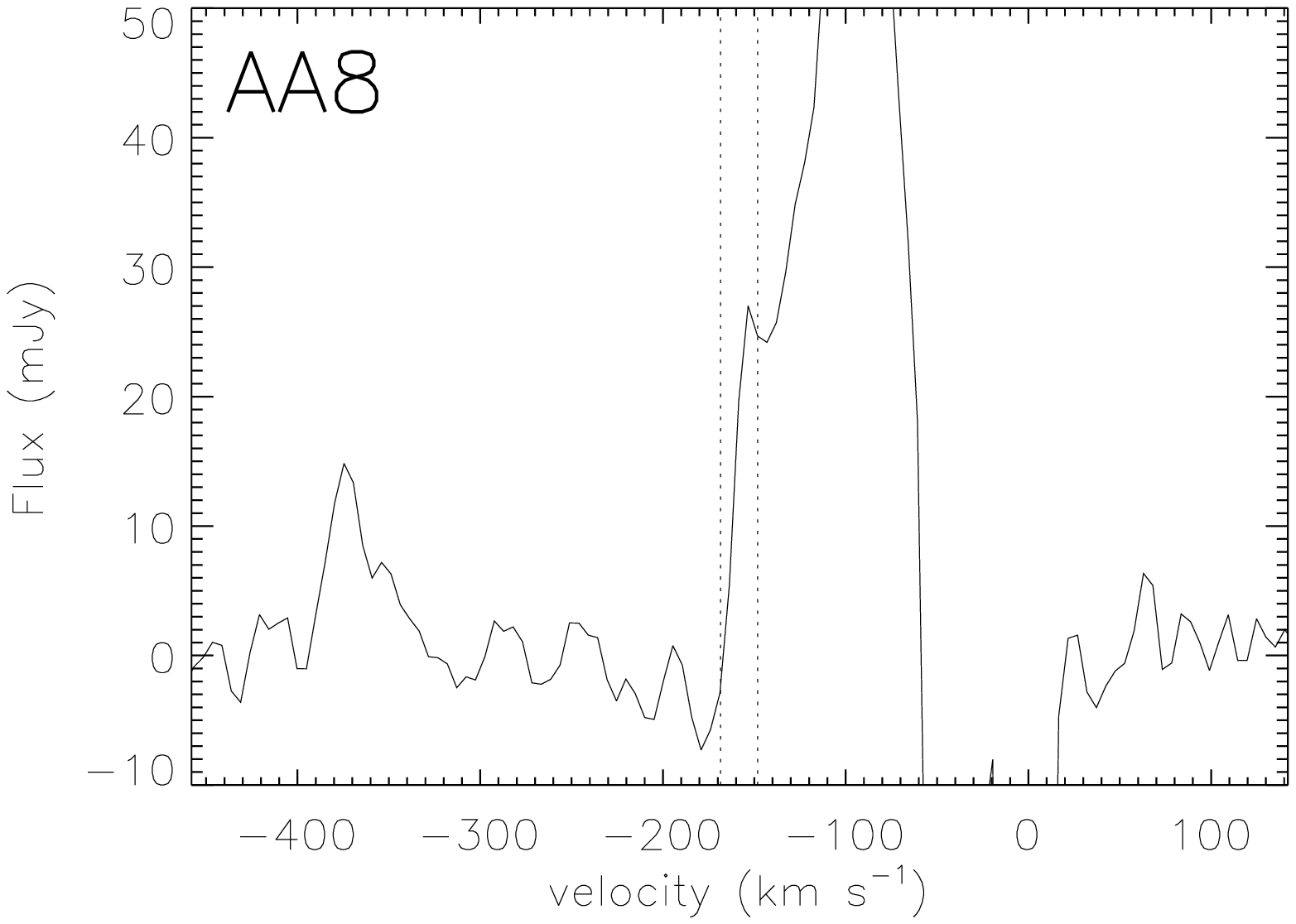}
\includegraphics[width=4cm]{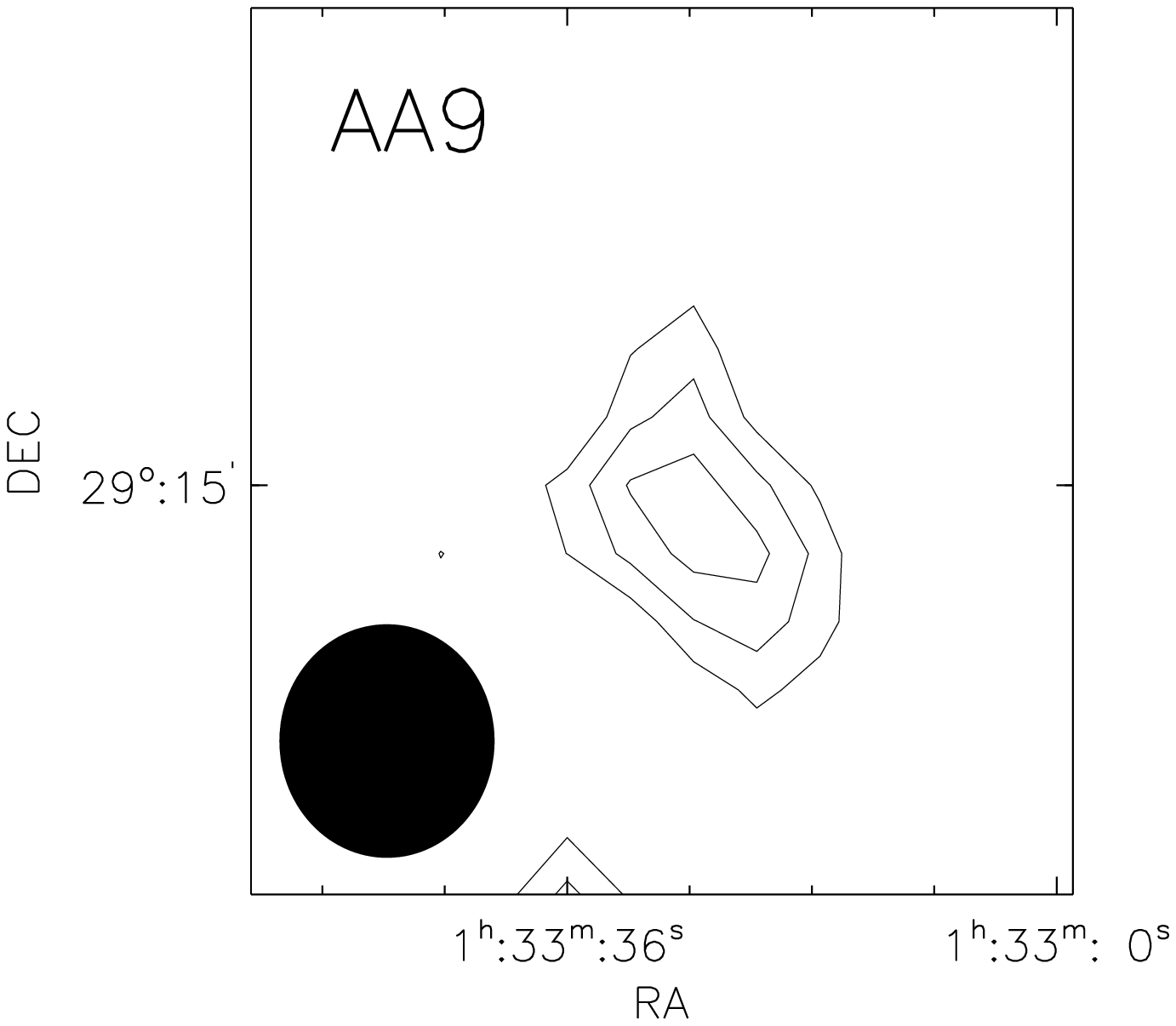}
\includegraphics[width=4cm]{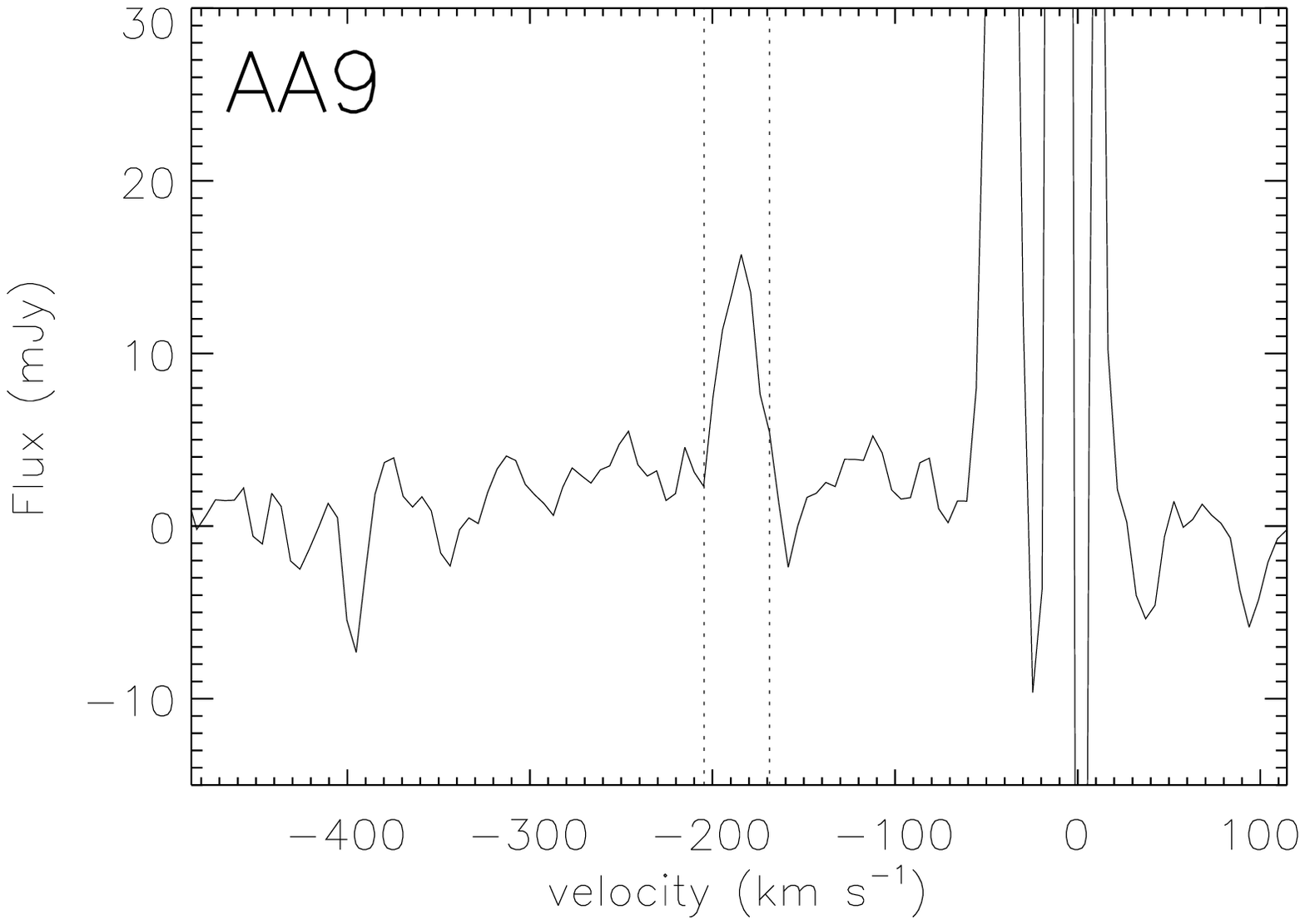}
\includegraphics[width=4cm]{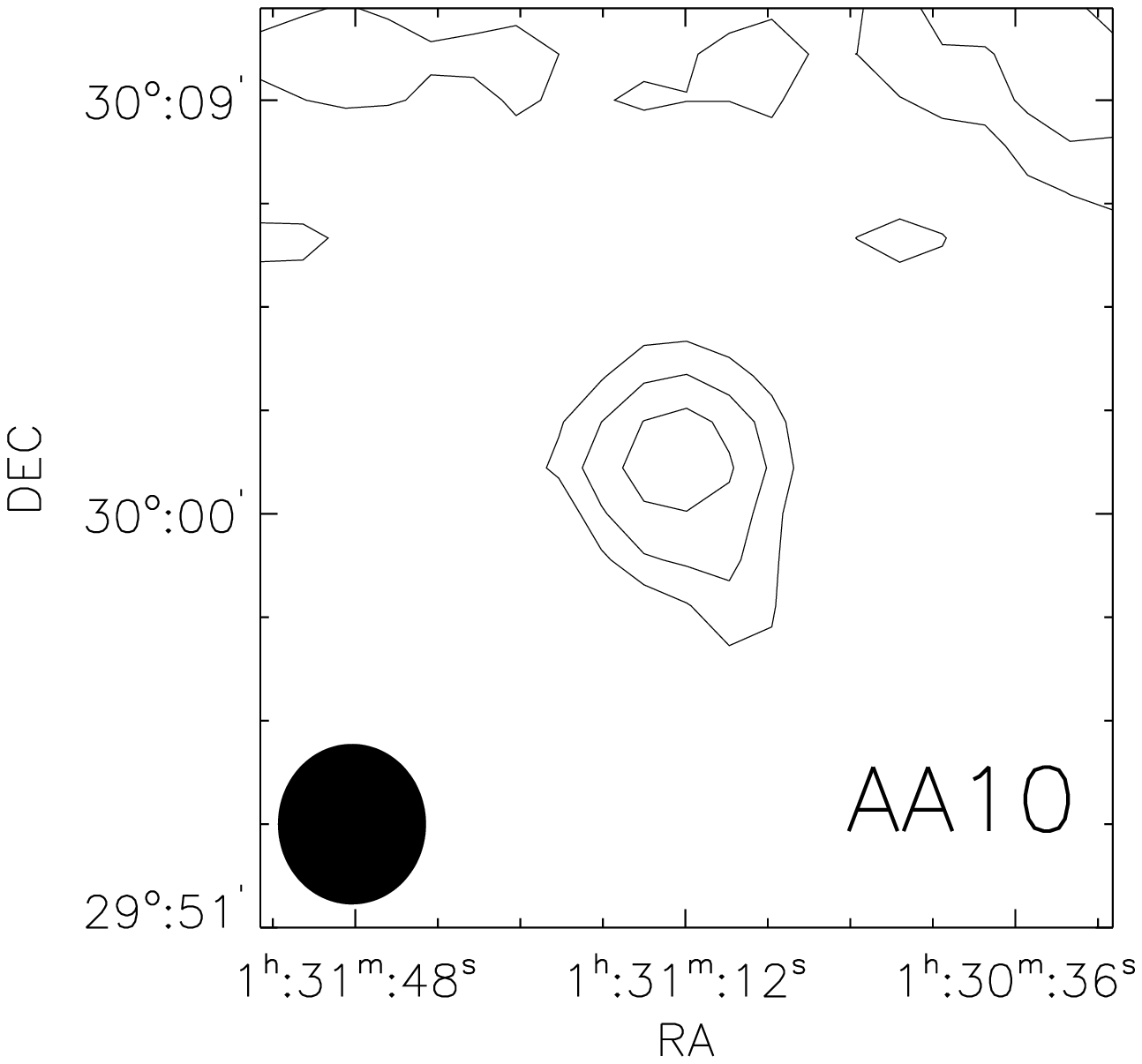}
\includegraphics[width=4cm]{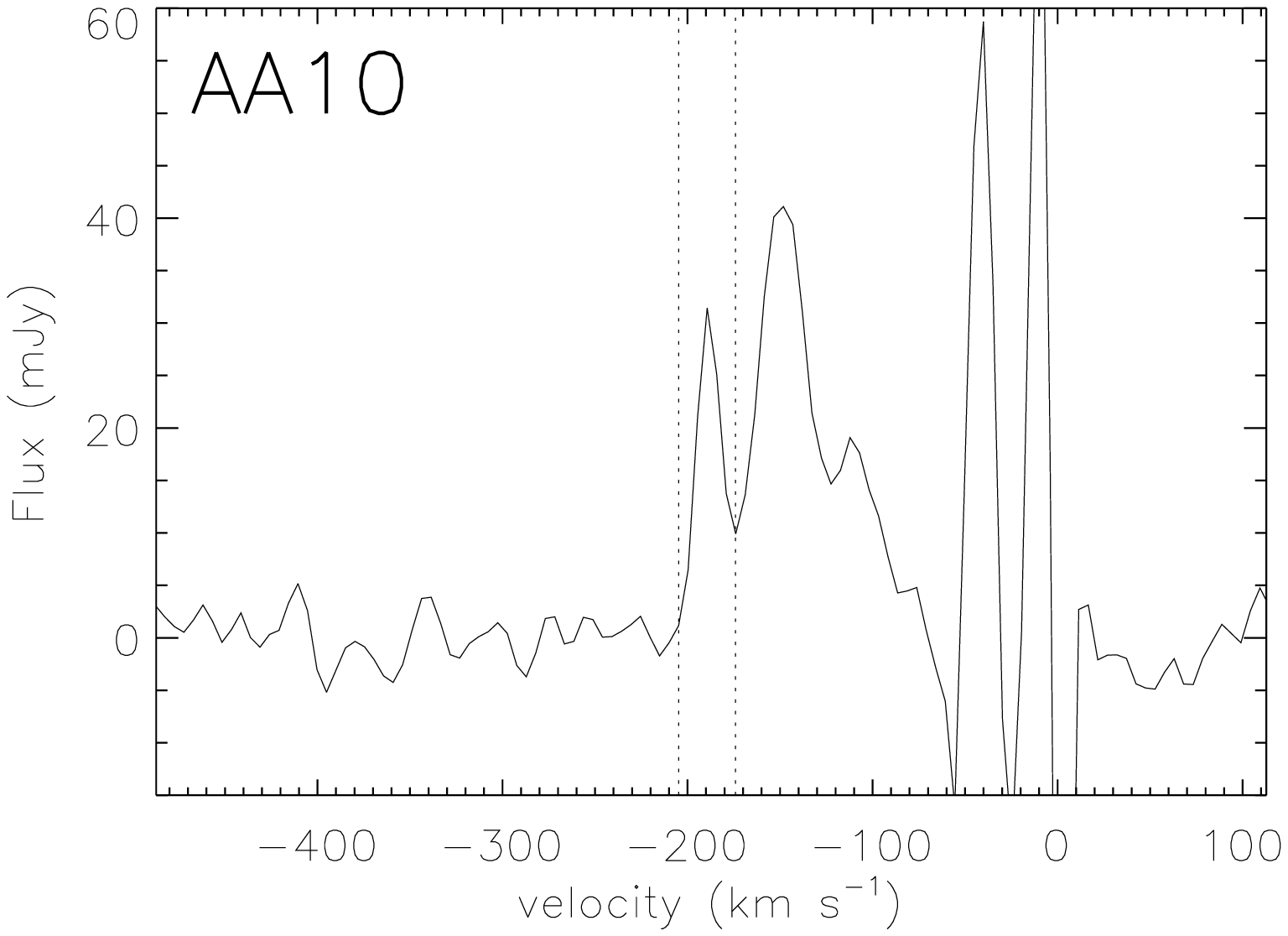}
\includegraphics[width=4cm]{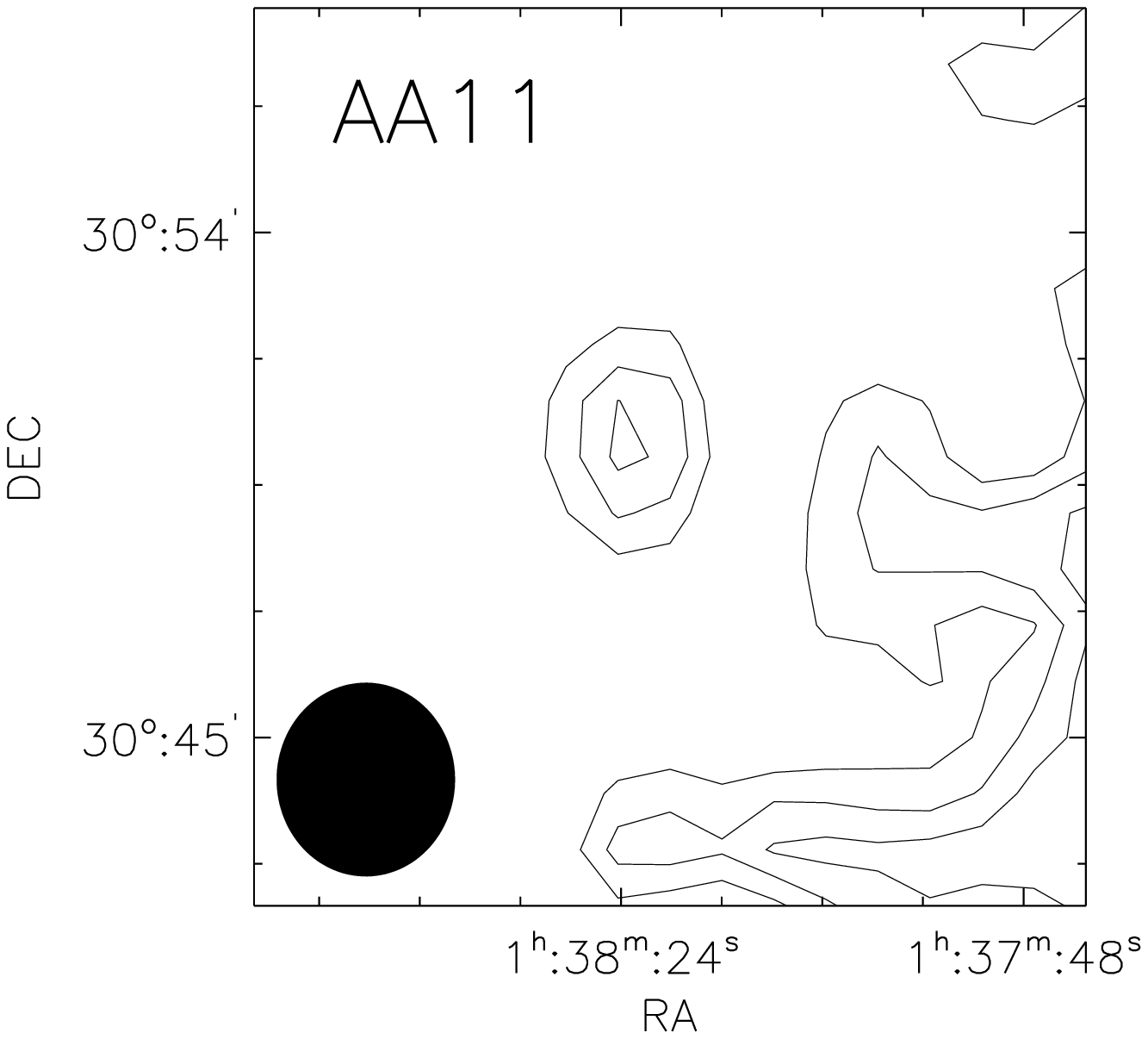}
\includegraphics[width=4cm]{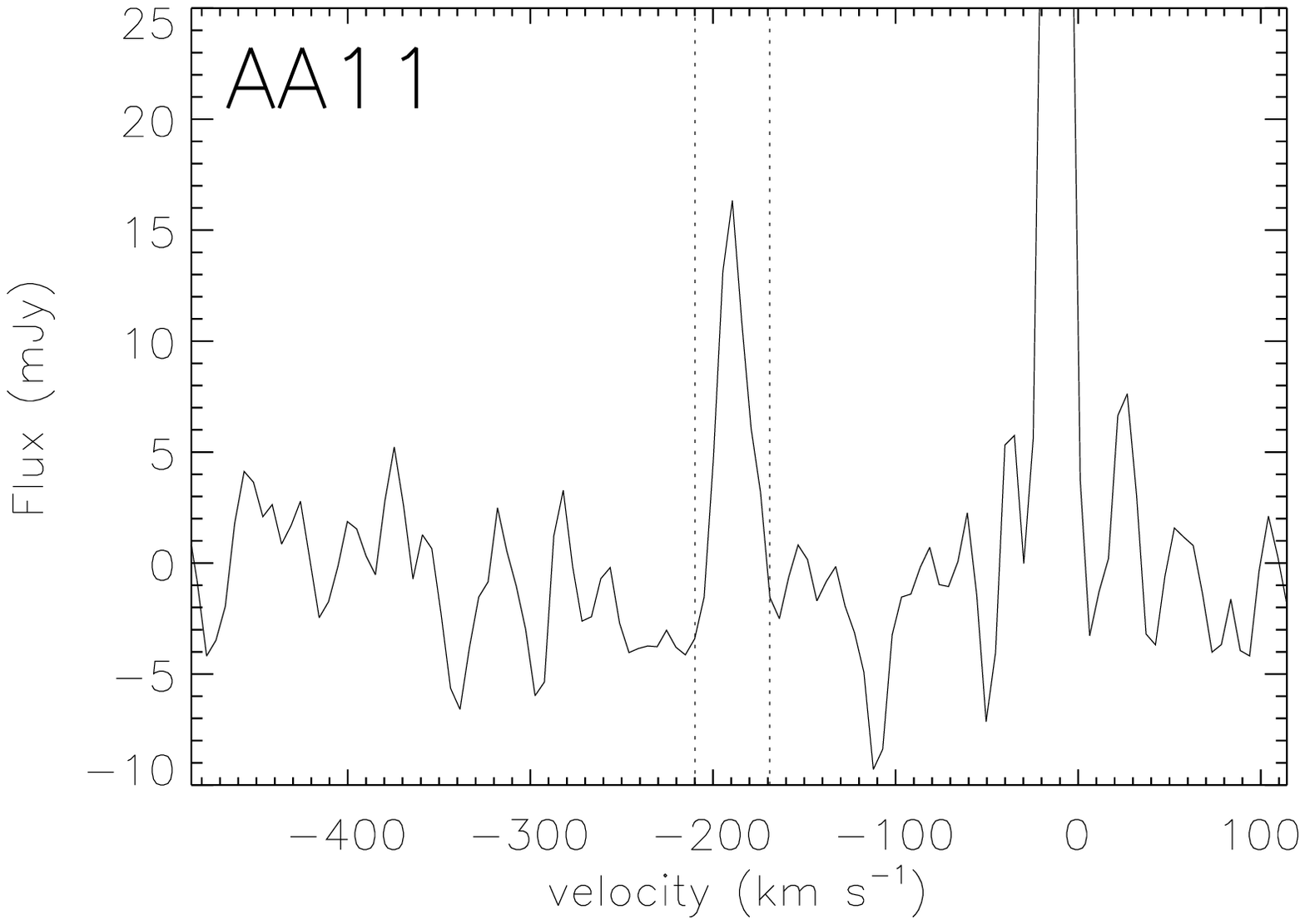}
\includegraphics[width=4cm]{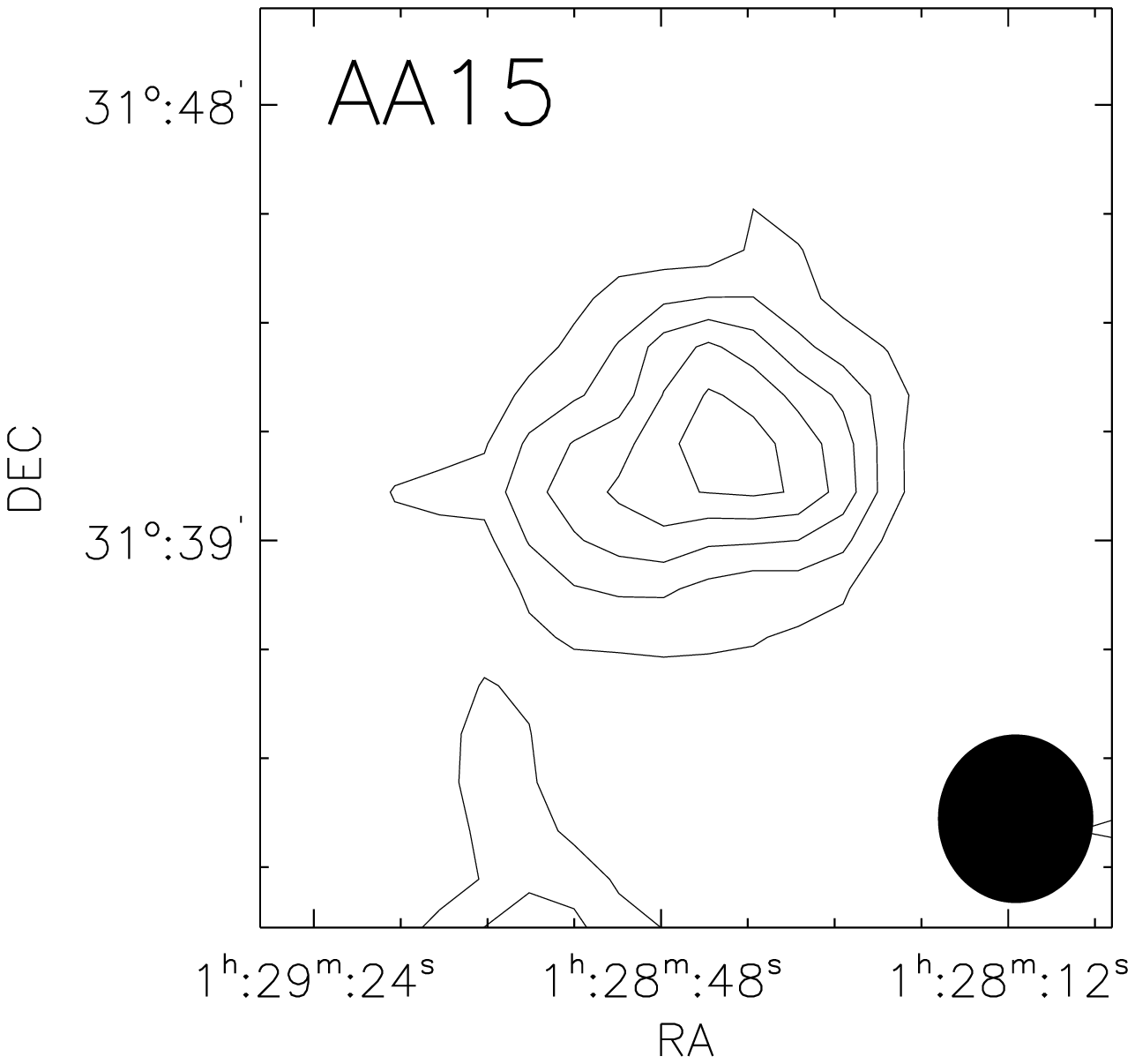}
\includegraphics[width=4cm]{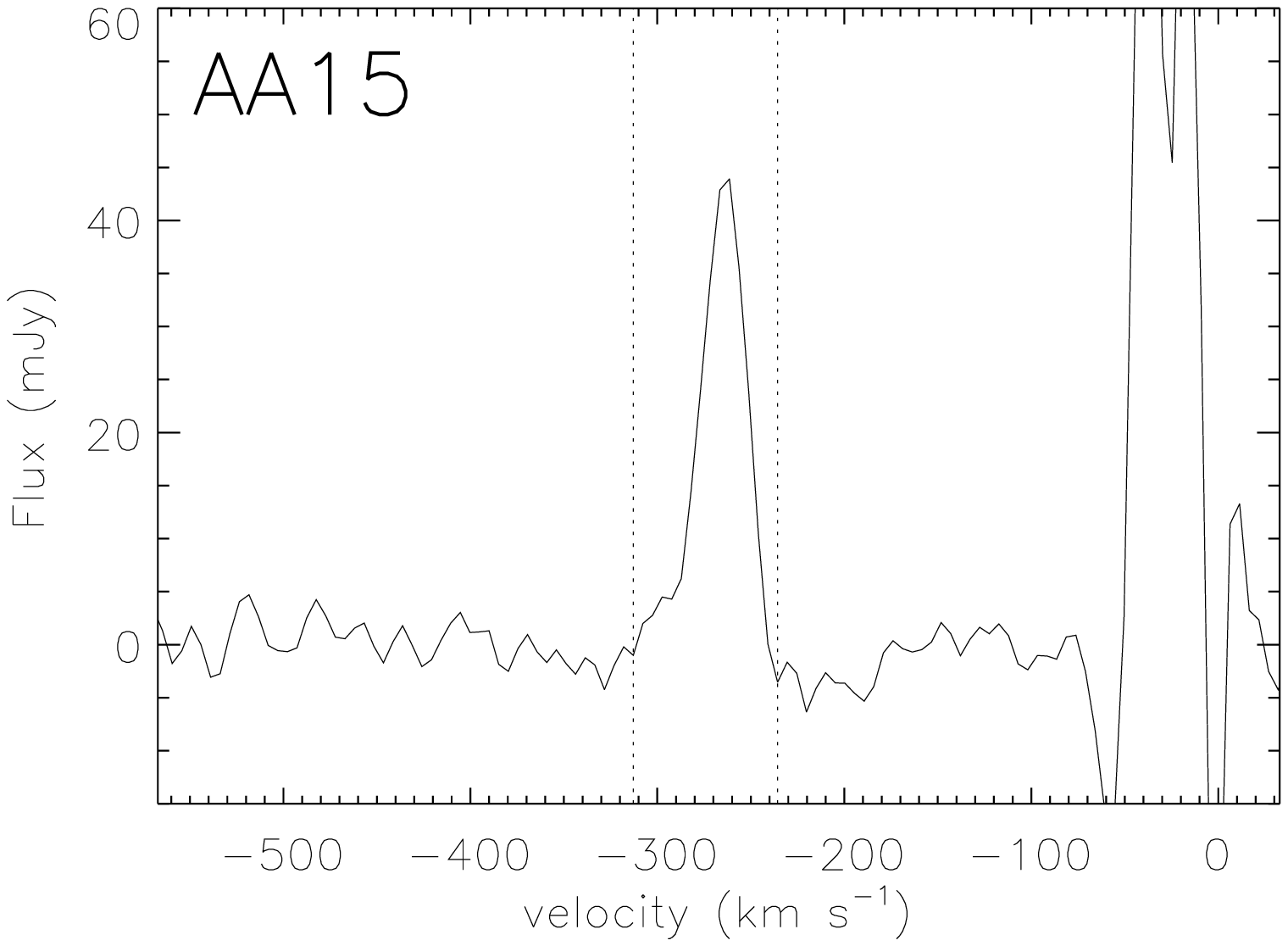}
\hfill
\includegraphics[width=4cm]{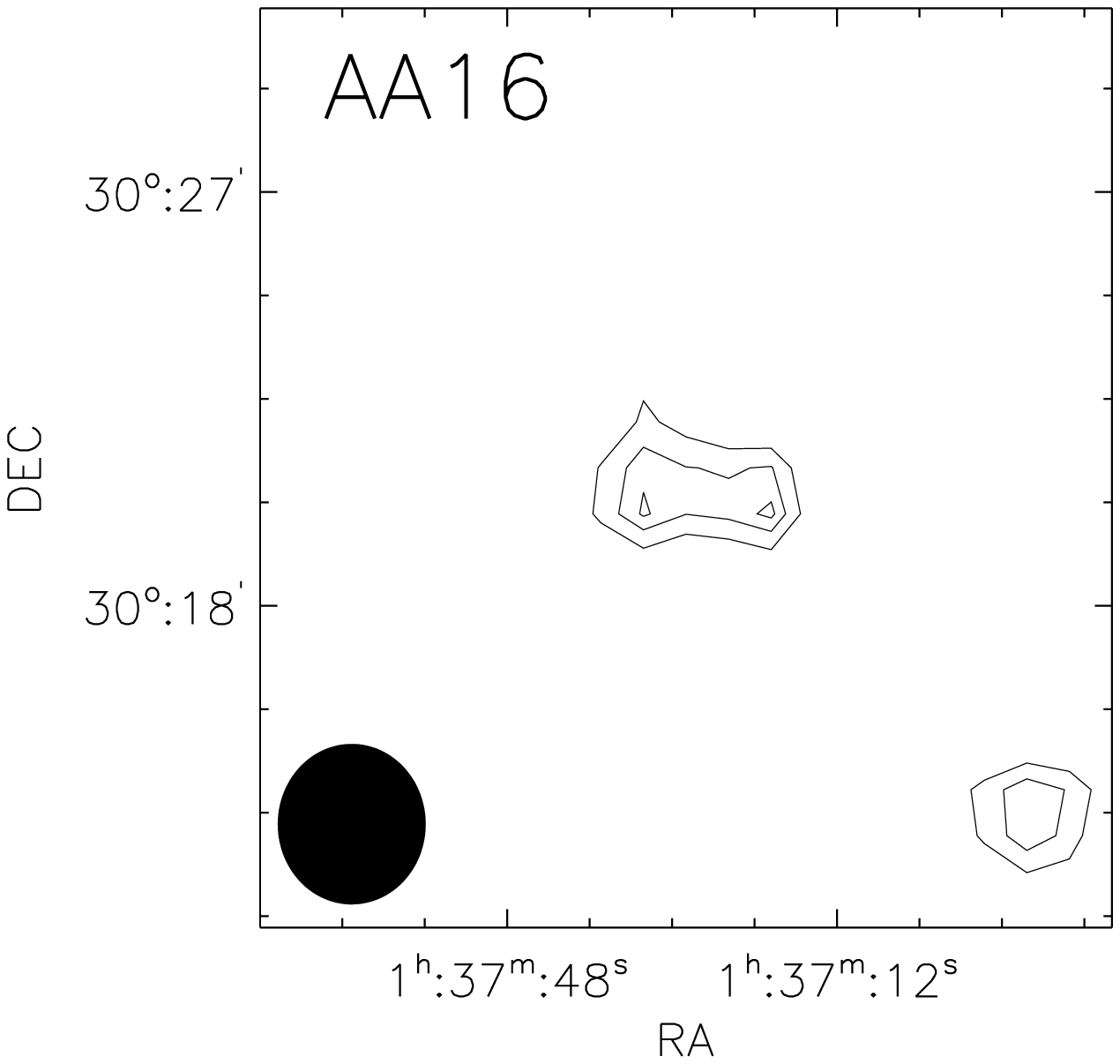}
\hfill
\includegraphics[width=4cm]{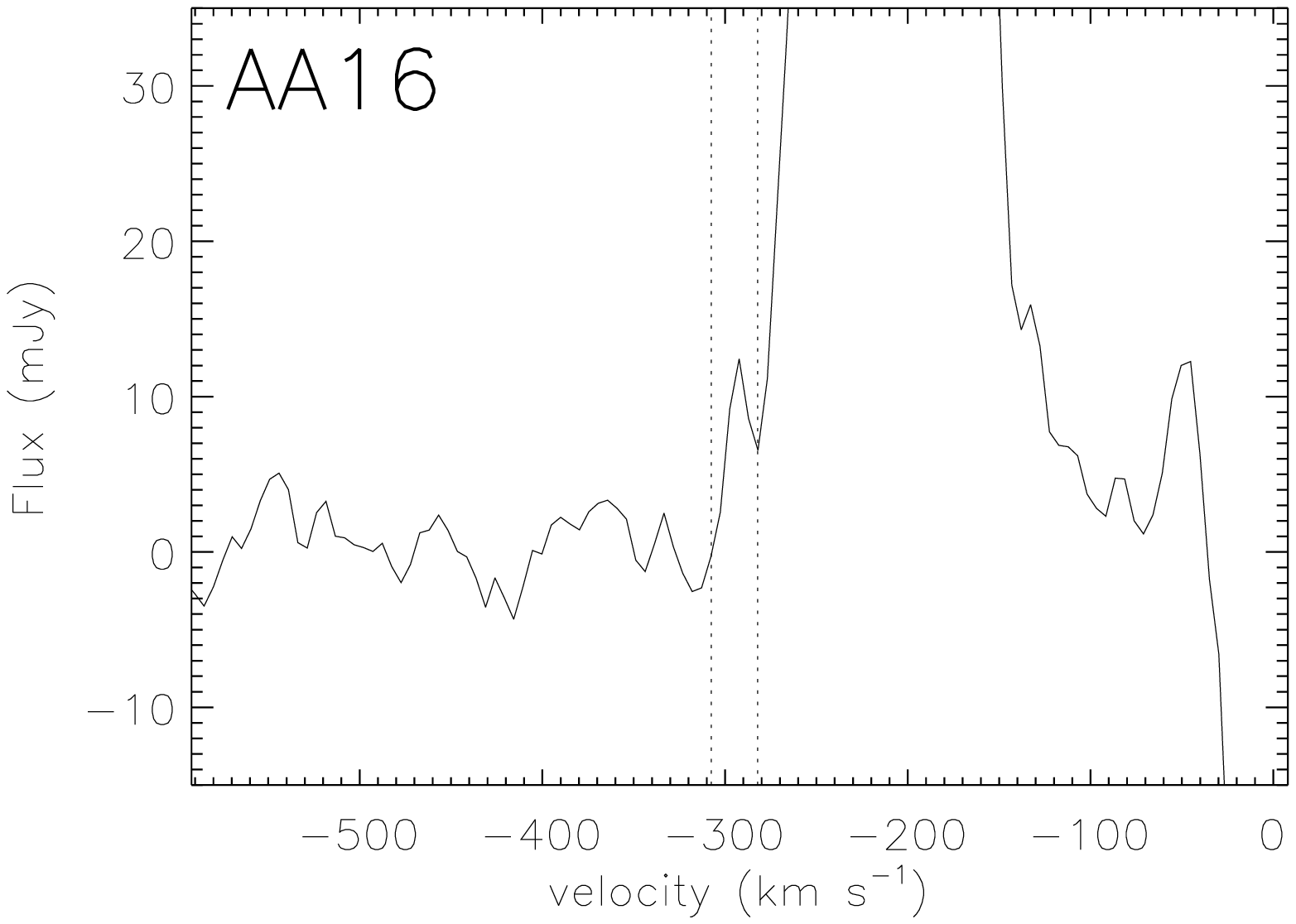}
\hfill
\includegraphics[width=4cm]{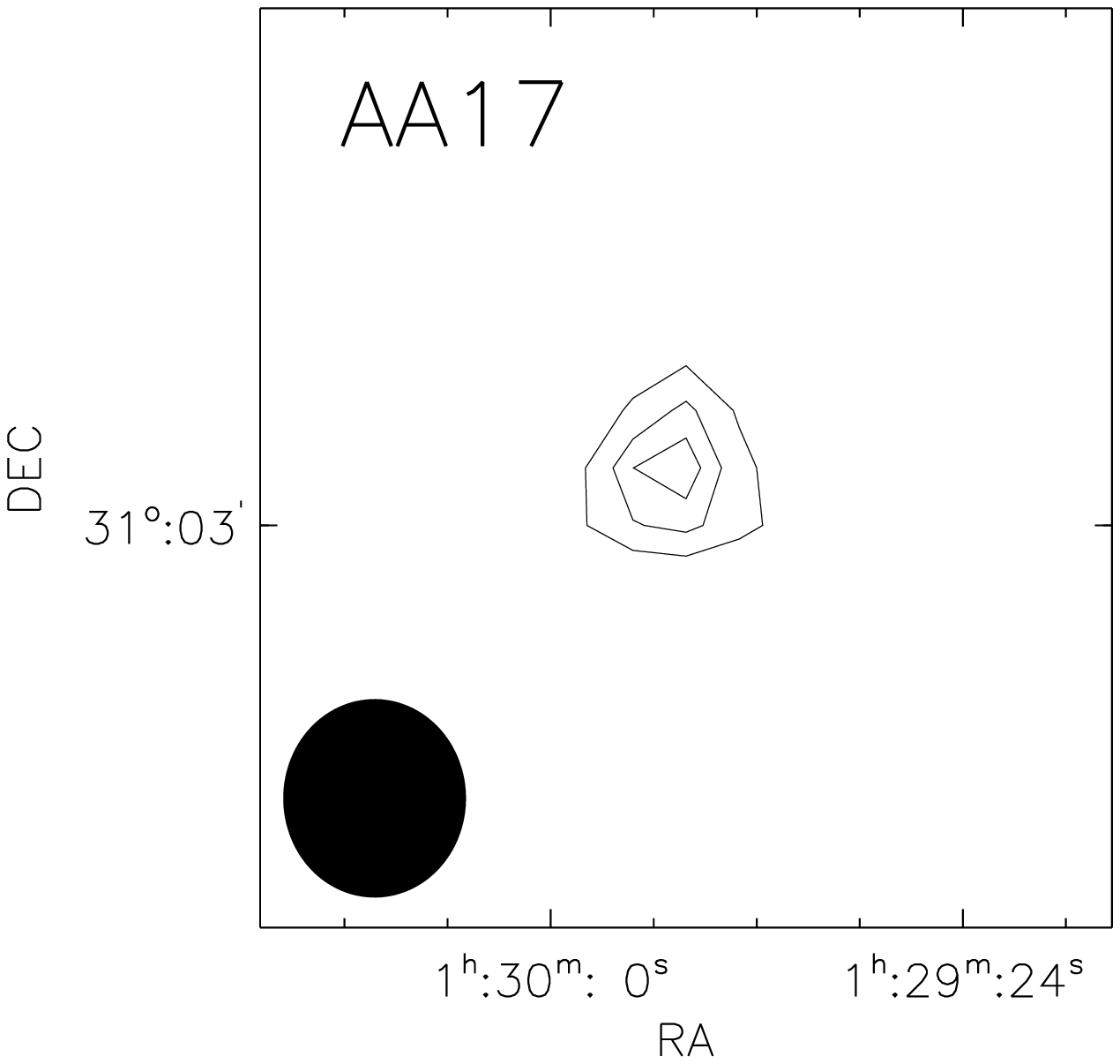}
\hfill
\includegraphics[width=4cm]{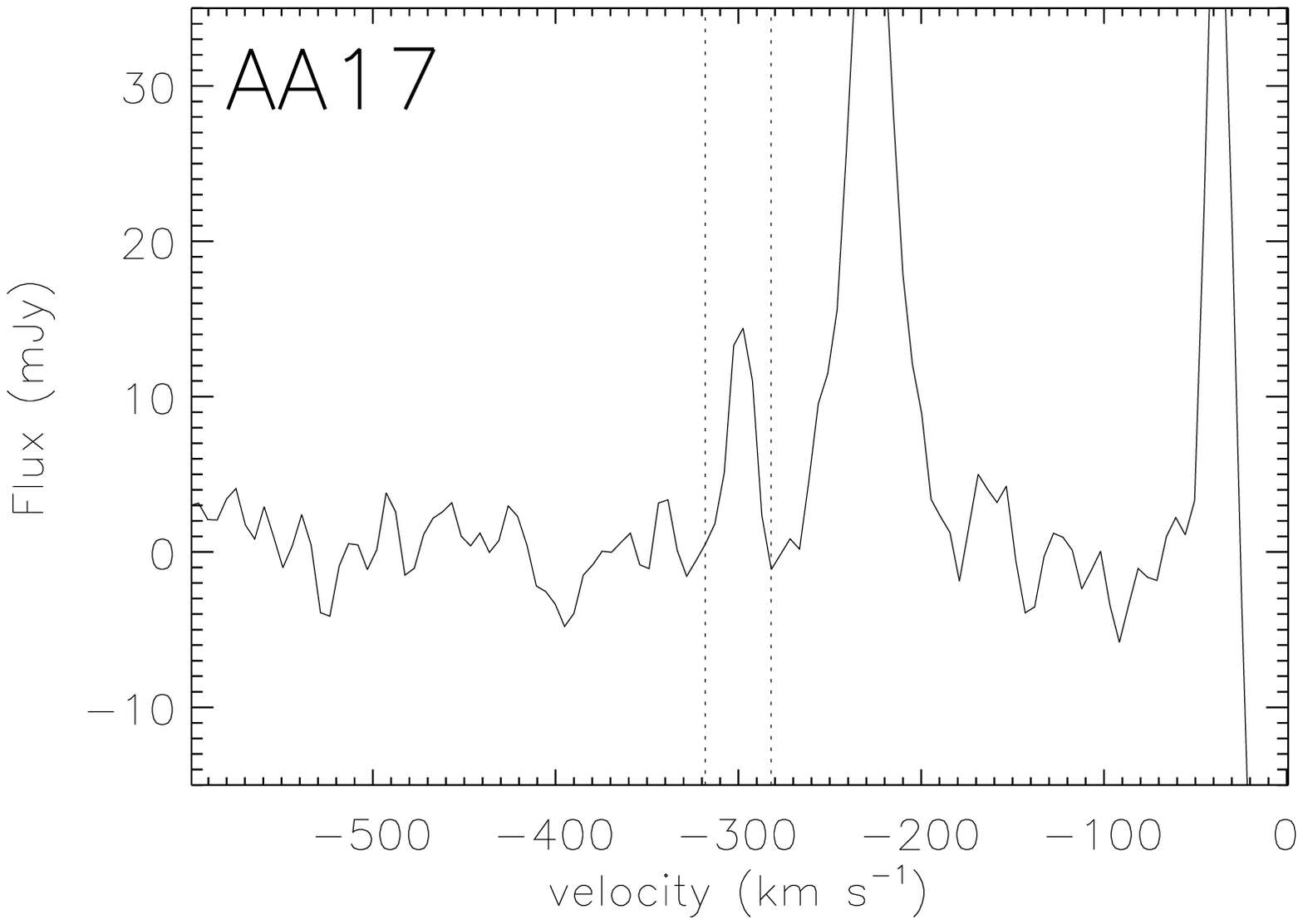}
\label{moment_maps_a} \caption{Column density maps and spectra  of the 'discrete' {\em Type 1} clouds  (see Section 4.2).}
\end{figure*}

\begin{figure*}
\includegraphics[width=4cm]{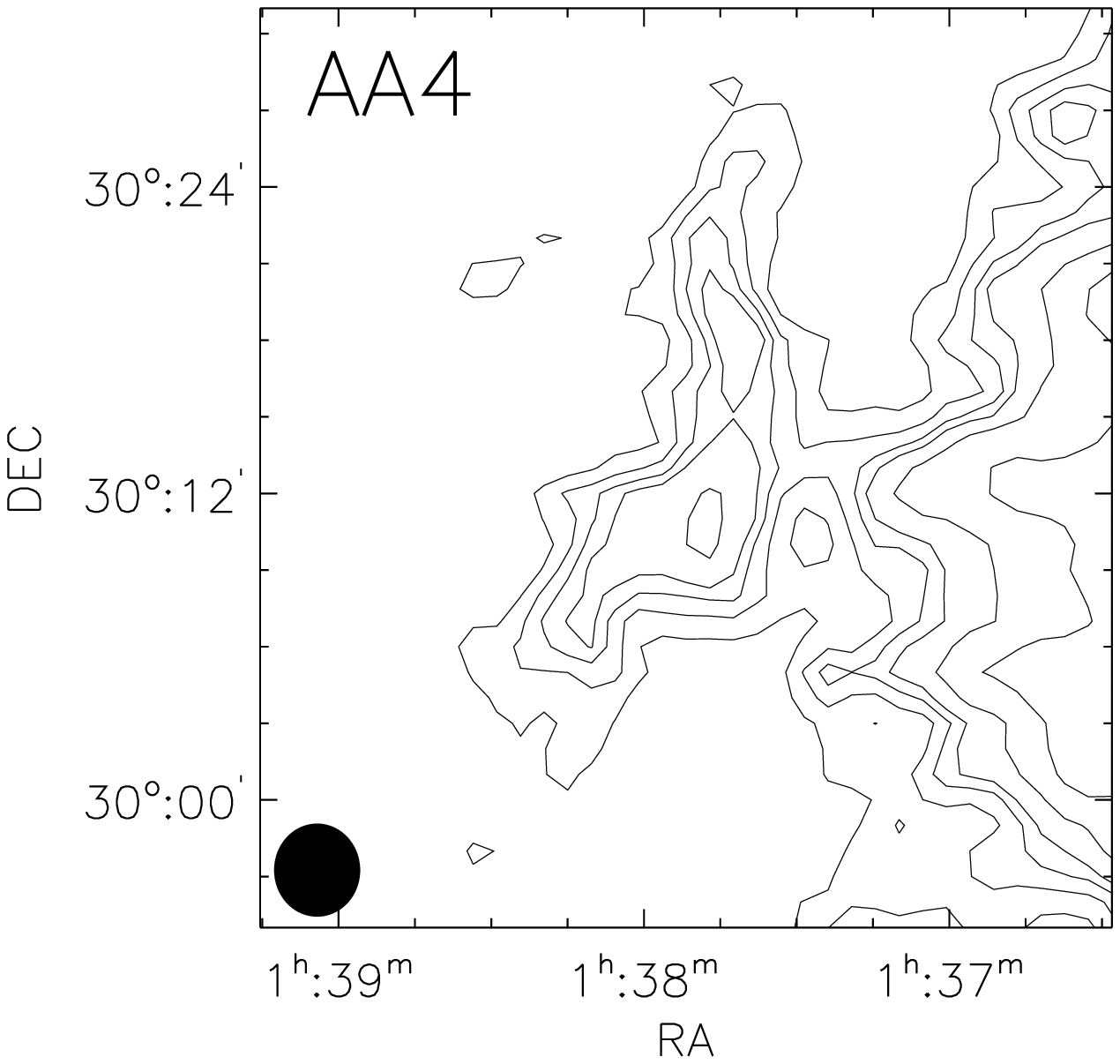}
\includegraphics[width=4cm]{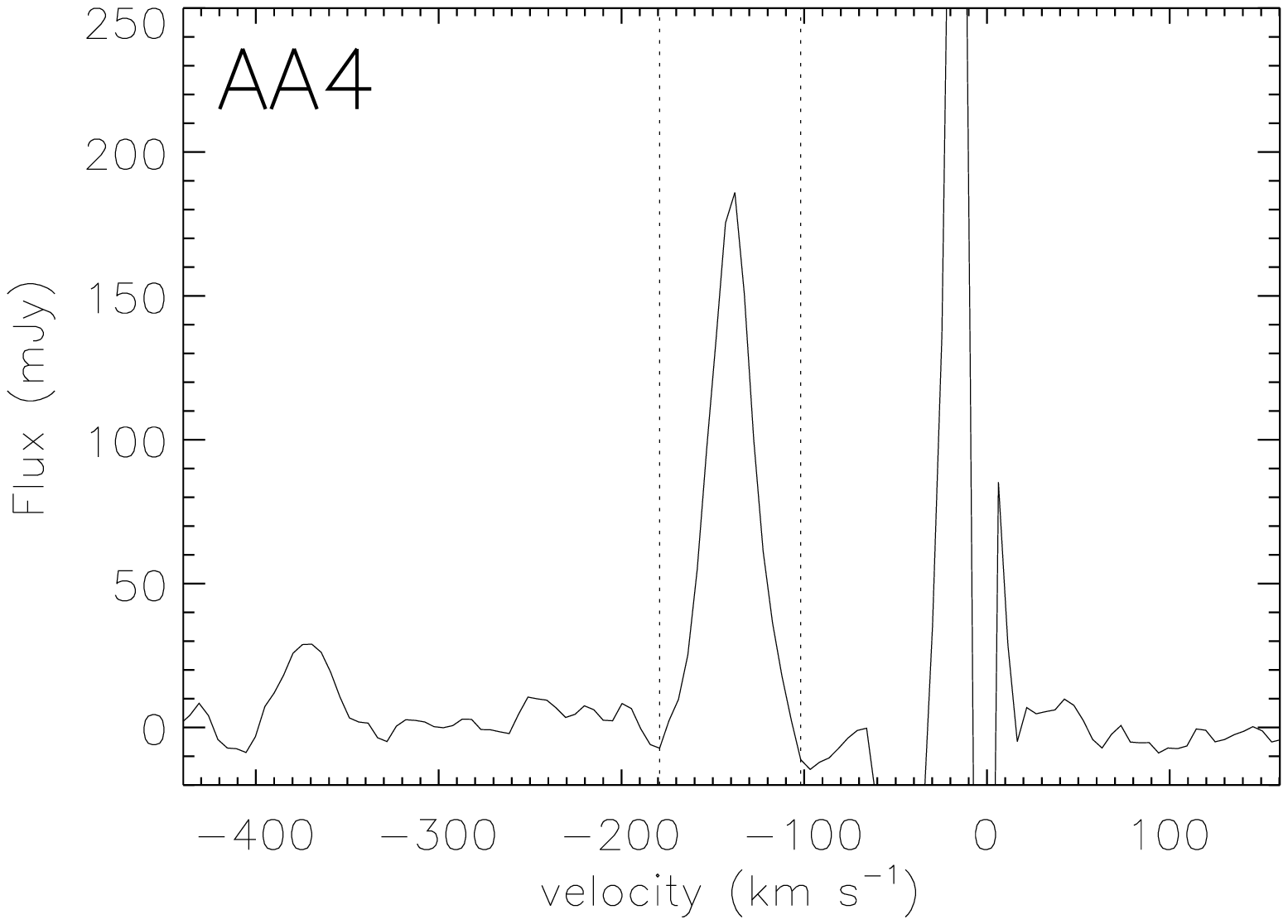}
\includegraphics[width=4cm]{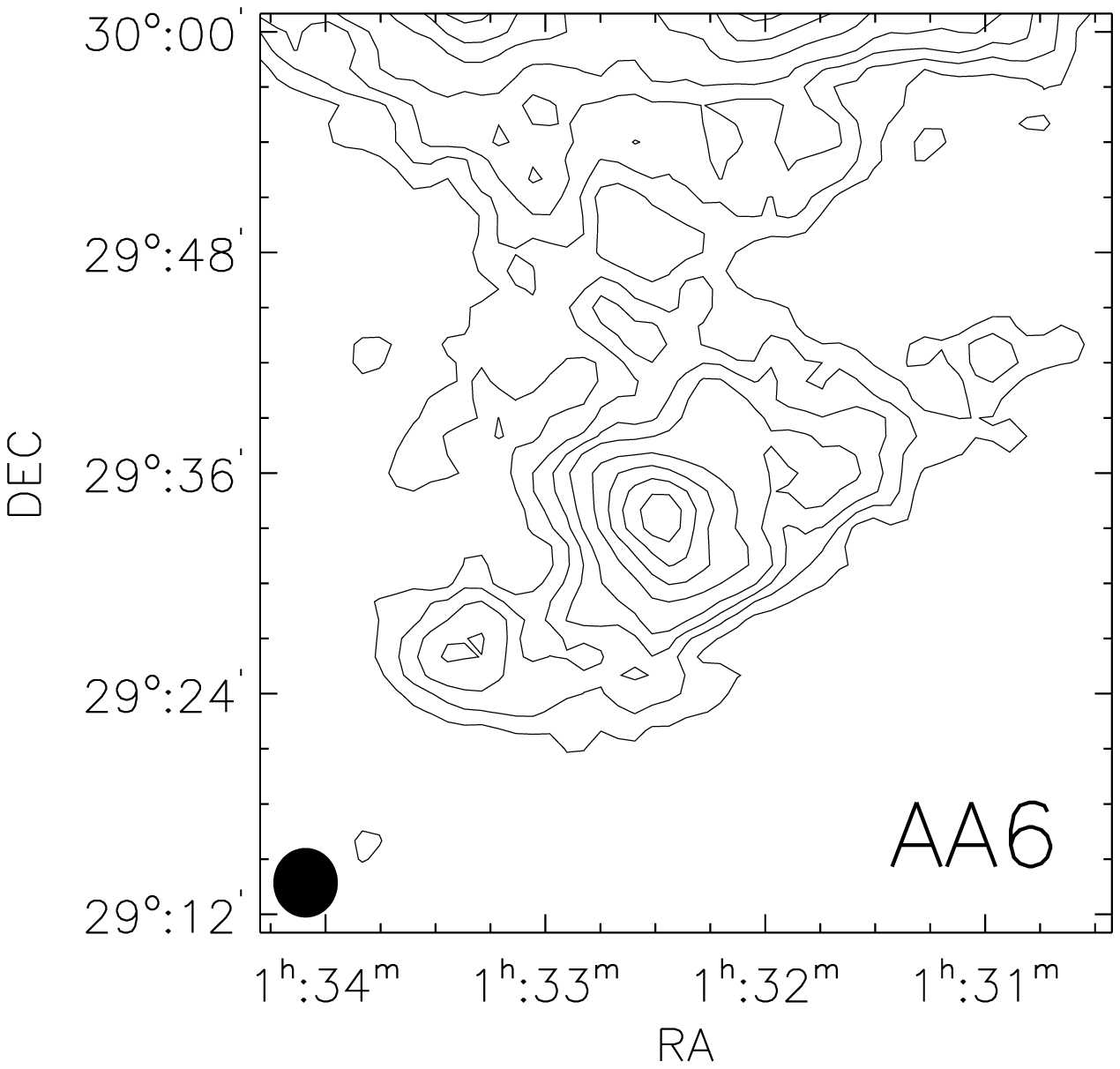}
\includegraphics[width=4cm]{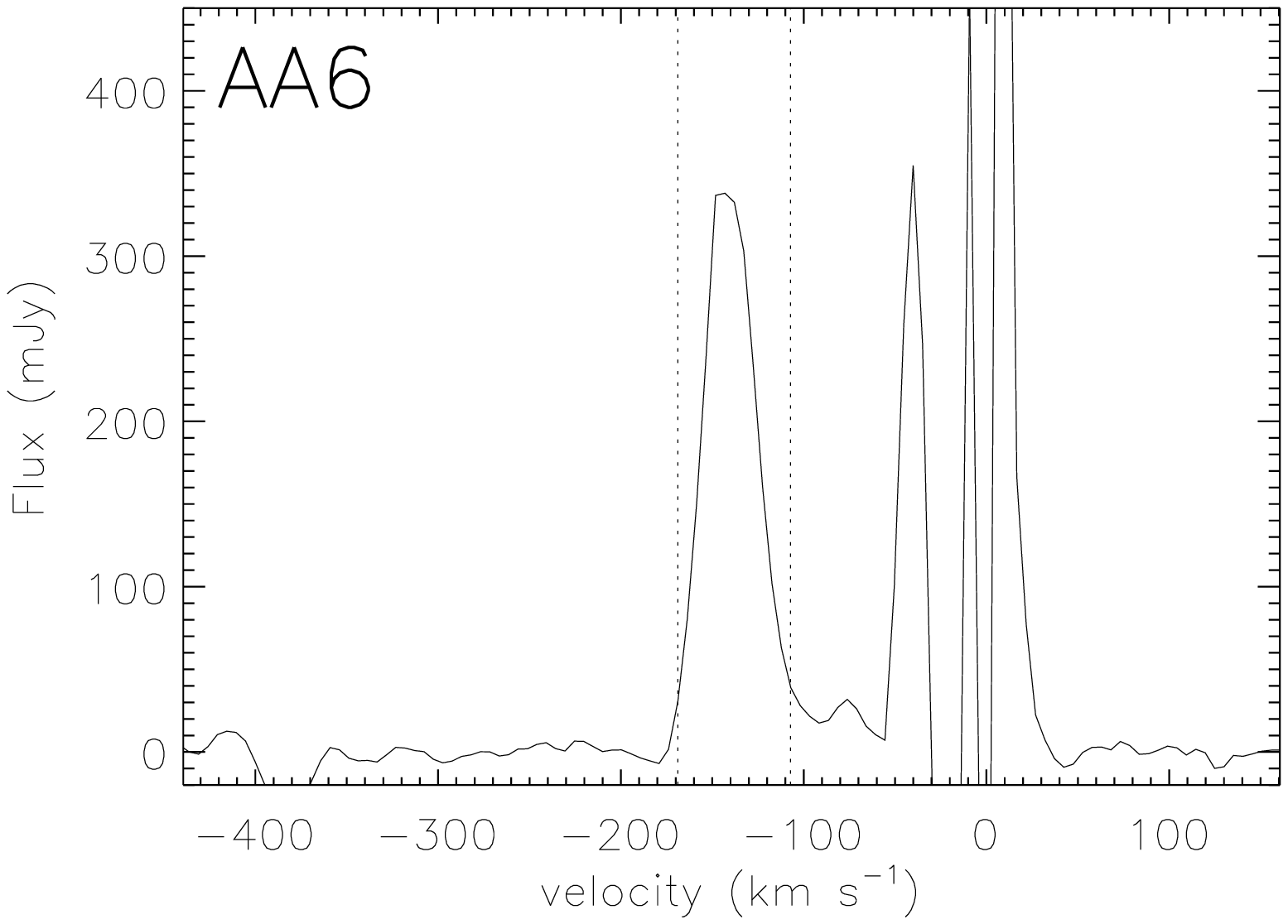}
\includegraphics[width=4cm]{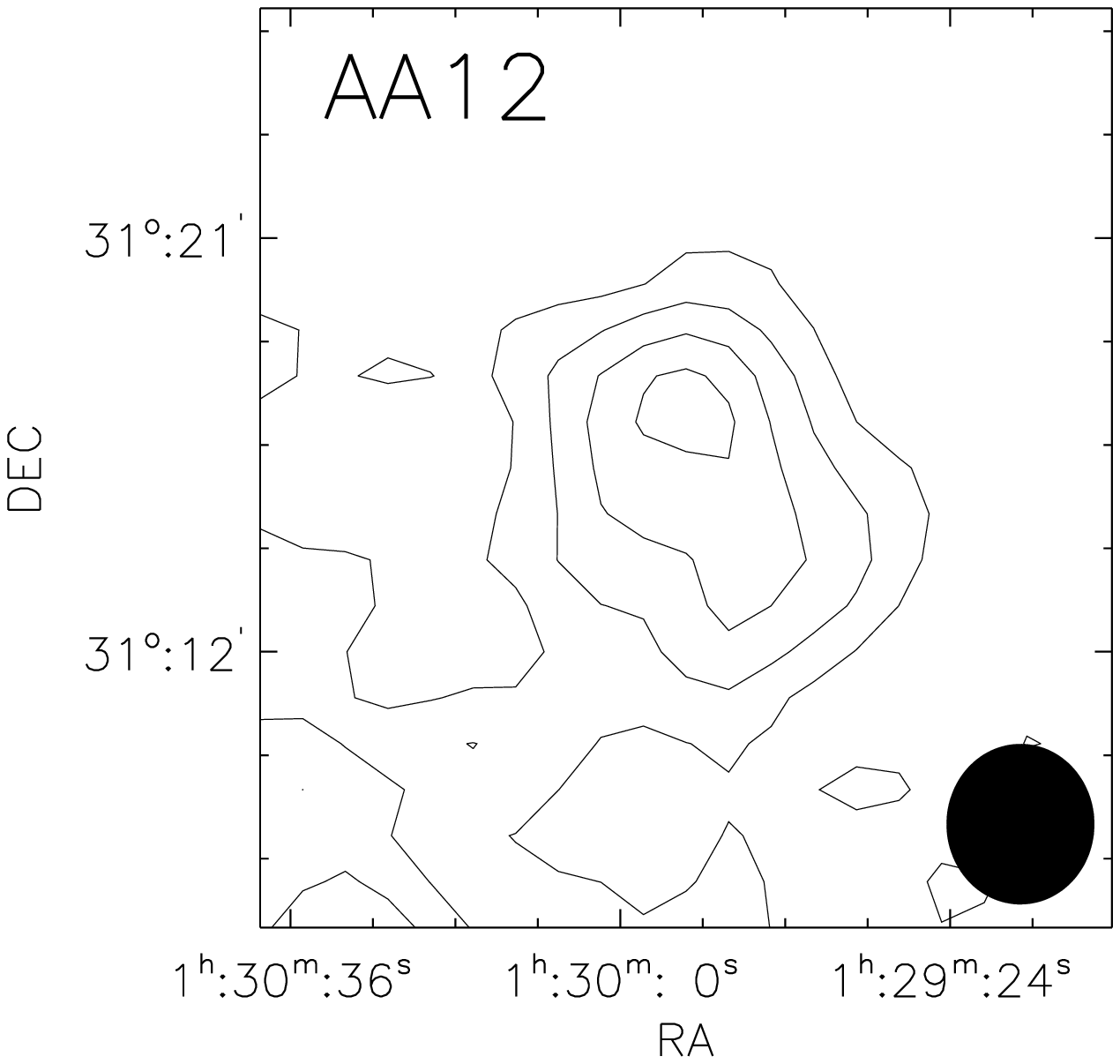}
\includegraphics[width=4cm]{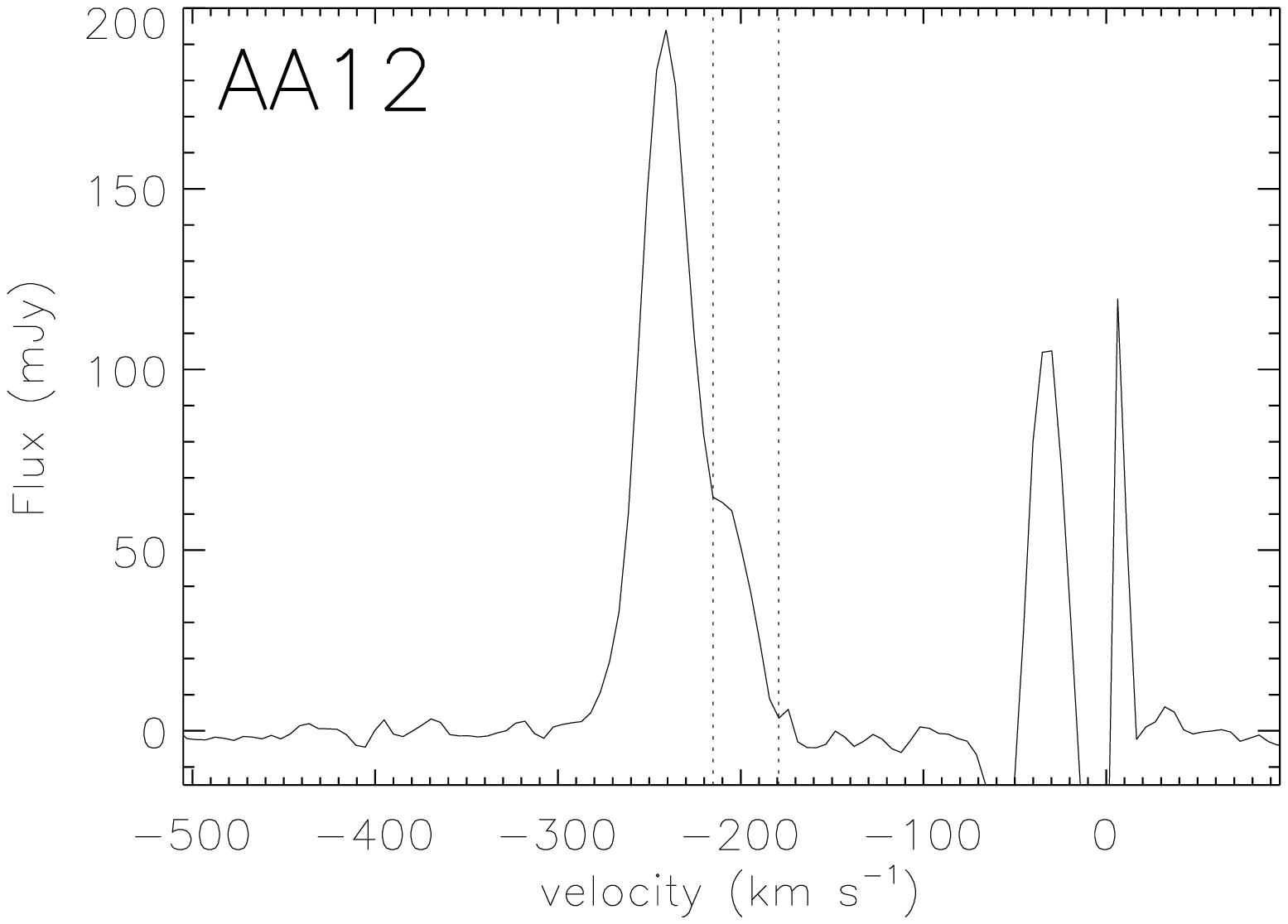}
\includegraphics[width=4cm]{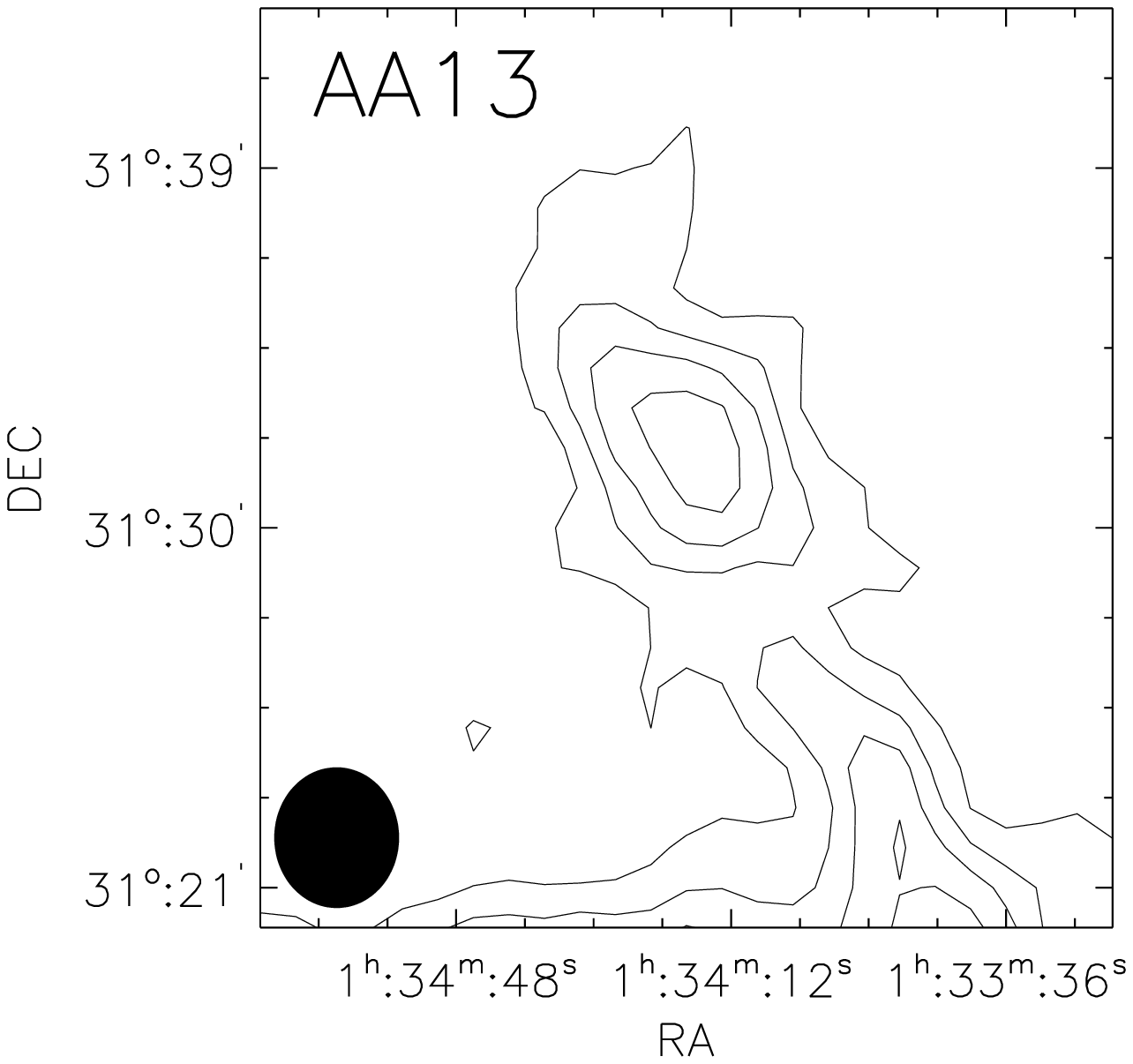}
\includegraphics[width=4cm]{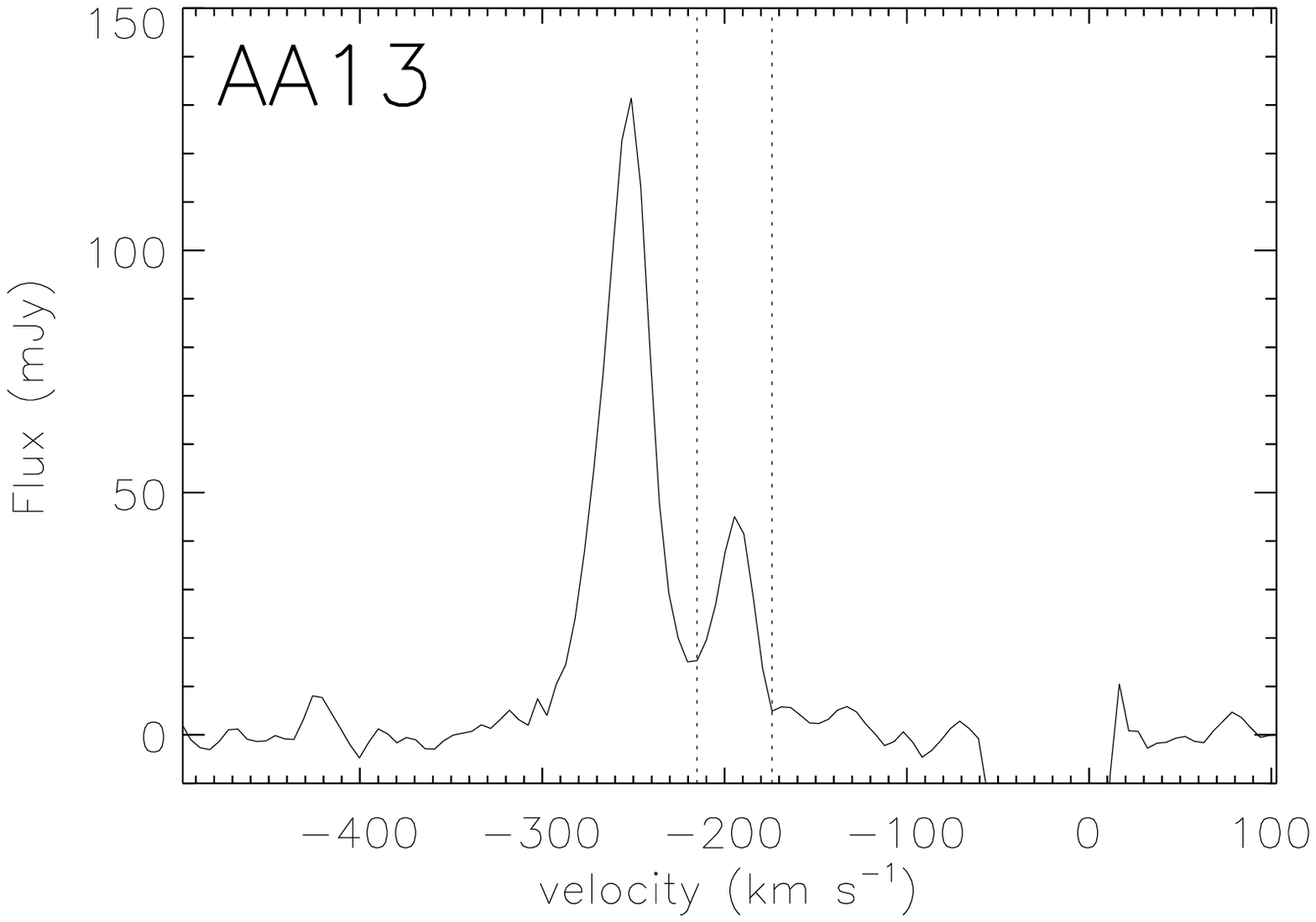}
\hfill
\includegraphics[width=4cm]{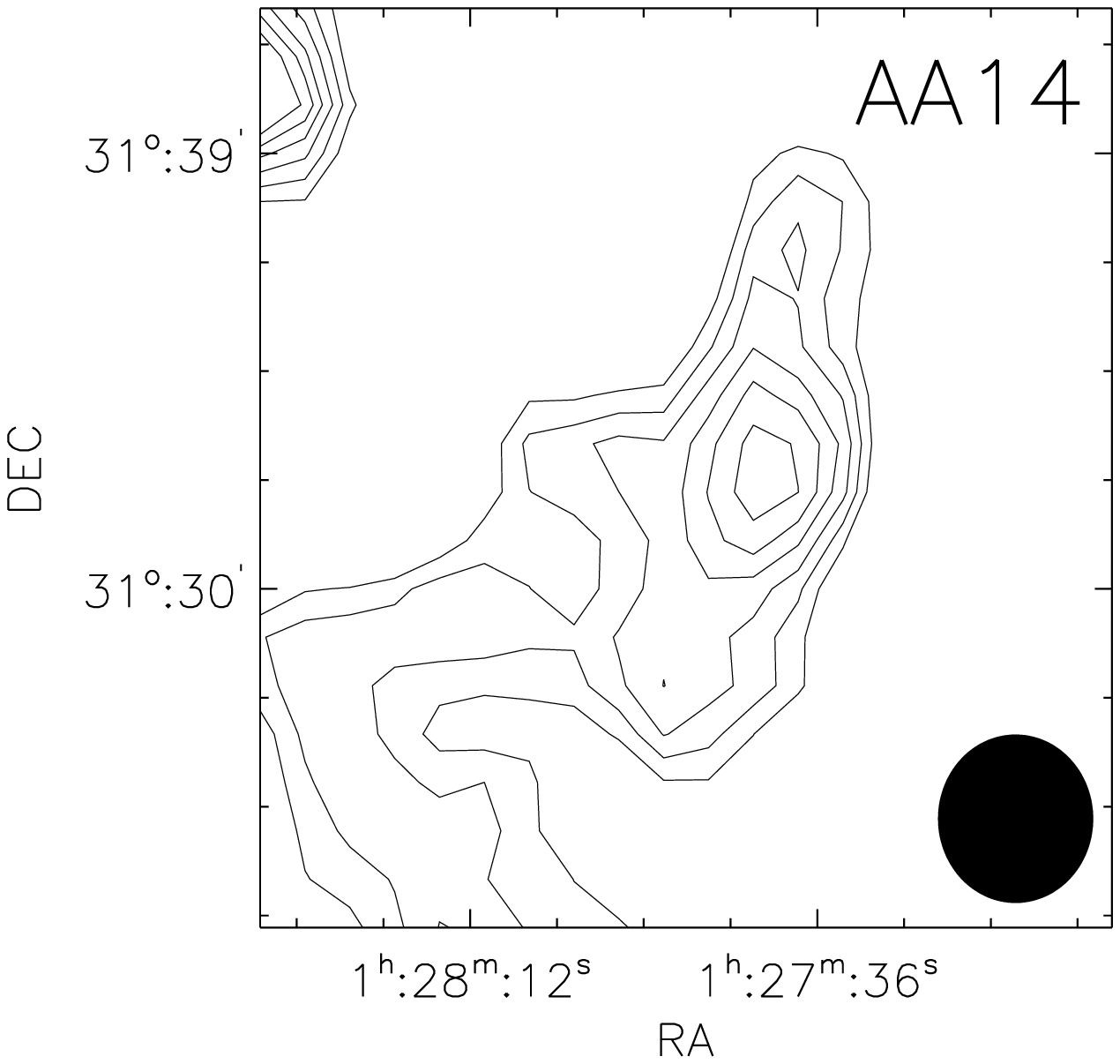}
\hfill
\includegraphics[width=4cm]{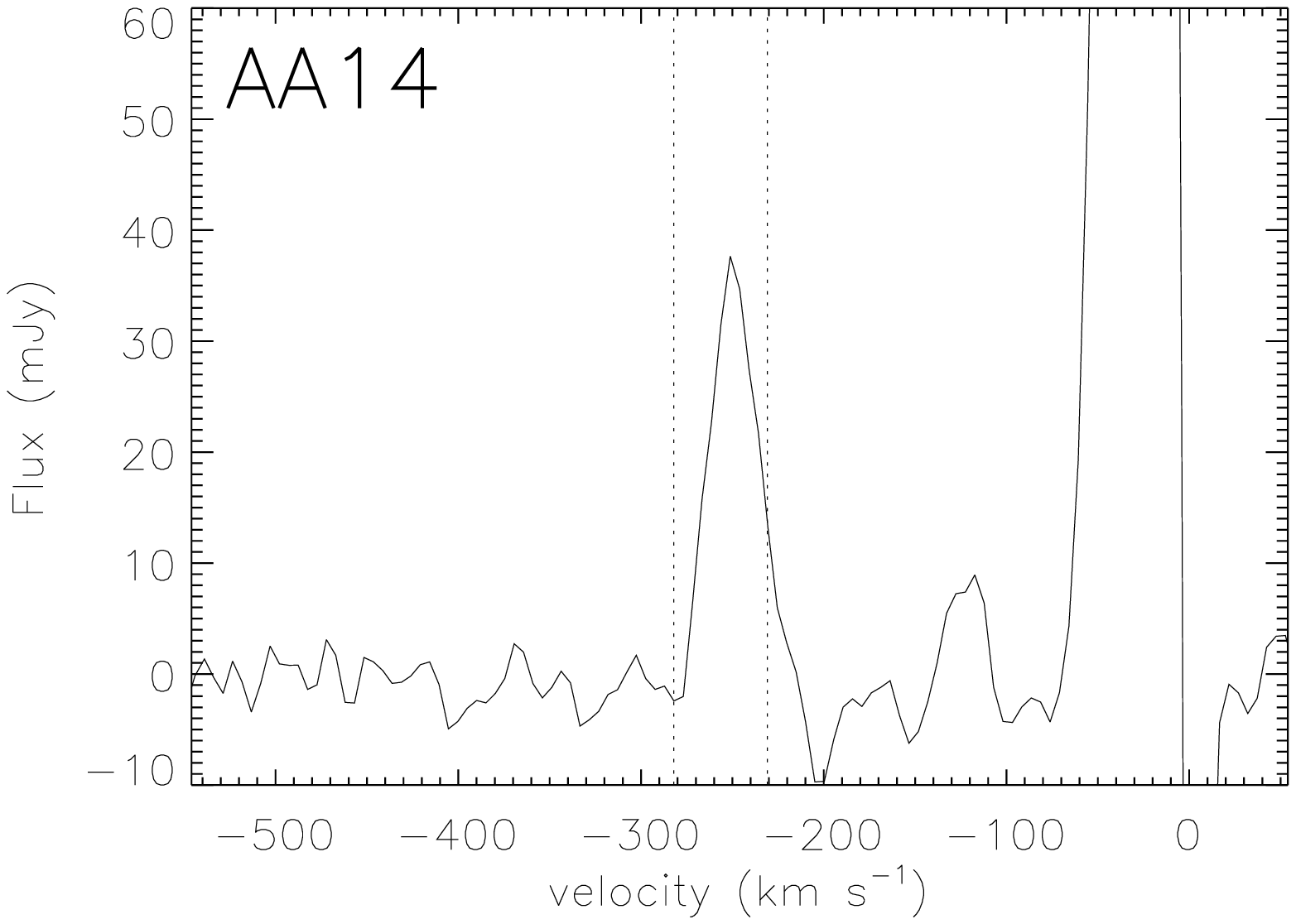}
\hfill
\hfill
\hfill
\hfill
\hfill
\hfill
\hfill
\hfill
\hfill
\hfill
\hfill
\hfill
\hfill
\label{moment_maps_b} \caption{Column density maps and spectra of {\em Type 1} clouds which are spatially connected to the disc of M33  (see Section 4.1).}
\end{figure*}

\begin{figure*}
\includegraphics[width=4cm]{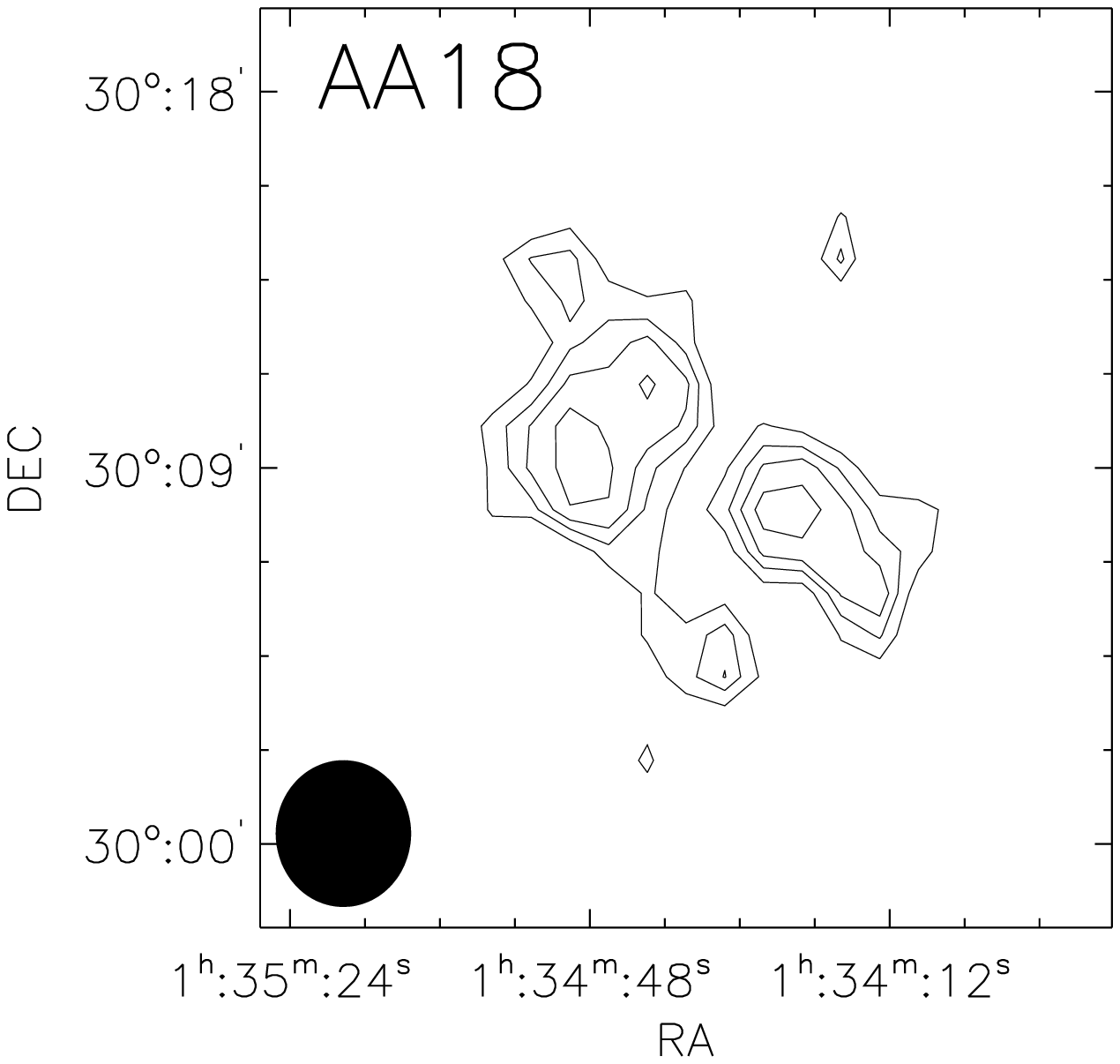}
\includegraphics[width=4cm]{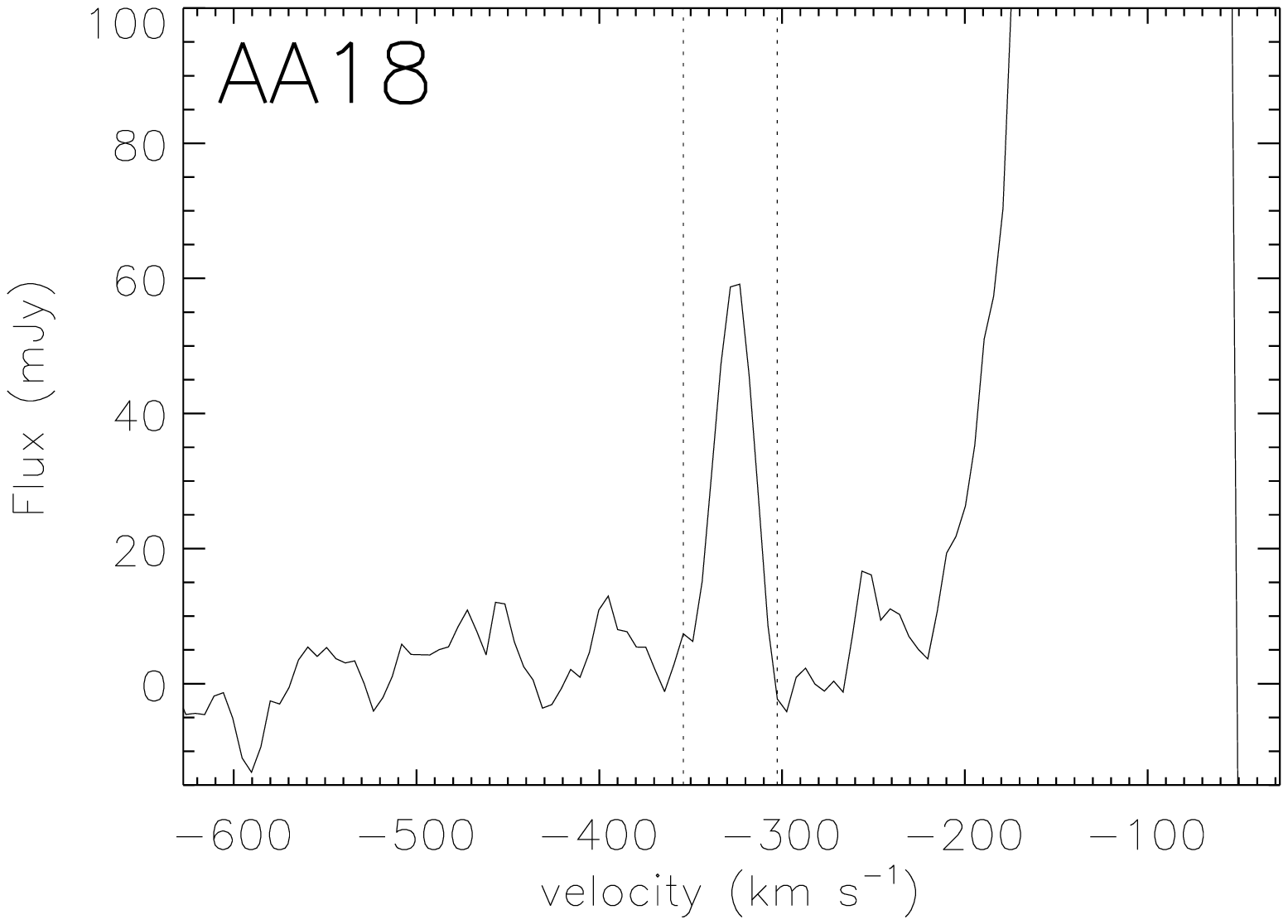}
\includegraphics[width=4cm]{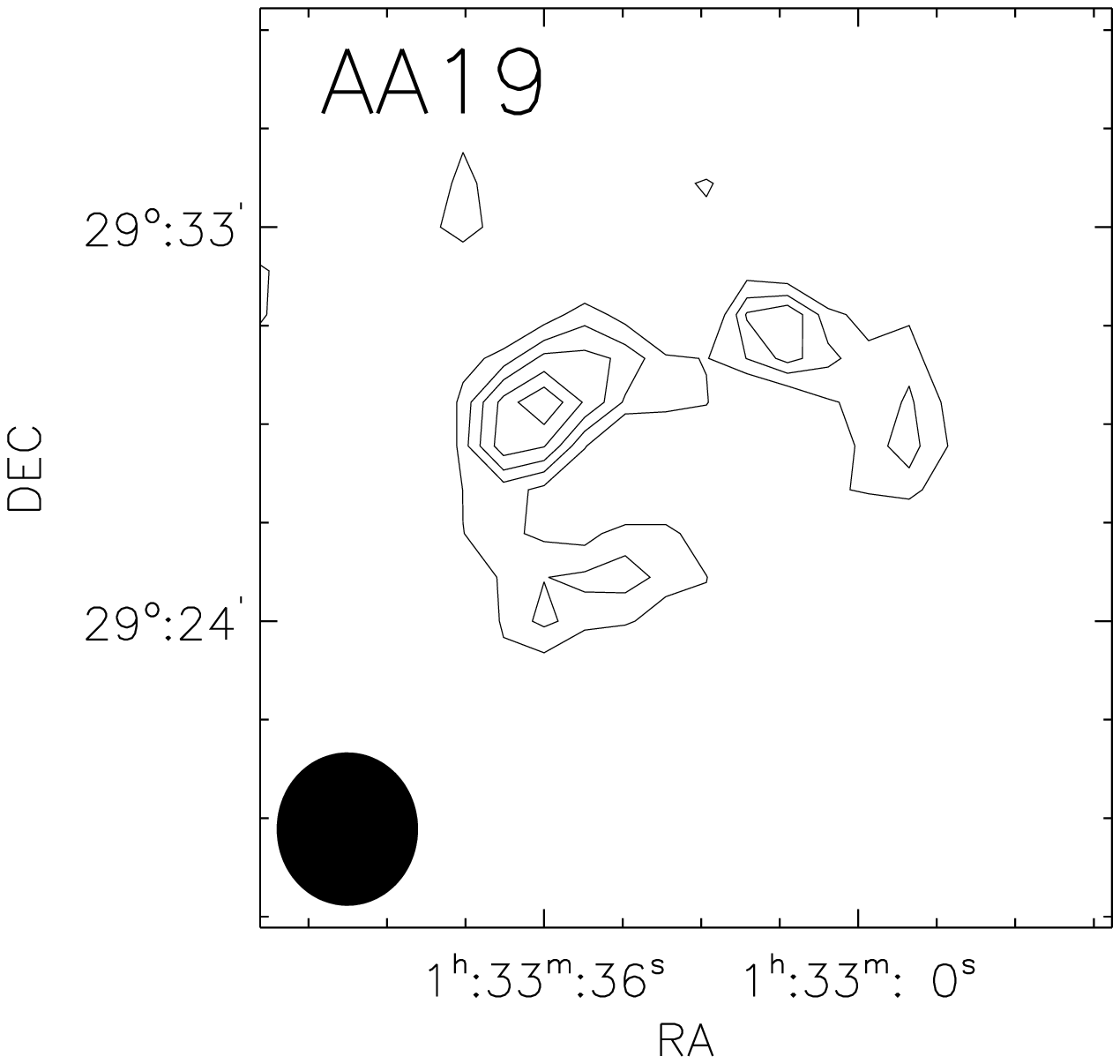}
\includegraphics[width=4cm]{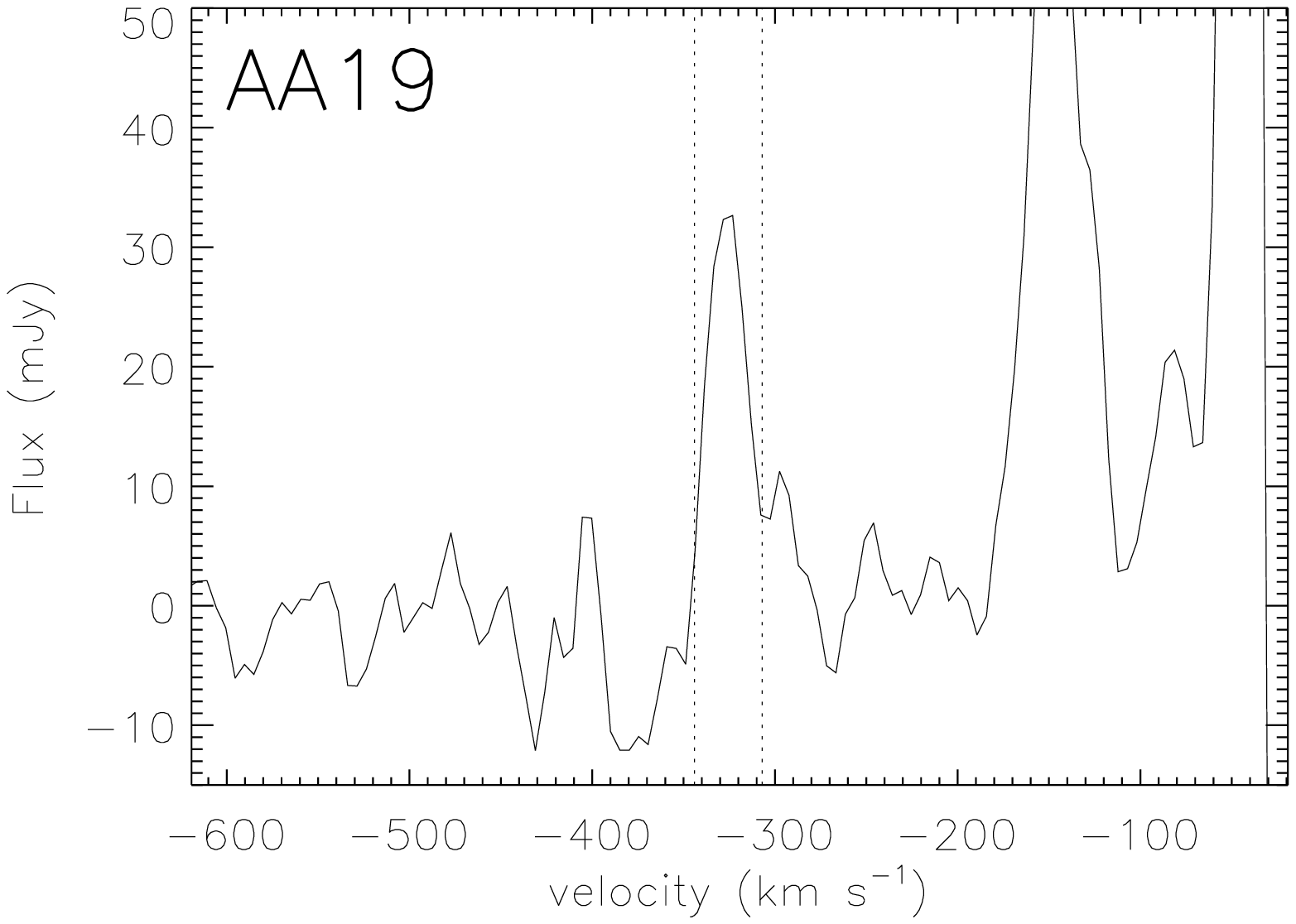}
\hfill
\includegraphics[width=4cm]{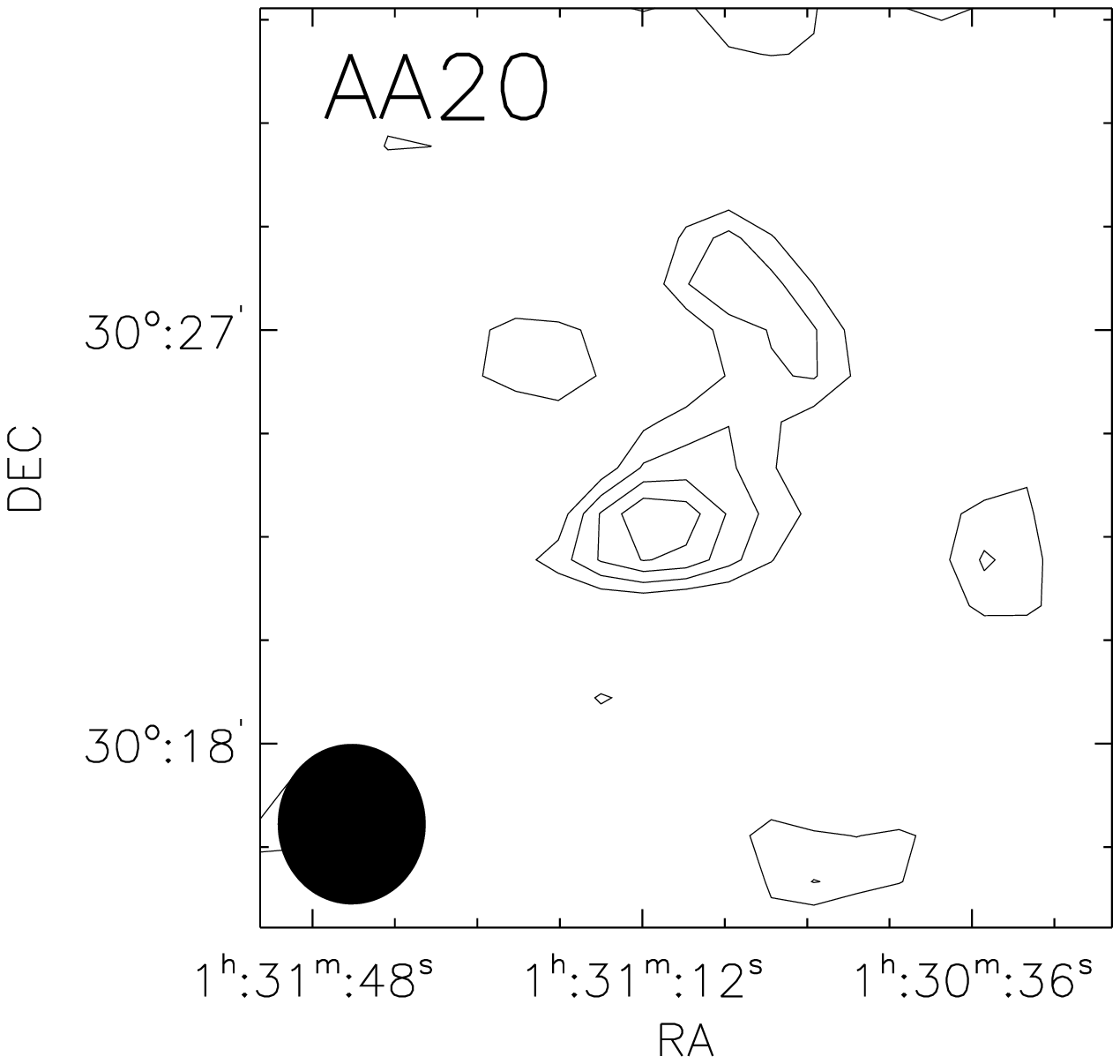}
\hfill
\includegraphics[width=4cm]{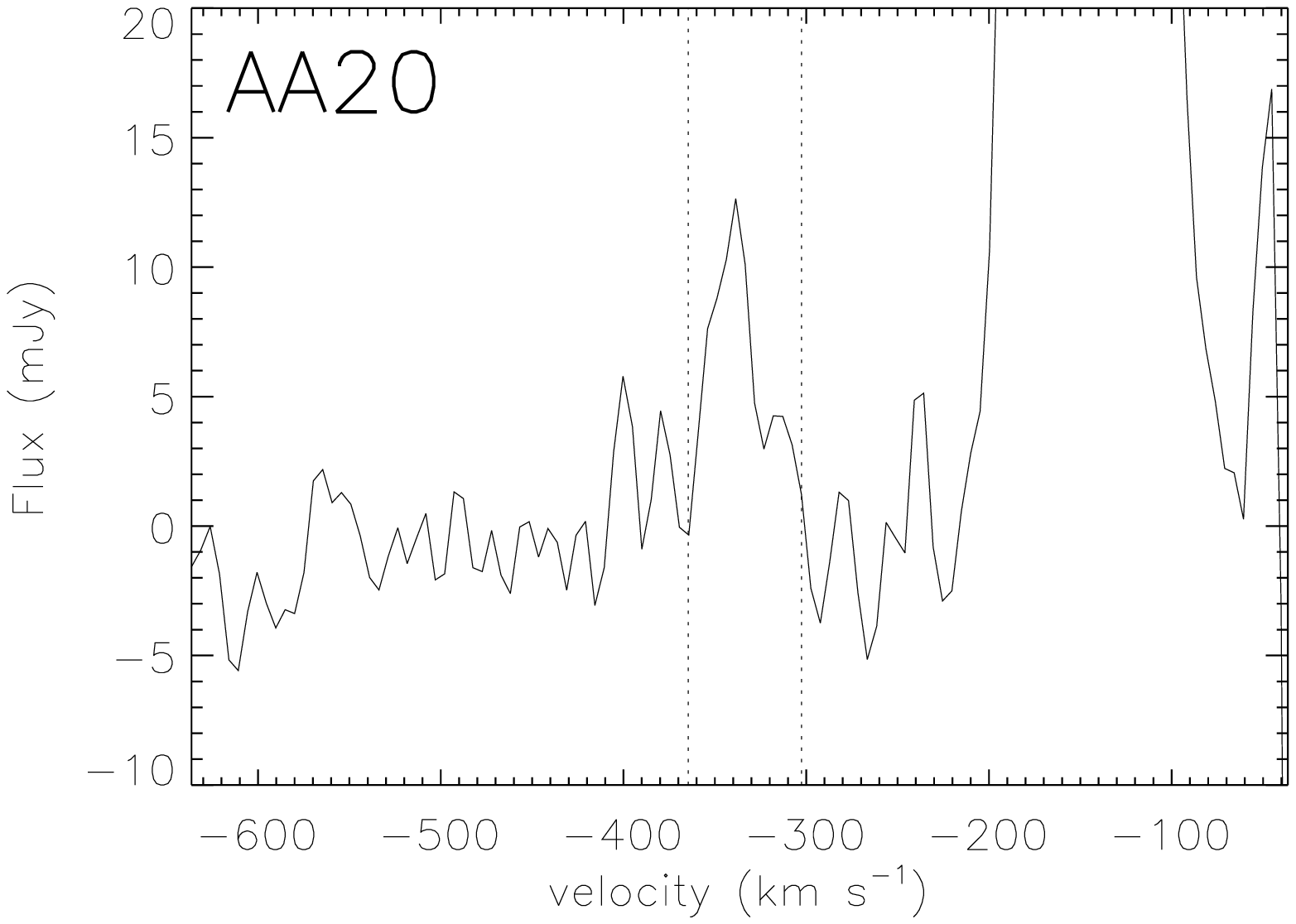}
\hfill
\includegraphics[width=4cm]{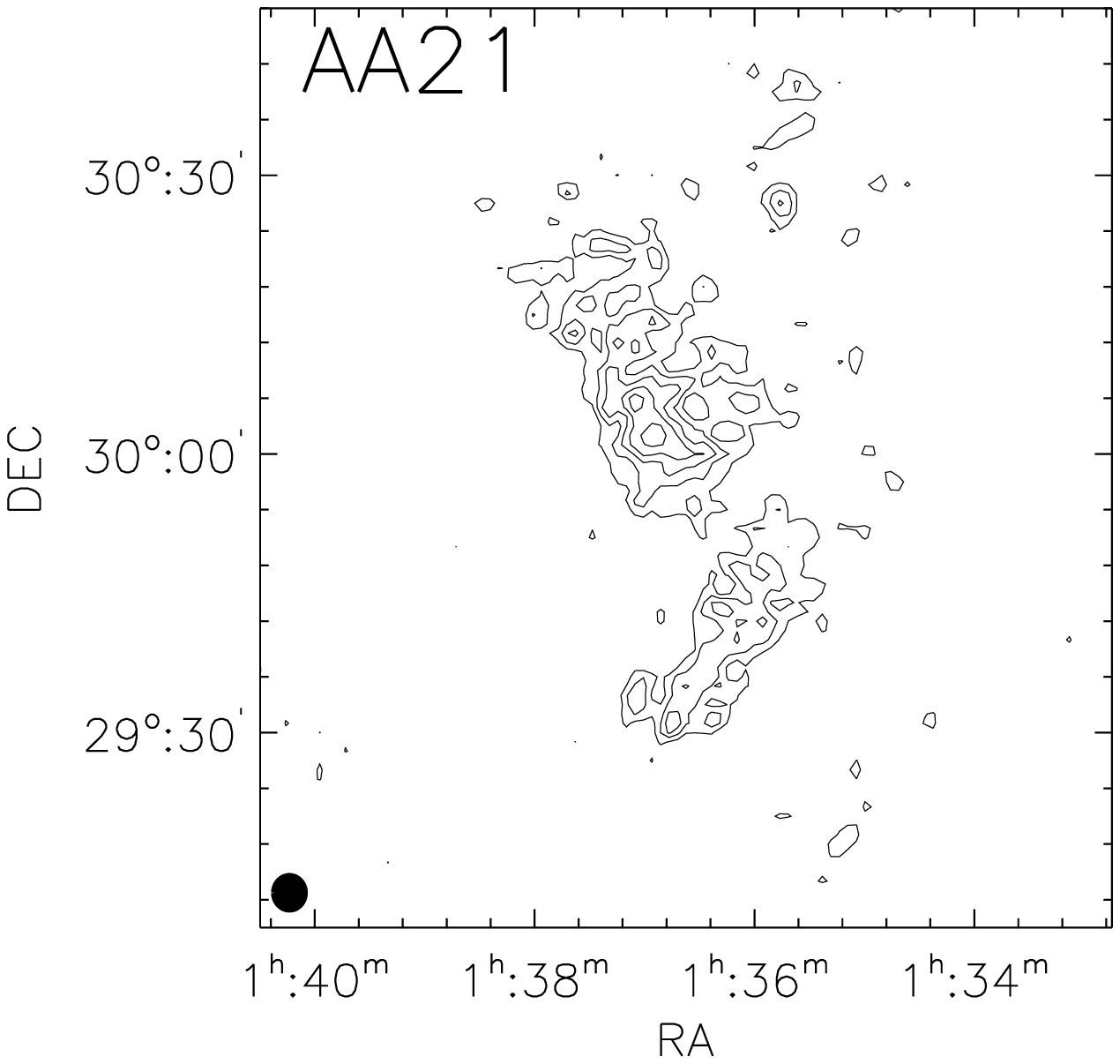}
\hfill
\includegraphics[width=4cm]{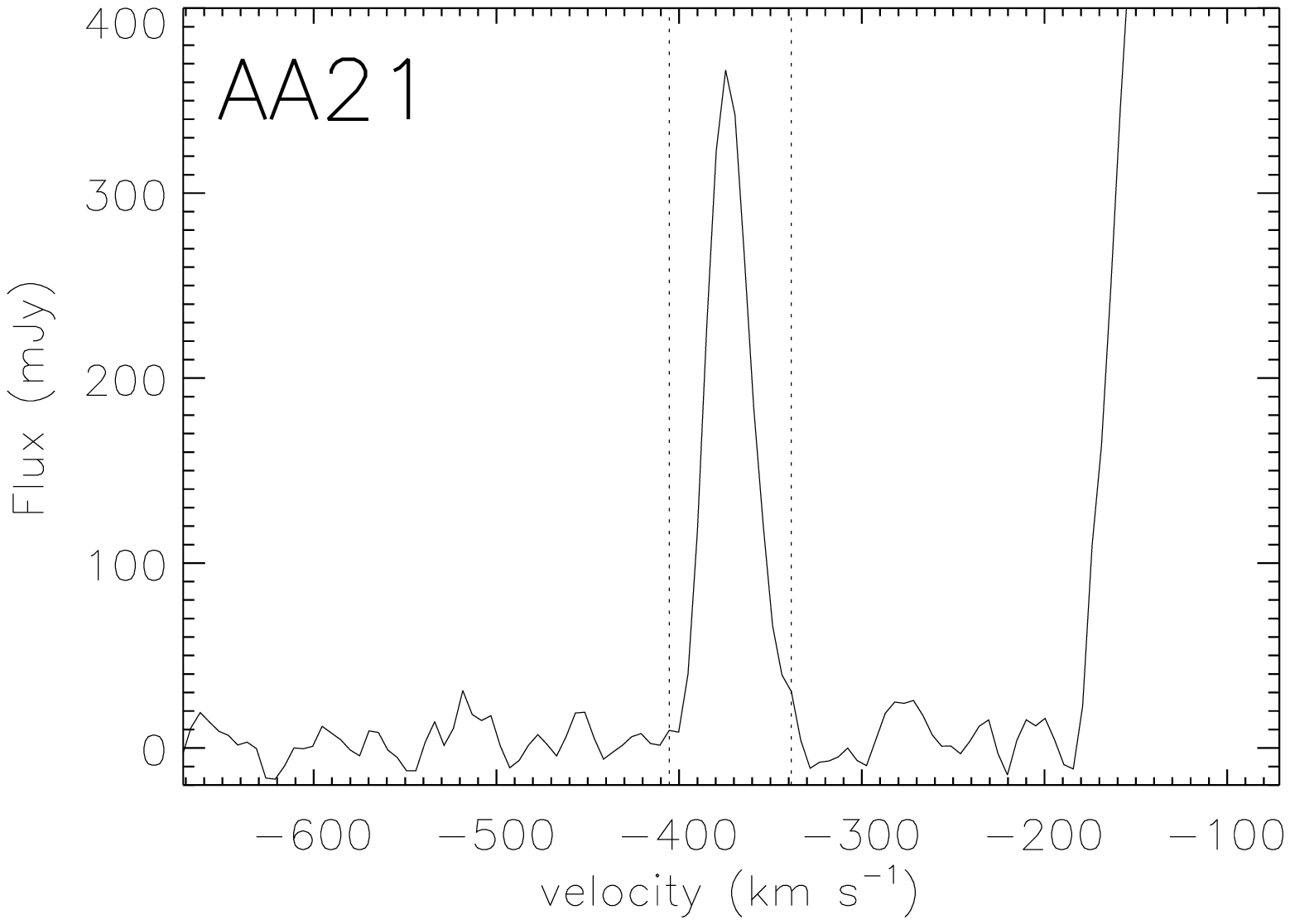}
\label{moment_maps_b} \caption{Column density maps and spectra of {\em Type 2} clouds (see Section 5).}
\end{figure*}

\begin{figure*}
\begin{center}
\includegraphics[width=7cm]{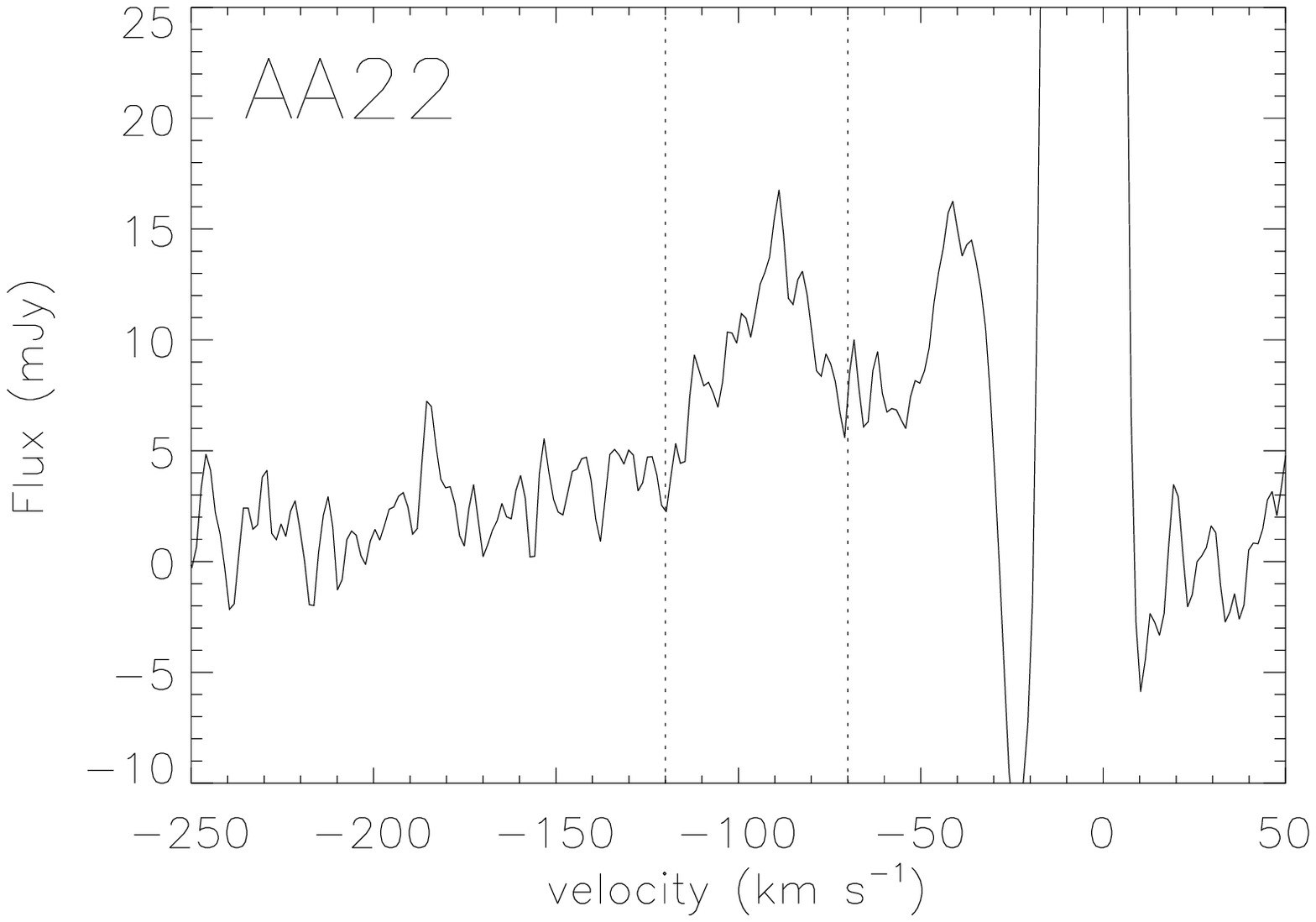}
\includegraphics[width=7cm]{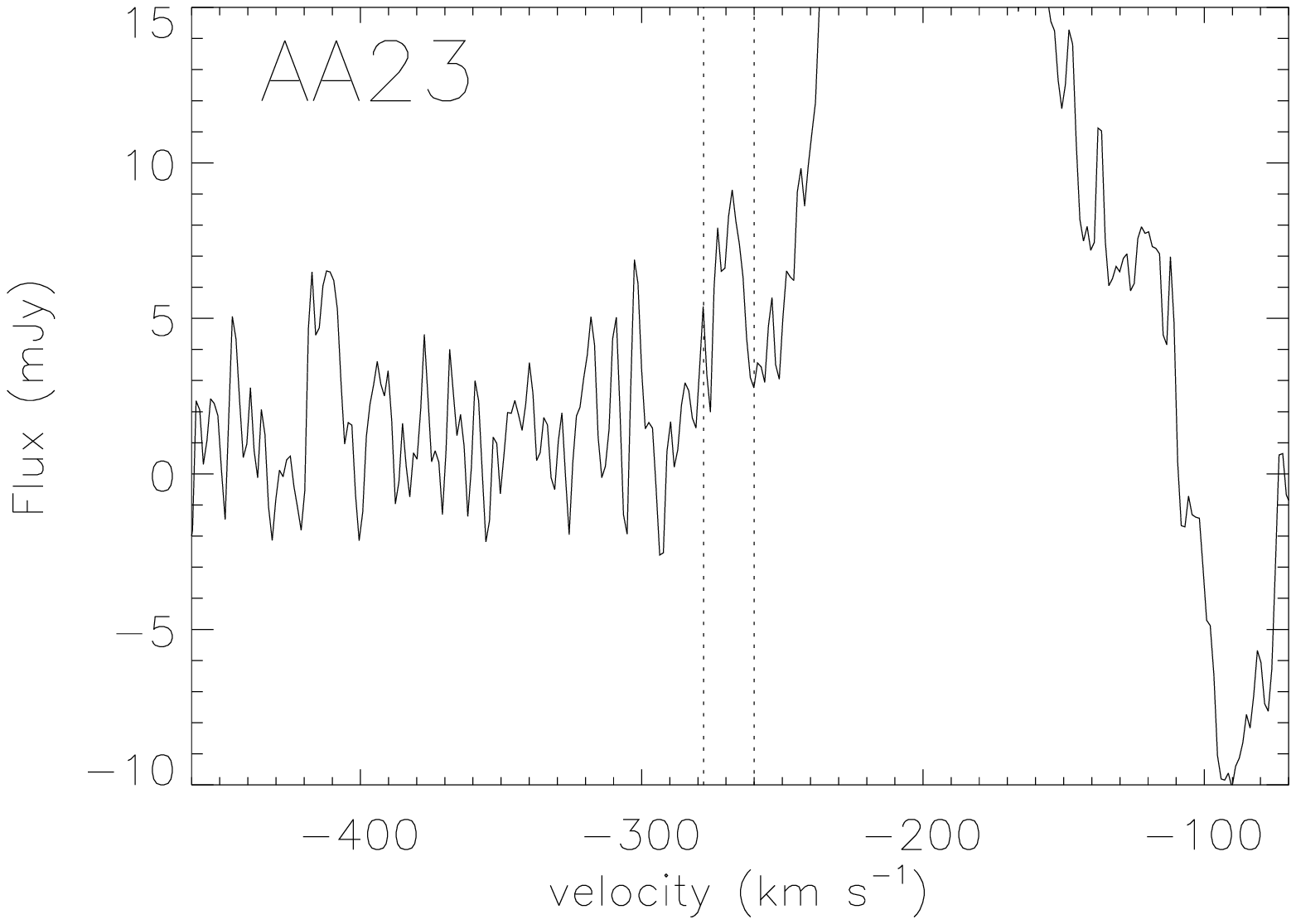}
\label{spectra_carlo} \caption{21-cm spectra of the additional clouds detected in the higher sensitivity data set (see Section 6).}
\end{center}
\end{figure*}

\end{document}